\documentclass[a4paper,11pt]{article}
\pdfoutput=1
\usepackage{jheppub}
\usepackage[utf8]{inputenc}
\usepackage{amsmath,amssymb,mathtools,bm}
\usepackage{graphicx}
\graphicspath{{figures/}}
\usepackage{tikz}
\usetikzlibrary{arrows.meta,calc,decorations.pathreplacing,patterns,positioning}
\usepackage{booktabs,array}
\usepackage{enumitem}
\usepackage{xcolor}
\usepackage{silence}
\usepackage{caption}
\usepackage{subcaption}
\usepackage{slashed}
\usepackage[nameinlink]{cleveref}
\hypersetup{
  colorlinks=true,
  linkcolor=blue!55!black,
  citecolor=red!65!black,
  urlcolor=blue!55!black,
  pdftitle={PEE threads and bit-threads in BTZ black brane and finite-cutoff AdS3},
  pdfauthor={Debarshi Basu, Ashish Chandra and Qiang Wen}
}
\numberwithin{equation}{section}
\allowdisplaybreaks
\newcommand{\dd}{\mathrm{d}}
\newcommand{\cI}{\mathcal{I}}
\newcommand{\cK}{\mathcal{K}}
\newcommand{\cL}{\mathcal{L}}

\newcommand{\cM}{\mathcal{M}}
\newcommand{\cN}{\mathcal{N}}

\newcommand{\eps}{\epsilon}
\newcommand{\zh}{z_h}
\newcommand{\zc}{z_{\mathrm c}}
\newcommand{\sech}{\operatorname{sech}}
\newcommand{\csch}{\operatorname{csch}}
\newcommand{\Htwo}{\mathbb H^2}

\newcommand{\grad}{\nabla}
\newcommand{\abs}[1]{\left|#1\right|}
\newcommand{\e}{\mathrm e}
\newcommand{\del}{\partial}
\newcommand{\Length}{\operatorname{Length}}
\DeclareMathOperator{\arccosh}{arccosh}
\DeclareMathOperator{\arcsinh}{arcsinh}
\DeclareMathOperator{\arctanh}{arctanh}

\definecolor{floworange}{RGB}{235,97,13}
\definecolor{flowblue}{RGB}{31,97,179}
\definecolor{flowpurple}{RGB}{122,74,179}
\definecolor{flowred}{RGB}{199,13,33}
\definecolor{flowgold}{RGB}{225,155,18}
\definecolor{flowgray}{RGB}{45,45,45}

\newcommand{\flowkey}[3]{%
	\tikz[baseline=-0.55ex]
	\draw[#1,line width=#2] (0,0)--(1.55em,0);%
	\hspace{0.35em}#3%
}

\title{PEE threads and bit-threads in BTZ black brane and finite-cutoff AdS$_3$}

\author[a]{Debarshi Basu}
\author[a]{Ashish Chandra}
\author[a]{and Qiang Wen}

\affiliation[a]{Shing-Tung Yau Center and School of Physics, Southeast University,\
Nanjing 210096, China}

\emailAdd{debarshi.128@gmail.com, achandrahep@gmail.com, wenqiang@seu.edu.cn}

\abstract{We develop PEE thread flows for the planar BTZ black brane and finite-cutoff holography. Exact geodesic-distance kernels define source-resolved PEE thread currents whose superposition gives divergenceless bit-thread fields. In BTZ, boundary--boundary and boundary--horizon sectors reproduce the entanglement contour and thermal horizon flux. At finite cutoff, the two-point PEE kernels are smooth and select macroscopic flows that are generically non-geodesic despite being built from geodesic elementary threads. We compare these PEE-selected flows with independent normal-geodesic bit-thread flows and show that they saturate the same RT bottleneck while differing in boundary calibration, endpoint pairing and off-bottleneck structure. The resulting constructions make explicit both the endpoint organization of finite-cutoff entanglement and the nonuniqueness of holographic max flows.}

\begin{document}
\maketitle
\flushbottom

\section{Introduction}
\label{sec:intro}

The AdS/CFT correspondence relates quantum gravity in an asymptotically
anti-de Sitter spacetime to a nongravitational conformal field theory
\cite{Maldacena:1997re,Gubser:1998bc,Witten:1998qj}.  One of its sharpest geometric
statements is the Ryu--Takayanagi (RT) prescription and its covariant
and quantum extensions
\cite{Ryu:2006bv,Ryu:2006ef,Hubeny:2007xt,Lewkowycz:2013nqa,Faulkner:2013ana,Engelhardt:2014gca}.
These results make entanglement central to bulk reconstruction and to the
quantum-error-correcting structure of holography
\cite{VanRaamsdonk:2010pw,Pastawski:2015qua,Hayden:2016cfa,Harlow:2016vwg}.
The RT formula, however, assigns a single number to a region.  It does not by
itself resolve the structure of the correlations contributing to
that number.

Bit threads replace the RT minimization by a Riemannian max-flow problem
\cite{Freedman:2016zud}.  In this language, the entropy of $A$ is the maximum flux of a
divergenceless vector field obeying $|v_A|\leq1/(4G)$.  Explicit geodesic
flows, multiflows, higher-curvature generalizations, and perturbative or
covariant extensions have clarified both the power and the nonuniqueness of
this representation
\cite{Headrick:2017ucz,Agon:2018lwq,Cui:2018dyq,Harper:2018sdd,Headrick:2020gyq,Agon:2020mvu,Headrick:2022nbe}.
The bottleneck fixes the maximal flux but generally does not select a unique
field away from it.

The partial entanglement entropy (PEE) and entanglement contour refine the
same entropy in an endpoint-sensitive way
\cite{Wen:2018whg,Wen:2019iyq,Wen:2020ech,Han:2019scu,Wen:2021qgx,Camargo:2022mme}.  In the vacuum holographic setting, the symmetric two-point PEE can be represented
by endpoint-labelled geodesics whose density is the mixed second derivative
of their length \cite{Lin:2023rxc,Lin:2024fze,Wen:2025gui}.  Fixing one endpoint then
defines a source-resolved PEE thread current, while integrating over sources in $A$
produces a distinguished macroscopic flow.  The microscopic geodesics may
intersect; the integral curves of the resulting vector field do not.

Integral geometry supplies the natural language for this construction.
Differential entropy and kinematic space reconstruct bulk curves, points,
and distances from geodesic data
\cite{Balasubramanian:2013lsa,Headrick:2014eia,Czech:2014ppa,Czech:2014wka,Czech:2015qta}.
For a gravitational subregion, the relevant objects are chords with
endpoints on the actual boundary of the subregion \cite{Basu:2026hbg}.  The
endpoint density obtained from their exact lengths is geometric data.  Once
the state, endpoint support and purification or hard-cutoff interpretation
are fixed as below, we refer to this density as the two-point PEE kernel.
This terminology does not assert a literal Hilbert-space factorization at a
horizon or finite Dirichlet wall, and a main purpose of the present paper is
to keep the exact geometric statement separate from that stronger
microscopic interpretation.

The planar BTZ black brane is the first nontrivial test.  Thermal
entanglement and the eternal-black-hole purification are standard
\cite{Maldacena:2001kr,Hartman:2013qma,Calabrese:2004eu,Calabrese:2009qy}.
Geodesic bit threads in black-brane backgrounds already isolate a horizon
flux equal to the thermal entropy \cite{Agon:2018lwq,Caggioli:2024uza}.  More recently,
two-sided BTZ entanglement threads were used to resolve correlations between
the two asymptotic systems and their relation to PEE
\cite{Lin:2025btz}.  Accordingly, the thermal two-point PEE kernels themselves are not
claimed here as new.  We instead derive an intrinsic push-forward from the
two-sided thermofield-double PEE kernel to the one-sided horizon kernel,
construct the local endpoint-resolved currents in BTZ coordinates, prove
that the boundary-returning and horizon-crossing chambers join into one
analytic field, and identify the separatrix of the \emph{superposed} bit-threads
flow.

The second setting is finite-cutoff holography.  The irrelevant but solvable
$T\bar T$ deformation has exact spectral and thermodynamic structures
\cite{Zamolodchikov:2004ce,Smirnov:2016lqw,Cavaglia:2016oda,Cardy:2018sdv,Datta:2018thy,Aharony:2018bad,Jiang:2019epa,Gu:2025tpy,Gu:2026rck}.
In the classical large-$c$ pure-gravity sector, the hard-cutoff boundary proposal
relates the deformation to a finite radial Dirichlet surface
\cite{McGough:2016lol,Kraus:2018xrn,Shyam:2017znq,Cottrell:2018skz,Hartman:2018tkw,Donnelly:2018bef,Apolo:2023vnm,Apolo:2023ckr,He:2024pbp,He:2019glx}.
The use of RT surfaces at finite cutoff and the role of holographic
counterterms were clarified in \cite{Murdia:2019fax}.  More general
formulations use mixed boundary conditions, and with generic bulk matter a
finite Dirichlet surface is not a universal definition of the complete
deformed theory\footnote{The mixed-boundary viewpoint is also closely tied to the deformed symmetry
	structure and Brown--York stress tensor, while its Chern--Simons formulation
	makes the deformed boundary WZW dynamics and surface-charge algebra explicit
	\cite{He:2019glx,He:2020hhm}.} \cite{Guica:2019nzm}.  Entanglement entropy has been studied
from the deformed field theory, the cutoff geometry and mixed-boundary conditions
formulations, including finite temperature and finite size
\cite{Chen:2018eqk,Jeong:2019ylz,Lewkowycz:2019xse,He:2019vzf,He:2020udl,He:2022xkh,He:2023xnb,Chang:2024voo,Apolo:2023ckr}.
These analyses also expose the ultraviolet nonlocality of the deformation,
for example through boosted strong-subadditivity tests
\cite{Lewkowycz:2019xse}, while recent work identifies a nonperturbative
entanglement scale proportional to $\sqrt\mu$ \cite{Lai:2025thy}.
Finite-cutoff thermofield-double states have an independent bulk construction
\cite{Coleman:2022tfd}.
At finite cutoff, the endpoints are genuine bulk points.  Differentiating
their exact distance changes the two-point PEE density, the normalization of every
elementary current, and the boundary calibration of the macroscopic flow.
It also exposes a distinction hidden in the asymptotic limit: the flow
selected by the two-point PEE kernel need not have geodesic integral curves.
This is not a contradiction, because summing geodesic tangent fields and
then integrating the resulting direction field are different operations.

The results established in this paper are:
\begin{enumerate}[leftmargin=2.1em]
  \item In asymptotic planar BTZ, the boundary--horizon PEE measure is the push-forward of the cross-boundary measure in the standard thermofield-double purification. Its factor of two is a Jacobian, not an orientation convention. The two endpoint chambers generate one smooth
  divergenceless current.
  \item The source superposition reproduces the known geodesic max-flow
  representative \cite{Agon:2018lwq,Caggioli:2024uza}.  Intrinsic BTZ Fermi coordinates give its stream function and prove that the horizon-ending congruence intersects the RT geodesic at
  $|x|=\frac{z_h}{2}\log\,\cosh(2b/z_h)$.  The intervening RT segment therefore carries the thermal flux.
  \item In finite-cutoff geometries, the exact two-point PEE kernels obtained from geodesic chords generate explicit elementary PEE thread currents and norm-bounded max flows. The PEE-calibrated flow is generically non-geodesic when its RT endpoints are finite.  A universal Fermi-coordinate description gives its exact integral curves and isolates this finite-focus bending.  Under smoothness, geodesicity, RT saturation, and single-crossing assumptions, flux conservation uniquely fixes the representative within the chosen RT-normal congruence.  It has a different cutoff-boundary density and endpoint map; its smooth completion contains entropy-invisible complement-to-complement spectator curves.
  \item In cutoff BTZ, the cutoff boundary and horizon fluxes reproduce their PEE
  contour densities pointwise. The two positive PEE kernels obey an exact local PEE-capacity identity. At fixed bulk $z_h$, the cutoff--horizon kernel tends distributionally to a delta function as $z_c\to z_h$, while the cutoff--cutoff sector vanishes.
\end{enumerate}

For orientation, table~\ref{tab:relation-prior-work} separates the ingredients
we use from the refinements developed here.  The distinction is particularly
important in the thermal section, where the existence of a horizon-carrying
max flow is known, while our emphasis is on its endpoint-resolved PEE
realization.
\begin{center}
  \footnotesize
  \begin{tabular}{>{\raggedright\arraybackslash}p{0.25\textwidth}
                  >{\raggedright\arraybackslash}p{0.31\textwidth}
                  >{\raggedright\arraybackslash}p{0.34\textwidth}}
    \toprule
    Setting & Established ingredient & Refinement developed here \\
    \midrule
    Vacuum PEE threads
    & Endpoint-labelled geodesics, two-point PEE measure and source currents
      \cite{Lin:2023rxc,Lin:2024fze}
    & Extension to finite geometric endpoints and explicit source-superposed
      max flows \\
    Planar BTZ
    & Geodesic thermal max flows and horizon flux \cite{Caggioli:2024uza}
    & Intrinsic PEE thread currents, exact RT separatrix and pointwise flux audit \\
    Two-sided thermal purification
    & Cross-boundary entanglement threads \cite{Lin:2025btz}
    & Measure-preserving TFD push-forward to the one-sided horizon PEE sector \\
    Finite-cutoff $T\bar T$ holography
    & Interval entropies, RT/replica descriptions and mixed-boundary formulations
      \cite{Donnelly:2018bef,Murdia:2019fax,Lewkowycz:2019xse,He:2023xnb,Chang:2024voo,Apolo:2023vnm,Apolo:2023ckr}
    & Two-point PEE kernels, elementary currents, non-geodesic PEE-selected
      max flows and local cutoff-BTZ PEE-capacity identities \\
    \bottomrule
  \end{tabular}
  \captionof{table}{Relation of the present constructions to ingredients
  already available in the literature.  The table is intended to delimit the
  novelty claims, not to be an exhaustive review.}
  \label{tab:relation-prior-work}
\end{center}

All statements are made on static slices of classical Einstein gravity with
AdS radius one, no bulk matter, and connected interval RT surfaces.  The
BTZ spatial direction is noncompact; compact quotients require image
minimization and can obstruct a global geodesic flow
\cite{Caggioli:2024uza}.  Once the purification and hard-cutoff
assumptions have been stated, we use the shorter terminology ``two-point PEE
kernel'' or ``PEE density'' for the corresponding endpoint measure.  The
underlying geometric identities remain valid independently of that
field-theory interpretation.

In section~\ref{sec:framework}, we fix these conventions and the logical status
of the construction.  Section~\ref{sec:BTZ-threads} treats the asymptotic
planar BTZ geometry, including the TFD push-forward, local currents, exact
separatrix, and contour fluxes.  Section~\ref{sec:finite-cutoff} proceeds
geometry-by-geometry through finite-cutoff Poincar\'e AdS$_3$ and BTZ black brane.  Section~\ref{sec:summary} isolates the new results,
limitations, and future directions.  Appendix~\ref{AppA} records the
covering-space embedding formulae used for the geodesic distances,
appendix~\ref{app:geodesic-curvature} derives the invariant curvature test,
and appendix~\ref{app:cutoff-fermi-nongeodesic} places the PEE-selected and
normal-geodesic finite-cutoff flows in one universal Fermi chart.  The
remaining appendices collect the asymptotic-BTZ normal geodesic construction and the
finite-cutoff global-AdS specialization.

\section{Review and framework}
\label{sec:framework}

This section reviews the ingredients that enter the constructions below and,
at the same time, fixes the distinctions that will be important in the
thermal and finite-cutoff examples.  There are three logically separate
objects.  The RT surface is a geometric bottleneck that fixes the entropy.
A bit-thread field is a macroscopic max-flow representative of that entropy.
A PEE thread is instead an endpoint-labelled microscopic chord weighted by a
two-point correlation density.  In the examples studied here the weighted
chords can be superposed to select a particular max flow, but the individual
PEE threads are not themselves the integral curves of the superposed vector
field.  Keeping these levels separate prevents endpoint data, flow lines and
purification-dependent interpretations from being conflated.

\subsection{RT surfaces and the Riemannian flow formulation}
\label{subsec:review-bit-threads}

Let $\cM$ be a static bulk time slice and let $A$ be a region of an allowed
boundary component.  In classical Einstein gravity, with the AdS radius set
to one throughout this paper, the Ryu--Takayanagi prescription reads
\begin{equation}
	S_A=\frac{\Length(\gamma_A)}{4G},
	\qquad
	\partial\gamma_A=\partial A,
	\qquad
	\gamma_A\sim A,
	\label{eq:review-RT}
\end{equation}
where the last relation denotes the appropriate relative homology condition
\cite{Ryu:2006bv,Ryu:2006ef}.  The surface $\gamma_A$ is codimension one in
$\cM$ and therefore separates the $A$ side of the slice from its complement
in all connected phases considered below.

The Riemannian max-flow/min-cut theorem gives the equivalent bit-thread
formulation \cite{Freedman:2016zud,Headrick:2017ucz}.  A feasible flow is a
vector field satisfying
\begin{equation}
	\nabla_\mu v^\mu=0,
	\qquad
	|v|\leq\frac{1}{4G},
	\label{eq:review-feasible-flow}
\end{equation}
with vanishing normal component on boundary pieces that are not declared to
be sources or sinks.  Its flux out of $A$ is
\begin{equation}
	\Phi_A(v)=\int_A v\cdot n_A\,\dd\Sigma,
	\qquad
	S_A=\max_v\Phi_A(v)
	=\frac{\Length(\gamma_A)}{4G}.
	\label{eq:review-MFMC}
\end{equation}
The norm bound is a local capacity constraint.  Since every unit of flux
leaving $A$ must cross a homologous cut, the RT surface provides the narrowest
available channel.  A maximizing field is normal to $\gamma_A$ and saturates
the bound there almost everywhere.  Away from this bottleneck, however, the
field can be redistributed while remaining divergenceless and norm bounded.
The entropy consequently fixes a maximal flux, not a unique bulk vector
field.

The familiar thread picture is obtained by following integral curves of a
feasible vector field and assigning them a transverse density.  The curves
carry an orientation inherited from the sign of $v$; reversing every curve
changes the bookkeeping direction but not the unoriented flux capacity.  A
regular vector field has nonintersecting integral curves because the local
streamline equation has a unique solution.  This statement is distinct from
the behavior of a collection of independently specified geodesics, which may
intersect freely.  Zeros of the field, boundary sources and deliberately
introduced support boundaries require the usual weak or distributional
qualification.

\subsection{Partial entanglement entropy and the entanglement contour}
\label{subsec:review-pee-bit}

The partial entanglement entropy refines a total entropy into additive
contributions assigned to ordered subsets
\cite{Wen:2018whg,Wen:2019iyq,Wen:2020ech}.  For a decomposition
$A=A_L\cup\alpha\cup A_R$ along a one-dimensional spatial slice, the additive
linear-combination prescription gives
\begin{equation}
	s_A(\alpha)
	=\frac12\left[
	S_{A_L\alpha}+S_{\alpha A_R}-S_{A_L}-S_{A_R}
	\right].
	\label{eq:review-ALC}
\end{equation}
When $\alpha$ is refined into adjacent pieces, the intermediate entropies
cancel telescopically.  This makes additivity manifest and ensures that the
sum over a complete partition returns $S_A$.  In the continuum one writes
$s_A(\alpha)=\int_\alpha s_A(u)\,\dd u$, where $s_A(u)$ is the entanglement
contour in the chosen spatial coordinate.  The contour is a density rather
than a scalar: under $u\mapsto\tilde u(u)$, the invariant quantity is
$s_A(u)\dd u$.

For a pure state on $\Sigma=A\cup\bar A$, the same refinement can be encoded
by a symmetric two-point PEE density,
\begin{equation}
	\cI(A,B)=\int_A\dd u\int_B\dd v\,\cI(u,v),
	\qquad
	\cI(u,v)=\cI(v,u),
	\qquad
	S_A=\cI(A,\bar A).
	\label{eq:PEE-additivity}
\end{equation}
The contour is then the amount paired with the complement,
\begin{equation}
	s_A(u)=\int_{\bar A}\dd v\,\cI(u,v),
	\qquad
	S_A=\int_A\dd u\,s_A(u).
	\label{eq:review-contour-from-two-point}
\end{equation}
Neither $\cI(u,v)$ nor $s_A(u)$ should be interpreted as the entropy of an
infinitesimal site.  Only the measures $\cI(u,v)\dd u\dd v$ and
$s_A(u)\dd u$ are invariant, and the two-point PEE is not generally equal to
the mutual information of two infinitesimal regions.  Its physical role is
to resolve which ordered endpoint pairs contribute to the additive entropy
budget.  Balanced and mixed-state variants extend this bookkeeping beyond a
single pure-state bipartition, but they will not be needed explicitly here
\cite{Wen:2021qgx,Camargo:2022mme}.

For later use, there is a particularly simple specialization in a
translationally invariant state.  If the entropy of a single interval of
length $\ell$ is $S(\ell)$, the continuum ALC prescription gives
\begin{equation}
  s_{[u,v]}(x)=\frac12\left[S'(x-u)+S'(v-x)\right].
  \label{eq:review-ALC-translational}
\end{equation}
When the same-boundary two-point PEE depends only on the separation
$r=|u-v|$, its density is
\begin{equation}
  \cI(r)=-\frac12S''(r).
  \label{eq:review-PEE-second-derivative}
\end{equation}
These formulas provide a field-theory check on the one-point PEE contour
whenever $S(\ell)$ is known independently.  In thermal mixed states an
additional purifier sector can contribute to the full contour, as will occur
for BTZ.

\subsection{Kinematic space, Crofton data and PEE threads}
\label{subsec:review-kinematic-space}

Integral geometry supplies the bridge between endpoint data and bulk chords
\cite{Balasubramanian:2013lsa,Headrick:2014eia,Czech:2014ppa,Czech:2014wka,Czech:2015qta}.
Let $\cL(u,v)$ be the length of the geodesic with ordered endpoints $(u,v)$ on
the relevant geometric boundary.  Its kinematic two-form and the associated
Crofton density are
\begin{equation}
	\omega_{\cK}
	=\left|\frac{\partial^2\cL(u,v)}{\partial u\,\partial v}\right|
	\dd u\wedge\dd v,
	\qquad
	\cI_{\rm C}(u,v)
	=\frac{1}{8G}
	\left|\frac{\partial^2\cL(u,v)}{\partial u\,\partial v}\right|.
	\label{eq:PEE-kernel-general}
\end{equation}
The factor $1/(8G)$ compensates for the two orientations of the same
unoriented chord in ordered endpoint space.  In the vacuum asymptotic-CFT
examples, $\cI_{\rm C}$ agrees with the two-point PEE and a geodesic weighted
by $\cI_{\rm C}(u,v)\dd u\dd v$ is called a PEE thread
\cite{Lin:2023rxc,Lin:2024fze}.  The orientation $u\to v$ specifies which
endpoint is treated as the source in a local current; it does not make the
underlying correlation asymmetric.

For disjoint endpoint sets,
\begin{equation}
	\cI(R_1,R_2)
	=\int_{R_1}\dd u\int_{R_2}\dd v\,\cI(u,v).
	\label{eq:regional-pee}
\end{equation}
If every relevant chord crosses a test curve $\Sigma$ once, its length is
recovered from the Crofton formula
\begin{equation}
	\Length(\Sigma)
	=\frac14\int_{\cK_\cM}
	\#(\gamma\cap\Sigma)\,\omega_{\cK}.
	\label{eq:Crofton-length}
\end{equation}
The intersection number is essential.  In disconnected phases it cannot be
replaced by a simple endpoint pairing, which is why the present paper
restricts the explicit flow constructions to connected interval phases
\cite{Lin:2023rxc}.

The geometric statement in \eqref{eq:PEE-kernel-general} is broader than its
field-theory interpretation.  It remains meaningful when one or both
endpoints lie on a horizon or a finite Dirichlet wall.  Interpreting the
resulting invariant endpoint measure as a two-point PEE additionally requires
a choice of state, endpoint support and, when relevant, purification or a
hard-cutoff boundary interpretation.  Once those assumptions have been
stated, we use the shorter language ``two-point PEE kernel'' or ``PEE
density'' throughout.

\subsection{From endpoint-resolved currents to a macroscopic flow}
\label{subsec:framework-flows}

Fix one endpoint $u$ and let $Q$ lie on the chord $\gamma_{uv}$.  The
source-resolved current is defined by
\begin{equation}
	V_u^\mu(Q)=|V_u(Q)|\,\tau_{uv}^\mu(Q),
	\qquad
	g_{\mu\nu}\tau_{uv}^\mu\tau_{uv}^\nu=1,
	\label{eq:elementary-current-general}
\end{equation}
where the magnitude is fixed by local flux matching,
\begin{equation}
	|V_u|\,(n\cdot\tau_{uv})\,\dd\Sigma
	=\cI(u,v)\,\dd v.
	\label{eq:flux-matching-general}
\end{equation}
Thus a small transverse segment intercepts precisely the PEE weight carried
by the corresponding endpoint interval.  In the open bulk $V_u$ is
divergenceless.  Its source and target appear only as boundary distributions;
equivalently, one imposes $\nabla_\mu V_u^\mu=0$ on $\cM^\circ$ together with
the prescribed normal flux on $\partial\cM$.

The current associated with a finite region is the linear superposition
\begin{equation}
	v_A^\mu(Q)=\int_A\dd u\,V_u^\mu(Q).
	\label{eq:superposition-general}
\end{equation}
Linearity guarantees $\nabla\cdot v_A=0$ in the interior, but it does not by
itself prove the norm bound or saturation of the RT surface.  These are
independent max-flow tests and will be checked for every explicit field.

On a two-dimensional oriented slice, divergencelessness can be made
particularly transparent by introducing a stream function,
\begin{equation}
	v_A=\frac{1}{4G}\star\dd\Psi_A.
	\label{eq:review-stream-function}
\end{equation}
The integral curves are the level sets of $\Psi_A$.  For an interval whose
endpoints are the finite foci $p_-$ and $p_+$, source integration telescopes
to a distance-difference potential of the form
\begin{equation}
	\Psi_A(X)=\frac12\left[\cL(X,p_+)-\cL(X,p_-)\right],
	\label{eq:review-distance-difference}
\end{equation}
up to an overall orientation convention.

In \cite{BasuWen:BCFTMinPurification}, we develop the same source-resolved superposition for
extended state-supporting surfaces in AdS/BCFT and in the surface/state correspondence inspired minimal purification.  Appendix A of \cite{BasuWen:BCFTMinPurification} shows that, after source
integration, these currents admit a common endpoint reduction in terms of
differences of finite-point distance functions, with Busemann functions
replacing distances when an endpoint is ideal.  Thus
\eqref{eq:review-stream-function}--\eqref{eq:review-distance-difference} are
the two-finite-focus specialization of the more general geometric
construction developed there.

The norm bound then follows from the eikonal property of each distance
function, while the linear behavior of $\Psi_A$ along the RT segment gives
pointwise saturation.  The detailed coordinate realizations, including the
finite-endpoint Fermi form, are derived later and in
appendix~\ref{app:cutoff-fermi-nongeodesic}.



\subsection{Extended geometric boundaries and purification}
\label{subsec:review-subregion-kinematic}

For a connected gravitational subregion $\mathfrak a\subset\cM$, the natural
kinematic elements are chords contained in $\mathfrak a$ with endpoints on
its full geometric boundary $\partial\mathfrak a$ \cite{Basu:2026hbg}.  The
endpoint measure can be understood directly from ordinary integral geometry,
without assuming a microscopic interpretation of the finite boundary.  Let
$s_1,s_2$ be proper coordinates on a smooth convex component of
$\partial\mathfrak a$.  The first variation of a chord length gives
$\partial_{s_1}\cL_{\mathfrak{a}}=-\cos\vartheta_1$, where $\vartheta_1$ is the angle
between the chord and the positively oriented boundary tangent.  A second
endpoint variation gives the Jacobian from the endpoint coordinates
$(s_1,s_2)$ to the Liouville measure on oriented geodesics.  Equivalently,
the invariant chord measure is
\begin{equation}
  \dd\Gamma_{\mathfrak a}
  =\frac12\left|
  \frac{\partial^2\cL_{\mathfrak a}(s_1,s_2)}
  {\partial s_1\partial s_2}\right|
  \dd s_1\dd s_2.
  \label{eq:subregion-kinematic-measure}
\end{equation}
The factor $1/2$ removes the double counting of the two orientations of one
unoriented chord.  Thus \eqref{eq:subregion-kinematic-measure} is simply the
invariant geodesic measure written in endpoint variables; the subregion
kinematic construction explains how to organize it when several boundary
components or chord segments are present.  It treats an asymptotic boundary,
a finite cutoff boundary, an RT surface and a bifurcation horizon on the same
geometric footing, but it does not by itself assign those components
independent microscopic Hilbert spaces\footnote{In \cite{BasuWen:BCFTMinPurification}, we implement this extended-boundary viewpoint
		explicitly for AdS/BCFT and for the surface/state minimal purification.  EOW
		brane segments and selected RT-surface segments are treated as additional
		PEE endpoint supports; in the minimal-purification construction, the latter
		become purifying degrees of freedom and their PEE-thread superposition
		reproduces a max flow whose bottleneck is the entanglement-wedge cross
		section.  This motivates the conservative convention used here: the
		geometric endpoint support is specified first, while its microscopic PEE
		interpretation is tied to the chosen boundary condition or purification.}.

For the one-sided BTZ exterior we use a relative flow problem in which the
asymptotic or cutoff boundary and the bifurcation line are allowed flux
boundaries.  The horizon is a geometric sink after the two-sided slice has
been cut open.  Its field-theory meaning is supplied by the standard
reflection-symmetric thermofield-double purification.  Sections~\ref{subsec:BTZ-TFD-map} and \ref{subsec:cutoff-BTZ} show that a two-sided
cross-boundary chord folds to a one-sided boundary--horizon chord, including
the Jacobian of the endpoint measure.  In this precise sense a
horizon-ending thread represents correlations with the second asymptotic
system.  It need not be regarded as a microscopic thread literally created
at the horizon, and another purification need not preserve the same
endpoint decomposition.

A finite Dirichlet wall is different.  Its points are genuine bulk points at
finite mutual distance, so the two-point PEE kernel loses the short-distance
singularity of the asymptotic theory.  Within hard-cutoff holography we treat
the Dirichlet wall as the cutoff boundary supporting the deformed-theory PEE
contour.  This is an additional physical assumption, not a consequence of
the Crofton formula alone.  Related island-phase PEE-thread constructions use
finite cutoff spheres as geometric endpoint regulators while leaving the
underlying asymptotic PEE-thread distribution unchanged
\cite{Wen:2024uwr,Basu:2023wmv,Lin:2023ajt}.  That setup is conceptually adjacent but distinct
from the hard radial cutoff studied here, where differentiating the exact
finite-endpoint distance changes the two-point PEE kernel itself.

The same distinction controls spectator curves.  A smooth congruence defined
around the complete geodesic containing $\gamma_A$ may include streamlines
that never intersect the physical RT segment and have both endpoints in
$A^c$.  Such curves can complete a smooth divergenceless representative,
but they carry no flux out of $A$ and do not define an additional PEE sector.
They may be removed by restricting the flow to a support bounded by
streamlines.  The cutoff examples and appendix
\ref{app:cutoff-fermi-nongeodesic} make this support freedom explicit.

\subsection{Finite-cutoff holography and  \texorpdfstring{$T\bar T$}{}}
\label{subsec:review-ttbar}
The $T\bar T$ deformation provides a particularly useful laboratory for
irrelevant deformations of two-dimensional quantum field theories.  Despite
being generated by an operator of dimension four, the deformation retains a
remarkable amount of analytic control.  The basic reason is the special
factorization property of the composite stress-tensor operator discovered by
Zamolodchikov.  In a translationally invariant state one has, schematically,
\begin{equation}
	\big\langle T\bar T\big\rangle
	=
	\langle T\rangle\,\langle\bar T\rangle
	-
	\langle\Theta\rangle^2 ,
	\label{eq:TTbar-factorization}
\end{equation}
where $\Theta$ denotes the mixed component of the stress tensor.  The
coincident-point composite may therefore be defined without introducing the
usual state-dependent short-distance ambiguities, and expectation values of
the deforming operator can be expressed directly in terms of one-point
functions of the stress tensor
\cite{Zamolodchikov:2004ce,Smirnov:2016lqw,Cavaglia:2016oda}.  We use the
covariant normalization\footnote{Our operator is four times the normalization employed in the original
	finite-cutoff proposal of \cite{McGough:2016lol}.}
\begin{equation}
	\frac{\partial I_\mu}{\partial\mu}
	=
	\int\dd^2x\,\sqrt{\gamma}\,(T\bar T)_\mu,
	\qquad
	(T\bar T)
	=
	\frac12
	\left(
	T^{ij}T_{ij}-(T^i_{\ i})^2
	\right).
	\label{eq:TTbar-flow}
\end{equation}
A central manifestation of this solvability is the exact flow of the
finite-volume spectrum.  Writing
$t=4\mu$, an energy eigenvalue on a spatial circle of
circumference $L$ obeys the inviscid Burgers equation
\begin{equation}
	\frac{\partial E_n(L,t)}{\partial t}
	=
	E_n(L,t)\,\frac{\partial E_n(L,t)}{\partial L}
	+
	\frac{P_n^2}{L},
	\qquad
	P_n=\frac{2\pi k_n}{L},
	\label{eq:TTbar-Burgers}
\end{equation}
with the undeformed spectrum supplying the initial condition at $t=0$
\cite{Smirnov:2016lqw,Cavaglia:2016oda}.  When the seed theory is a CFT, the
flow can be integrated explicitly,
\begin{equation}
	E_n(L,t)
	=
	\frac{L}{2t}
	\left[
	\sqrt{
		1+\frac{4t}{L}E_n^{(0)}(L)
		+\frac{4t^2}{L^2}P_n^2
	}
	-1
	\right],
	\label{eq:TTbar-spectrum}
\end{equation}
where the branch continuously connected to the undeformed theory has been
chosen.  Thus the deformation reorganizes the entire spectrum
nonperturbatively even though the perturbing operator is irrelevant.  The
same structure extends to the torus partition function: modular properties
together with the spectral flow strongly constrain, and perturbatively fix,
the deformed theory
\cite{Datta:2018thy,Aharony:2018bad,He:2020udl}.
Perturbative correlation functions, including higher-point functions and
entanglement/chaos observables, have also been developed systematically
\cite{He:2019vzf,He:2022xkh}.  The square-root structure of
\eqref{eq:TTbar-spectrum} is also a reminder that the deformation is not an
ordinary UV-complete local QFT.  Depending on the sign of the coupling and
the branch under consideration, one encounters a limiting scale, complex
energies, or Hagedorn-type high-energy behavior.  These features are part of
the physics of the deformation rather than artifacts of perturbation theory
\cite{Aharony:2018bad,Cardy:2018sdv}.

For holographic CFTs, the stress-tensor form of the flow suggests a direct
geometric interpretation.  The Hamilton--Jacobi equation governing radial
evolution of the on-shell AdS$_3$ gravitational action has precisely the
quadratic stress-tensor structure required by the $T\bar T$ flow.  This led
to the proposal that, in the large-$c$ classical-gravity regime, deforming
the boundary CFT by $T\bar T$ is equivalent to removing the asymptotic
portion of AdS$_3$ and placing the theory on a timelike Dirichlet surface at
finite radial position
\cite{McGough:2016lol,Shyam:2017znq}.  The Brown--York tensor evaluated on
that surface then reproduces the deformed energy spectrum and thermodynamic
relations of the boundary theory.  Perturbative stress-tensor correlators
provide further checks of the correspondence in the pure-gravity sector
\cite{Kraus:2018xrn}.  A complementary holographic completion for the opposite
sign/branch is furnished by glue-on AdS, where an auxiliary AdS$_3^*$ patch
is attached to the asymptotic spacetime rather than terminating the geometry
at a Dirichlet surface \cite{Apolo:2023vnm}.  The associated signed-area
extremal-surface prescription has been analyzed in Poincar\'e AdS$_3$, global
AdS$_3$, and BTZ \cite{Apolo:2023ckr}.  We therefore regard the
hard-cutoff and glue-on constructions as complementary branches of
$T\bar T$ holography; only the hard-cutoff branch is used in the calculations
below.  For unit AdS radius and a Poincar\'e cutoff
$z=z_c$, our conventions give
\begin{equation}
	\mu
	=
	4\pi G\,z_c^2
	=
	\frac{6\pi}{c}\,z_c^2,
	\qquad
	z_c^2=\frac{c\mu}{6\pi},
	\qquad
	c=\frac{3}{2G}.
	\label{eq:TTbar-cutoff-dictionary}
\end{equation}
The conformal boundary is recovered as $\mu\to0$, or equivalently
$z_c\to0$.

It is important, however, not to interpret the finite radial cutoff as a
completely universal definition of holographic $T\bar T$.  A more precise
large-$c$ dictionary formulates the deformation as a change of the
asymptotic boundary conditions for the metric.  In pure AdS$_3$ gravity
these mixed boundary conditions can, on shell and on the appropriate
branch, be re-expressed as Dirichlet data on a finite radial slice.  Once
bulk matter is included this equivalence is generally lost: the matter
boundary conditions remain asymptotic, and reproducing a literal finite
bulk cutoff requires additional operator deformations
\cite{Cottrell:2018skz,Kraus:2018xrn,Guica:2019nzm,Hartman:2018tkw}.
Consequently, ``$T\bar T$ deformation'' and ``hard radial cutoff'' should not
be regarded as interchangeable notions in complete generality.  Their
identification is sharpest in the semiclassical, large-$c$, locally
AdS$_3$, pure-gravity sector that is relevant for the geometries considered
below.  The use of RT surfaces at finite cutoff, including the role of
holographic counterterms, was clarified in \cite{Murdia:2019fax}.  In the
mixed-boundary-condition formulation, BTZ entanglement entropy can also be
derived directly from the RT surface or from Chern--Simons/Wilson-line
methods \cite{He:2023xnb}.

The entanglement problem has been analyzed from several complementary
directions.  Perturbative and finite-cutoff computations were developed in
\cite{Chen:2018eqk,Jeong:2019ylz,He:2019vzf,He:2020udl}; all-orders large-$c$ comparisons and
subregion encoding were studied in \cite{Lewkowycz:2019xse}; entanglement and
modular-Hamiltonian corrections were further explored in
\cite{He:2022xkh}; and analytic finite-size and
finite-temperature formulas were developed in
\cite{He:2023xnb,Chang:2024voo,Apolo:2023ckr}.  These studies make clear that $T\bar T$ is nonlocal
in the ultraviolet.  In particular, relatively boosted regions can violate
the boosted strong-subadditivity criterion even though ordinary static
interval concavity remains intact \cite{Lewkowycz:2019xse}.  More recently,
replica analysis has identified a nonperturbative entanglement length
proportional to $\sqrt\mu$ \cite{Lai:2025thy}.  The finite-resolution scale
extracted from our positive cutoff PEE kernels will be compared with this
result, while keeping the two mechanisms conceptually distinct.

This qualification is particularly useful for entanglement observables.
In the hard-cutoff description, the endpoints of a boundary interval are
moved from the asymptotic boundary to genuine bulk points on the cutoff
surface, and the RT entropy is computed by the regulated geodesic joining
those finite-radius endpoints
\cite{Donnelly:2018bef,Chen:2018eqk,Jeong:2019ylz}.  The change is therefore
not merely a replacement of the UV regulator in the CFT formula: finite
endpoint separation in the bulk modifies the geodesic distance itself.
In particular, the asymptotic short-distance singularity is softened, and
the derivatives of the finite-endpoint geodesic length define a genuinely
cutoff-dependent two-point PEE kernel.  This is precisely the piece of the
$T\bar T$/finite-cutoff correspondence used in the present work.

Accordingly, throughout sections~\ref{sec:finite-cutoff} and the associated
appendices we make a deliberately restricted identification.  The
finite-radius Dirichlet surfaces define exact classical geometric problems
in AdS$_3$ and BTZ, while $T\bar T$ supplies their boundary interpretation
within the large-$c$ pure-gravity regime described above.  Our construction
of endpoint-resolved PEE kernels and bit-thread flows requires only the
finite-distance bulk geometry and does not assume that a hard cutoff provides
a universal nonperturbative definition of the deformed QFT.  This separation
will be important when interpreting the cutoff-dependent thread density,
the spectator sectors, and the finite-temperature BTZ results below.


\section{PEE threads and bit-threads in the planar BTZ black brane}
\label{sec:BTZ-threads}
In this section, we construct explicit PEE threads in the BTZ black brane geometry, extending the framework developed in the Poincar\'e AdS$_3$ patch in \cite{Lin:2023rxc} to a finite-temperature setting.
Finite temperature changes the PEE-thread network in an essential way.  A
constant-time slice of the planar BTZ exterior becomes a hyperbolic surface
with two geometric boundary components when it is cut open at the
bifurcation surface: the regulated asymptotic boundary and the horizon.
Consequently, a geodesic emitted from the asymptotic
boundary can cross the reference RT surface and either return to the same
asymptotic boundary or terminate on the horizon.  The second possibility is
geometrically well defined, but its interpretation as a PEE involving a
purifying system must not be assumed from the one-sided geometry alone.  We
will derive that interpretation from the thermofield-double purification in
subsection~\ref{subsec:BTZ-TFD-map}.

This section separates three logically distinct steps.  We first identify the
allowed endpoint sectors and their two-point PEE kernels.  We then show
that the boundary--horizon kernel is precisely the push-forward of the
two-sided thermofield-double kernel.  Finally, we turn the endpoint data into
local divergenceless PEE thread flow and superpose the boundary-sourced currents to
obtain the PEE-selected bit thread flow.  Note that the geodesics are microscopic PEE threads and may intersect, whereas the integral curves of the superposed bit-thread field constitute a
different, non-intersecting congruence wherever the field is regular and nonzero.

On a constant-time slice the planar BTZ geometry
\cite{Banados:1992wn,Maldacena:2001kr} has the following metric
\begin{equation}
  \dd s^2=\frac{1}{z^2}\left(\dd x^2+\frac{\dd z^2}{1-z^2/\zh^2}\right),
  \qquad 0<z\leq\zh,
  \label{eq:BTZ-slice}
\end{equation}
where the spatial direction $x$ is non-compact, the horizon is at $z=\zh$, and the inverse temperature of the dual field theory is given by
$\beta=2\pi\zh$.  

As in the Poincaré case described in \cite{Lin:2023rxc}, we wish to determine the PEE threads emanating from the origin $x = 0$ on the asymptotic boundary and extend the construction by translation along the $x$ direction. We consider the reference surface $\Sigma$ as the RT surface of the boundary interval $A = [-b, b]$, which has the profile
\begin{equation}
  \gamma_A:\qquad
  \sqrt{1-\frac{z^2}{\zh^2}}
  =\frac{\cosh\left(\frac{x}{\zh}\right)}
  {\cosh\left(\frac{b}{\zh}\right)},
  \label{eq:BTZ-RT-profile}
\end{equation}
The regulated length of this geodesic is given by
\begin{align}
	\cL_A=2\log\left[\frac{2\zh}{\eps}
	\sinh\left(\frac{b}{\zh}\right)\right]+O(\eps^2)
\end{align}
so that the RT result agrees with the universal finite-temperature CFT
entropy \cite{Calabrese:2004eu,Calabrese:2009qy,Ryu:2006ef},
\begin{equation}
  S_A=\frac{1}{2G}\log\left[\frac{2\zh}{\eps}
  \sinh\left(\frac{b}{\zh}\right)\right]
  =\frac{c}{3}\log\left[\frac{\beta}{\pi\eps}
  \sinh\left(\frac{2\pi b}{\beta}\right)\right].
  \label{eq:BTZ-entropy}
\end{equation}

For later purposes, it is useful to fix the orientation and normalization
of its tangent and normal vectors once and for all.  Writing the curve as
$\Phi(x,z)=z-z_\Sigma(x)=0$, a unit tangent and an outward-pointing unit normal
covector are obtained as
\begin{equation}
	\tau_\Sigma^\mu
	=\frac{z\sqrt{f(z)}}
	{\sqrt{f(z)+z_\Sigma'(x)^2}}
	\bigl(1,z_\Sigma'(x)\bigr),
	\qquad
	n_{\Sigma,\mu}
	=\frac{\bigl(-z_\Sigma'(x),1\bigr)}
	{z\sqrt{f(z)+z_\Sigma'(x)^2}}.
	\label{eq:BTZ-RT-frame}
\end{equation}
From the metric \eqref{eq:BTZ-slice}, the normalization is found to be
\begin{align}
	\cN^2=g^{\mu\nu}\del_\mu\Phi\del_\nu\Phi=z^2\left(\frac{z_h\sinh\left(\frac{2x}{z_h}\right)}{2z\cosh^2\left(\frac{b}{z_h}\right)}\right)^2+z^2\left(1-\frac{z^2}{z_h^2}\right)=(z_h^2-z^2)\tanh^2\left(\frac{b}{z_h}\right)\,.
\end{align}
Therefore, the unit normal covector to $\Sigma$ may be written compactly as
\begin{align}
	n^{}_{\Sigma,\mu}&=\left(\sqrt{\frac{1}{z^2}-\frac{1}{z_h^2}\coth ^2\left(\frac{b}{z_h}\right)},\frac{\coth \left(\frac{b}{z_h}\right)}{z_h \sqrt{1-\frac{z^2}{z_h^2}}}\right)=\frac{\left(\left(1-\frac{z^2}{z_h^2}\right) \sinh \left(\frac{x}{z_h}\right),\frac{z }{z_h}\cosh \left(\frac{x}{z_h}\right)\right)}{z \sqrt{1-\frac{z^2}{z_h^2}} \sqrt{\sinh ^2\left(\frac{x}{z_h}\right)+\frac{z^2}{z_h^2}}}\,.\label{normal-Sigma-BTZ}
\end{align}
wherein the last equality, we have eliminated $b$ utilizing \eqref{eq:BTZ-RT-profile}. The sign is chosen to point out of the entanglement wedge of $A$; reversing it reverses all subsequent thread orientations but leaves the positive PEE measures unchanged.

\subsection{Endpoint sectors and thermal two-point PEE}
\label{subsec:BTZ-two-point}
\begin{figure}[t]
	\centering
	\includegraphics[width=0.82\textwidth]{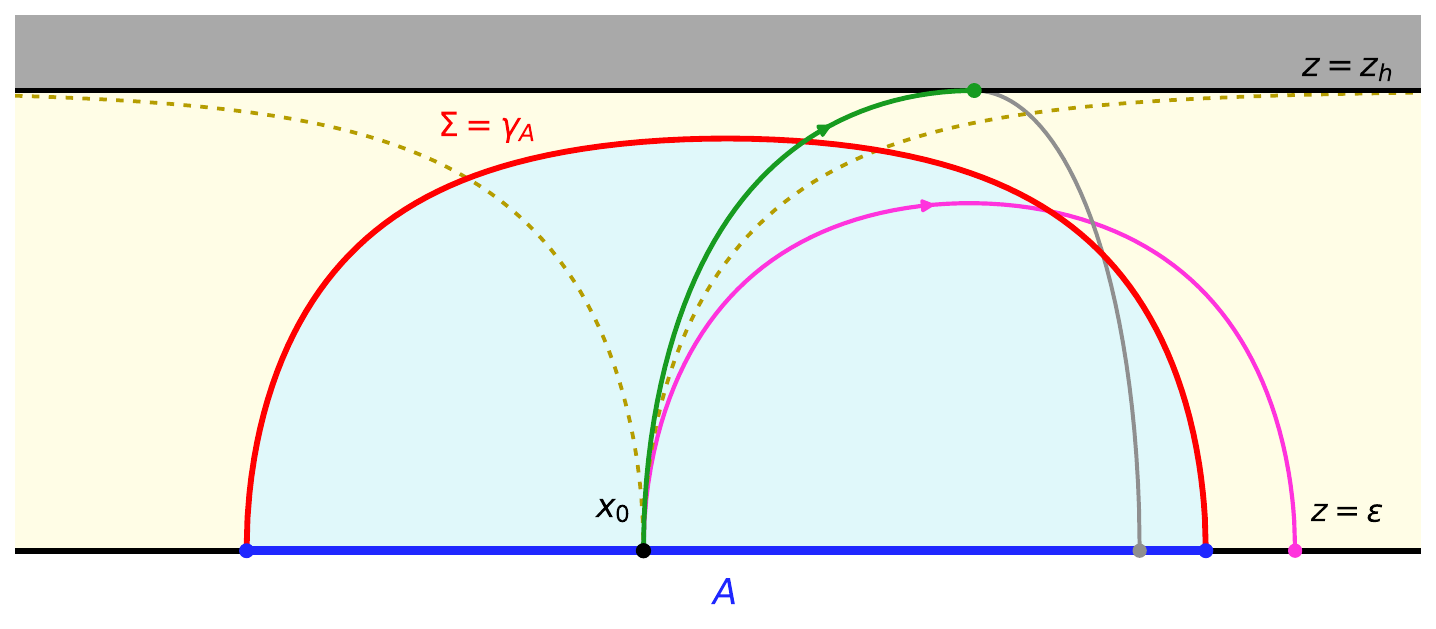}
	\caption{Schematic endpoint sectors (not to scale) for a representative
		noncentral source $x_0$ on the regulated asymptotic boundary of the
		planar BTZ exterior.  The lower black line is $z=\epsilon$, the upper
		black line is the horizon $z=z_h$, and the red curve is the RT geodesic
		$\Sigma=\gamma_A$ of $A=[-b,b]$, whose boundary segment is highlighted
		in blue.  The magenta solid curve is a boundary--boundary PEE thread,
		the green solid curve is a physical boundary--horizon PEE thread, and
		the gray curve is its auxiliary companion in the continued description
		discussed in subsection~\ref{subsec:BTZ-auxiliary}.  The two golden
		dashed curves are the limiting separatrices obtained when the second
		endpoint is pushed to spatial infinity; they asymptote to the horizon
		at the ideal ends of the planar geometry.  Arrows indicate the chosen
		thread orientation.}
	\label{fig:BTZ-chambers}
\end{figure}
As advertised earlier, in the presence of the horizon the PEE network changes fundamentally. Fix a source $x=x_0$ on the regulated asymptotic boundary.  The geodesics
through $x_0$ that cross $\Sigma$ fall into two open physical sectors,
distinguished by the location of their second endpoint.  In the outer
sector the second endpoint lies on the same asymptotic boundary; in the
central sector it lies on the horizon.  The two sectors meet when the second
endpoint is pushed to spatial infinity.  This limiting curve is a
measure-zero separatrix, not an independent contribution to the PEE.

\begin{itemize}
	\item \textbf{Class I: boundary-boundary.} The geodesics crossing $\Sigma$ and ending on the same asymptotic boundary at $z = \epsilon$ at finite spatial separation fall in this category. For two distinct asymptotic endpoints $x_1$ and $x_2$, the same-boundary
	geodesic has the following profile
	\begin{equation}
		\sqrt{1 - \frac{z^2}{z_h^2}} = \frac{\cosh \left( \frac{2x - x_1 + x_2}{2 z_h} \right)}{\cosh \left( \frac{x_2 - x_1}{2 z_h} \right)}\,.
		\label{eq:class-I-geodesic}
	\end{equation}
	Its regulated length may be obtained from \eqref{eq:length-BTZ-boundary} as
	\begin{equation}
		\cL_{\partial\partial}(x_1,x_2)
		=2\log\left[\frac{2\zh}{\eps}
		\sinh\left(\frac{|x_1-x_2|}{2\zh}\right)\right]\,.
		\label{eq:BTZ-bb-length}
	\end{equation}
	Applying the mixed-derivative rule \eqref{eq:PEE-kernel-general} gives the positive
	two-point PEE,
	\begin{equation}
		\cI_{\partial\partial}(x_1,x_2)
		=\frac{1}{16G\zh^2}
		\csch^2\left(\frac{x_1-x_2}{2\zh}\right).
		\label{eq:BTZ-bb-kernel}
	\end{equation}
	At short separation this kernel reduces to the vacuum inverse-square density, while at separations much larger than $\zh$ it is exponentially suppressed.
	\item \textbf{Class II: boundary-horizon.} A geodesic joining the regulated asymptotic point $(x_1,\eps)$ to the
	horizon point $(x_2,\zh)$ obeys
	\begin{equation}
		\sqrt{1-\frac{z^2}{\zh^2}}
		=\frac{\sinh\left(\frac{x-x_2}{\zh}\right)}
		{\sinh\left(\frac{x_1-x_2}{\zh}\right)}.
		\label{eq:class-II-geodesic}
	\end{equation}
	It has only one asymptotic UV divergence, and its regulated length is
	\begin{equation}
		\cL_{\partial h}(x_1,x_2)
		=\log\left[\frac{2\zh}{\eps}
		\cosh\left(\frac{x_1-x_2}{\zh}\right)\right]
		+O\!\left(\frac{\eps^2}{\zh^2}\right).
		\label{eq:BTZ-bh-length}
	\end{equation}
	With the endpoint orientation chosen so that the measure is positive, the
	same mixed-derivative prescription \eqref{eq:subregion-kinematic-measure} yields
	\begin{equation}
		\cI_{\partial h}(x_1,x_2)
		=\frac{1}{8G\zh^2}
		\sech^2\left(\frac{x_1-x_2}{\zh}\right).
		\label{eq:BTZ-bh-kernel}
	\end{equation}
	Unlike the same-boundary kernel, this density is regular at coincident
	spatial coordinates and is localized on the thermal scale $\zh$.  It is the
	endpoint sector that will carry the extensive thermal contribution after
	the source superposition.
	
	There is also a set of \emph{auxiliary} geodesics which share their endpoints on the horizon with geodesics in class-II:
	\begin{align}
		\sqrt{1-\frac{z^2}{z_h^2}}=\frac{\sinh\left(\frac{x-x_2}{z_h}\right)}{\sinh\left(\frac{x_2+x_1}{z_h}\right)}\,.
	\end{align}
	reaching the horizon at $x=x_2$ with the other endpoint on the boundary at $x=x_1+2x_2$.
	These auxiliary geodesics will also be useful when we discuss the PEE currents and the corresponding bit-thread flow.  However, this family should not be
	counted as an additional positive boundary--horizon PEE kernel.
	
	\item \textbf{Limiting class.} The boundary--boundary and boundary--horizon families meet when their second
	endpoint is taken to $+\infty$ or $-\infty$.  For a source at $x_0$, the two
	limiting profiles are most transparently written as
	\begin{equation}
		\sqrt{1-\frac{z^2}{\zh^2}}
		=\exp\left(-\frac{|x-x_0|}{\zh}\right),
		\qquad
		z=\zh\sqrt{1-
			\exp\left(-\frac{2|x-x_0|}{\zh}\right)}.
		\label{eq:BTZ-limiting-geodesic}
	\end{equation}
	The separatrix determines where the global endpoint of an elementary thread
	changes from the asymptotic boundary to the horizon.  It carries no
	independent measure.  As shown in subsection~\ref{subsec:BTZ-currents}, the local PEE thread flow
	is nevertheless analytic across it: the two endpoint sectors are different
	global completions of a single local current rather than unrelated bulk
	flows.
\end{itemize}
At this stage, $\cI_{\partial\partial}$ and
$\cI_{\partial h}$ are exact PEE densities associated with the
two geometric boundary components of the cut-open exterior
\cite{Czech:2015qta,Basu:2026hbg}.  The first has the standard thermal-CFT
interpretation.  The second should not yet be read as evidence for an
autonomous horizon Hilbert space.  In the following, we show
instead that it is the one-sided image of the ordinary cross-boundary kernel
in the thermofield-double purification, including the endpoint measure and
entropy sum rule.

\subsection{Thermofield-double origin of the horizon sector}
\label{subsec:BTZ-TFD-map}

The boundary--horizon sector of the cut-open BTZ exterior should not be
interpreted as postulating an independent set of microscopic degrees of
freedom localized on the bifurcation line.  Its origin is instead
transparent in the standard thermofield-double purification of the thermal
state.  The following derivation makes this statement precise at the level
of the complete geodesics, their endpoint measure, the entropy sum rule, and
the associated local contour.

Let $\mathcal H_R$ and $\mathcal H_L$ denote two identical copies of the CFT
Hilbert space.  The reflection-symmetric purification selected by the
eternal black-hole geometry is
\begin{equation}
  \left|\Psi(\beta)\right\rangle
  =
  \frac{1}{\sqrt{Z(\beta)}}
  \sum_n
  e^{-\frac{\beta E_n}{2}}
  |n\rangle_R\otimes|n\rangle_L,
  \label{eq:BTZ-TFD-state}
\end{equation}
Tracing out the left system gives the thermal state on the right system:
\begin{equation}
  \rho^{}_R
  =
  \operatorname{Tr}_{\mathcal H_L}
  \left|\Psi(\beta)\right\rangle
  \left\langle\Psi(\beta)\right|
  =
  \frac{1}{Z(\beta)}e^{-\beta H_R}.
  \label{eq:BTZ-TFD-reduced-state}
\end{equation}
The bulk dual of \eqref{eq:BTZ-TFD-state} is the two-sided eternal planar BTZ
black hole \cite{Maldacena:2001kr}.  Consequently, if
$A_R\subset R$ is an interval in the right CFT, then its complement in the
purified system is $\overline{A_R}=(R\setminus A_R)\cup L$.
Under the Crofton/PEE interpretation, additivity over the
two components of the complement gives
\begin{equation}
  S(A_R)
  =
  \cI(A_R,R\setminus A_R)
  +
  \cI(A_R,L).
  \label{eq:BTZ-TFD-PEE-normalization}
\end{equation}
The first term is the same-boundary channel, whereas the second records
correlations with the purifying left CFT.  We now show that the one-sided
boundary--horizon measure is precisely the geometric push-forward of this
second contribution.

\paragraph{Two-sided bridge in Fermi coordinates.}

Introduce a signed proper-distance coordinate $\rho$ on the complete
time-reflection-symmetric bridge,
\begin{equation}
  \dd s_\Sigma^2
  =
  \dd\rho^2+\cosh^2\rho\,\dd \left(\frac{x}{z_h}\right)^2~~,~~
  z=\zh\sech\rho.
  \label{eq:BTZ-TFD-Fermi-metric}
\end{equation}
The right and left exteriors correspond respectively to $\rho>0$ and
$\rho<0$, while the bifurcation line $h$ is at $\rho=0$.  The sign of
$\rho$ distinguishes the two exterior copies even though the
Schwarzschild radial coordinate $z=\zh\sech\rho$ is even.  We choose the spatial coordinates
$x$ and $\tilde x$ to increase in the same bulk spatial orientation on
the two asymptotic boundaries.

For two points $P_i=(\rho_i,x_i)$, the geodesic distance in the metric 
\eqref{eq:BTZ-TFD-Fermi-metric} obeys
\begin{equation}
  \cosh \cL(P_1,P_2)
  =
  \cosh\rho_1\cosh\rho_2\cosh\left(\frac{x_1-x_2}{z_h}\right)
  -
  \sinh\rho_1\sinh\rho_2.
  \label{eq:BTZ-TFD-Fermi-distance}
\end{equation}
We place the regulated right and left boundaries at
$\rho=\rho_\eps$ and $\rho=-\rho_\eps$, with
\begin{equation}
  \cosh\rho_\eps=\frac{\zh}{\eps},
  \qquad
  \rho_\eps
  =
  \log\left(\frac{2\zh}{\eps}\right)
  +O\left(\eps^2\right).
  \label{eq:BTZ-TFD-rho-regulator}
\end{equation}
For the right endpoint
$P_R=(\rho_\eps,x)$ and the left endpoint
$P_L=(-\rho_\eps,\tilde x)$, equation
\eqref{eq:BTZ-TFD-Fermi-distance} gives
\begin{align}
  \cL_{\partial\tilde\partial}
  (x,\tilde x)
  &=
  2\arccosh\left[
  \frac{\zh}{\eps}
  \cosh\left(\frac{x-\tilde x}{2\zh}\right)
  \right]=2\log\left[
  \frac{2\zh}{\eps}
  \cosh\left(\frac{x-\tilde x}{2\zh}\right)
  \right]
  +O\left(\eps^2\right).
  \label{eq:BTZ-TFD-cross-boundary-length-exact}
\end{align}
The same dependence follows from the equal-time TFD correlator
\begin{equation}
  \left\langle
  \mathcal O_R(x)\mathcal O_L(\tilde x)
  \right\rangle_{\Psi}
  \propto
  \left[
  \frac{\pi/\beta}
  {\cosh\left(\frac{\pi(x-\tilde x)}{\beta}\right)}
  \right]^{2\Delta},
  \qquad
  \beta=2\pi\zh.
  \label{eq:BTZ-TFD-cross-correlator}
\end{equation}

Applying the Kinematic space prescription \eqref{eq:subregion-kinematic-measure}, the extended two-point PEE between the left and right boundaries may now be obtained as
\begin{equation}
  \cI_{\partial\tilde\partial}(x,\tilde x)
  =
  \frac{1}{16G\zh^2}
  \sech^2\left(\frac{x-\tilde x}{2\zh}\right).
  \label{eq:BTZ-cross-boundary-kernel}
\end{equation}
This is the two-sided ancestor of the one-sided kernel
\eqref{eq:BTZ-bh-kernel}.

\paragraph{Cutting at the bifurcation line.}

Cut the complete bridge along $h$ and let $y$ be the spatial coordinate at
which a candidate cross-boundary geodesic meets the cut.  The exact
regulated lengths of its right and left segments are
\begin{align}
  \cL_{\partial h}(x,y)
  &=
  \arccosh\left[
  \frac{\zh}{\eps}
  \cosh\left(\frac{x-y}{\zh}\right)
  \right]~~,~~
  \cL_{\tilde\partial h}(\tilde x,y)
  =
  \arccosh\left[
  \frac{\zh}{\eps}
  \cosh\left(\frac{\tilde x-y}{\zh}\right)
  \right].
  \label{eq:BTZ-TFD-left-half-length}
\end{align}
Their leading asymptotic forms agree with
\eqref{eq:BTZ-bh-length}.  The correct crossing point is obtained by
minimizing the sum $\mathcal L_(y)=\cL_{\partial h}(x,y)+\cL_{\tilde\partial h}(\tilde x,y)$. The unique minimum occurs at the point
\begin{equation}
  y=\frac{x+\tilde x}{2},
  \qquad
  \tilde x=2y-x.
  \label{eq:BTZ-TFD-crossing-map}
\end{equation}
where the two segments have equal length and
\begin{equation}
  \cL_{\partial\tilde\partial}
  (x,\tilde x)
  =
  2\cL_{\partial h}
  \left(x,\frac{x+\tilde x}{2}\right).
  \label{eq:BTZ-TFD-half-length}
\end{equation}
Thus the complete two-sided chord is cut into two equal geodesic segments;
the horizon is a coordinate seam of the complete bridge, not a reflecting
or terminal surface.

\paragraph{Push-forward of the endpoint measure.}

The invariant statement concerns the endpoint measure rather than the
numerical value of a kernel in isolation.  At fixed $x$, the midpoint map
\eqref{eq:BTZ-TFD-crossing-map} gives $\dd\tilde x=2\,\dd y$.
Indeed, differentiating the exact half-length relation gives
\begin{equation}
  \cI_{\partial h}(x,y)
  =
  2\,
  \cI_{\partial\tilde\partial}
  (x,2y-x)=\frac{1}{8G\zh^2}
  \sech^2\left(\frac{x-y}{\zh}\right)\,,
  \label{eq:BTZ-TFD-exact-pushforward}
\end{equation}
which reproduces \eqref{eq:BTZ-bh-kernel}.  Equivalently,
\begin{equation}
  \cI_{\partial\tilde\partial}(x,\tilde x)
  \,\dd x\wedge\dd\tilde x
  =
  \cI_{\partial h}(x,y)\,\dd x\wedge\dd y.
  \label{eq:BTZ-TFD-kernel-pushforward}
\end{equation}
The relative factor of two between the cross-boundary and
boundary--horizon kernels is thus fixed by the Jacobian of the crossing-point
map.  It is not an arbitrary normalization and does not arise from changing
between oriented and unoriented geodesic conventions.

The preservation of the endpoint measure can also be checked directly:
\begin{equation}
  \int_{-\infty}^{\infty}
  \dd\tilde x\,
  \cI_{\partial\tilde\partial}(x,\tilde x)
  =
  \int_{-\infty}^{\infty}
  \dd y\,
  \cI_{\partial h}(x,y)
  =
  \frac{1}{4G\zh}.
  \label{eq:BTZ-horizon-kernel-direct}
\end{equation}

\paragraph{Global entropy sum rule.}

For $A_R=[-b,b]$, the cross-boundary contribution is
\begin{align}
  \cI(A_R,L)
  &=
  \int_{-b}^{b}\dd x
  \int_{-\infty}^{\infty}\dd\tilde x\,
  \cI_{\partial\tilde\partial}(x,\tilde x)
 =
  \int_{-b}^{b}\dd x
  \int_{-\infty}^{\infty}\dd y\,
  \cI_{\partial h}(x,y)
  =
  \frac{2b}{4G\zh}.
  \label{eq:BTZ-horizon-PEE-direct}
\end{align}
Using the Brown--Henneaux relation \cite{Brown:1986nw} and
$\beta=2\pi\zh$, this becomes
\begin{equation}
  \cI(A_R,L)
  =
  s_{\mathrm{th}}|A_R|,
  \qquad
  s_{\mathrm{th}}
  =
  \frac{1}{4G\zh}
  =
  \frac{\pi c}{3\beta}.
  \label{eq:BTZ-TFD-thermal-density}
\end{equation}
The two components of the complement on the right boundary instead give,
for $x\in(-b,b)$,
\begin{align}
  \left(\int_b^\infty\dd y+\int_{-\infty}^{-b}\right)
  \cI_{\partial\partial}(x,y)
  &=
  \frac{1}{8G\zh}
  \left[
  \coth\left(\frac{b-x}{2\zh}\right)+\coth\left(\frac{b+x}{2\zh}\right)-2
  \right],
  \label{eq:BTZ-complement-integral}
\end{align}
Appropriate regularization of the $x$ integration as
$x\in[-b+\delta,b-\delta]$ yields
\begin{align}
  \cI(A_R,R\setminus A_R)
  &=
  \frac{1}{2G}
  \log\left[
  \frac{2\zh}{\eps}
  \sinh\left(\frac{b}{\zh}\right)
  \right]
  -
  \frac{b}{2G\zh}\,.
  \label{eq:BTZ-boundary-PEE-direct}
\end{align}
Combining
\eqref{eq:BTZ-horizon-PEE-direct} and
\eqref{eq:BTZ-boundary-PEE-direct} gives
\begin{equation}
  \cI(A_R,R\setminus A_R)+\cI(A_R,L)
  =
  \frac{1}{2G}
  \log\left[
  \frac{2\zh}{\eps}
  \sinh\left(\frac{b}{\zh}\right)
  \right]
  =
  S(A_R),
  \label{eq:BTZ-TFD-complete-normalization}
\end{equation}
which proves \eqref{eq:BTZ-TFD-PEE-normalization}.  After the left exterior
is removed, the same identity is represented one-sidedly as
\begin{equation}
  S(A_R)
  =
  \cI(A_R,A^c_\partial)
  +
  \cI(A_R,h).
  \label{eq:BTZ-one-sided-extended-normalization}
\end{equation}
The horizon term has not been added as an independent contribution: it is
the one-sided image of $\cI(A_R,L)$.

\paragraph{Local contour and horizon endpoint density.}

The decomposition is local on $A_R$.  At every $x\in A_R$, the horizon
channel contributes
\begin{equation}
  s_{A_R}^{(h)}(x)
  :=
  \int_{-\infty}^{\infty}\dd y\,
  \cI_{\partial h}(x,y)
  =
  \frac{1}{4G\zh}.
  \label{eq:BTZ-TFD-local-horizon-contour}
\end{equation}
The same-boundary contribution is
\begin{align}
  s_{A_R}^{(\partial)}(x)
  &:=
  \int_{R\setminus A_R}\dd y\,
  \cI_{\partial\partial}(x,y)
  =
  \frac{1}{8G\zh}
  \left[
  \coth\left(\frac{b-x}{2\zh}\right)
  +
  \coth\left(\frac{b+x}{2\zh}\right)-2
  \right].
  \label{eq:BTZ-TFD-local-boundary-contour}
\end{align}
The two constants cancel in the sum, giving the usual thermal interval
contour,
\begin{equation}
  s_{A_R}(x)
  =
  s_{A_R}^{(\partial)}(x)+s_{A_R}^{(h)}(x)
  =
  \frac{1}{8G\zh}
  \left[
  \coth\left(\frac{b-x}{2\zh}\right)
  +
  \coth\left(\frac{b+x}{2\zh}\right)
  \right].
  \label{eq:BTZ-TFD-complete-contour}
\end{equation}

Although \eqref{eq:BTZ-TFD-local-horizon-contour} is constant along
$A_R$, the distribution of the corresponding endpoints on the horizon is
not uniform.  It is
\begin{align}
  h_{A_R}(y)
  &:=
  \int_{-b}^{b}\dd x\,
  \cI_{\partial h}(x,y)
  =
  \frac{1}{8G\zh}
  \left[
  \tanh\left(\frac{b-y}{\zh}\right)
  +
  \tanh\left(\frac{b+y}{\zh}\right)
  \right],
  \label{eq:BTZ-TFD-horizon-endpoint-density}
\end{align}
and satisfies
\begin{equation}
  \int_{-\infty}^{\infty}\dd y\,h_{A_R}(y)
  =
  \frac{2b}{4G\zh}.
  \label{eq:BTZ-TFD-horizon-density-normalization}
\end{equation}
Thus a constant contour density on the boundary interval should not be
confused with a uniform endpoint density on $h$.  For finite $b$, the latter
is a smooth function concentrated around the portion of the horizon beneath
the interval.

\paragraph{Flow interpretation and qualifications.}

In the complete two-sided geometry, a physical TFD thread crosses the
bifurcation line and continues smoothly into the left exterior.  Restricting
the geometry to the right exterior retains only its right half and turns
$h$ into an allowed flux boundary.  The outward horizon flux of the
one-sided problem is therefore the restriction of the cross-wormhole flux
into the purifying CFT.  The horizon is a sink only because the left
exterior has been excised; it is not a terminal surface in the complete
eternal geometry.  This also explains why the boundary-returning and
horizon-crossing families are two global endpoint chambers of one local
current, as will be demonstrated explicitly in
subsection~\ref{subsec:BTZ-currents} \cite{Lin:2025btz,Caggioli:2024uza}.

Several qualifications are essential.  First, the crossing-point map
\begin{equation}
  \tilde x
  \longmapsto
  y=\frac{x+\tilde x}{2}
  \label{eq:BTZ-TFD-relational-map}
\end{equation}
depends explicitly on the right endpoint $x$.  A horizon point is therefore
a relational label for a complete two-ended chord, rather than a
source-independent relabelling of a local left-CFT degree of freedom.

Second, a different purification,
\begin{equation}
  |\Psi_U\rangle
  =
  \left(\mathbf 1_R\otimes U_L\right)
  |\Psi(\beta)\rangle,
  \label{eq:BTZ-TFD-unitary-purification}
\end{equation}
produces the same reduced state $\rho_R$ but need not preserve the simple
geometric localization of the left endpoint.  The horizon decomposition is
therefore natural in the reflection-symmetric TFD selected by the eternal
black-hole geometry, but it is not purification-independent.  In
particular, no literal pointwise factorization of the gravitational Hilbert
space into horizon-localized degrees of freedom is assumed.

Third, although
\begin{equation}
  \cI(A_R,L)=s_{\mathrm{th}}|A_R|
  \label{eq:BTZ-TFD-extensive-sector}
\end{equation}
is exact for every $b/\beta$ within our construction, it
should not be identified with the total thermal correction
$S_{A_R}(\beta)-S_{A_R}(\infty)$.  The same-boundary sector is itself
temperature-dependent and compensates the linear TFD contribution,
especially for intervals shorter than the thermal scale.  The term
\emph{thermal horizon sector} refers to the TFD origin of this channel and
to its density $s_{\mathrm{th}}$, not to a purification-independent split
of the entropy into vacuum and thermal pieces.

Finally, the distance decomposition, midpoint relation, and equality of
endpoint measures are exact classical geometric statements.  Their
interpretation as microscopic PEEs additionally assumes the semiclassical
Crofton/PEE dictionary.  Subject to that assumption, the horizon sector is
the measure-preserving one-sided encoding of correlations between $A_R$ and
the second asymptotic CFT.  This physical positive sector should not be
confused with the auxiliary signed continuation discussed in
subsection~\ref{subsec:BTZ-auxiliary}, whose flux through the physical RT
surface is not an entropy contribution.

The same push-forward persists at finite radial cutoff.  The corresponding
finite-cutoff boundary lengths, kernels, midpoint map, and exact equality of measures
are derived in subsection~\ref{subsec:cutoff-BTZ}.  Thus the finite-cutoff boundary
horizon sector is the one-sided encoding of the cutoff TFD purification
\cite{Coleman:2022tfd}, rather than an artifact of the asymptotic limit.

\subsection{PEE thread flow}
\label{subsec:BTZ-currents}
We now construct the PEE thread-flow vector field in the BTZ black brane geometry. At a fixed boundary source $x_0$, the geodesic PEE threads define the source-resolved current $V_{x_0}$  emanating from the asymptotic boundary. Only afterwards do we integrate over $x_0\in A$ to obtain the macroscopic flow $v_A$. We will construct the flows for physical and auxiliary branches separately.

Recall that the unit normal covector to the reference RT surface \eqref{eq:BTZ-RT-profile} is given in \eqref{normal-Sigma-BTZ}. The induced line element on the same geodesic is
\begin{equation}
  \dd\Sigma
  =\frac{\tanh\left(\frac{b}{\zh}\right)}
  {\zh\left[1-\sech^2\left(\frac{b}{\zh}\right)
  \cosh^2\left(\frac{x}{\zh}\right)\right]}\,\dd x.
  \label{eq:BTZ-RT-line-element}
\end{equation}
These results will be used below to normalize the elementary currents
directly against the two-point PEEs discussed earlier.

\subsubsection{Class-I: boundary--boundary PEE thread flow}
\label{subsec:BTZ-geodesic-chambers}
We first determine the PEE thread flow vector $V^\mu_O$ emanating from the origin $O$ (at $x = 0$), and extend to arbitrary boundary points under a translation in the $x$ direction. A geodesic emitted from $x=0$ and crossing $\gamma_A$
belongs to one of the chambers summarized in figure~\ref{fig:BTZ-chambers}.  The distinction is global: it is determined by the second endpoint of the same locally smooth congruence. Thus the chamber decomposition organizes the endpoint data of a single congruence; it should not be interpreted as introducing a bulk source localized on the separatrix.

\begin{figure}[t]
  \centering
  \includegraphics[width=0.82\textwidth]{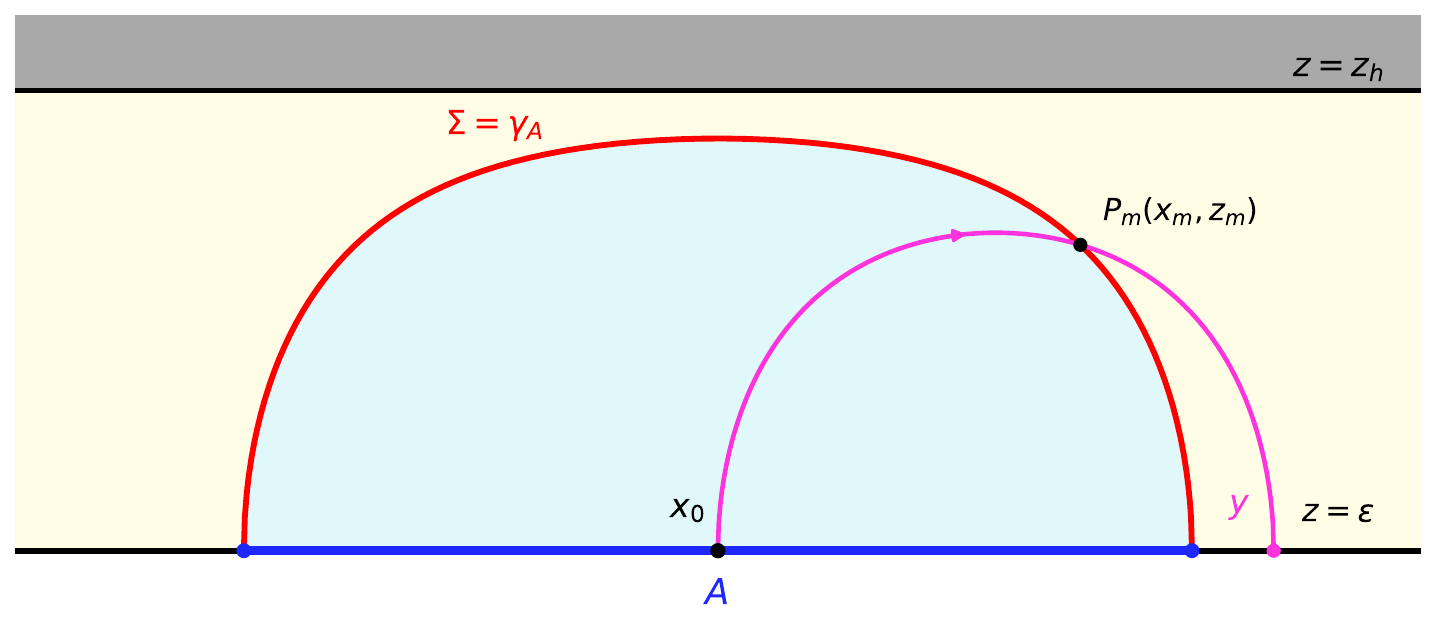}
  \caption{Schematic boundary--boundary (Class-I) PEE thread in the BTZ
  exterior (not to scale).  The solid magenta geodesic starts at the source
  $x_0=0$, crosses the RT geodesic $\Sigma=\gamma_A$ at
  $P_m=(x_m,z_m)$, and reaches the second asymptotic-boundary endpoint $y$.
  The blue segment denotes $A$ and the cyan region its entanglement wedge;
  the horizon at $z=z_h$ is shown only to locate the exterior region.  The
  arrow indicates the chosen thread orientation.}
  \label{fig:BTZ-class-I}
\end{figure}

From \eqref{eq:class-I-geodesic},  a geodesic from $x=0$ to a second boundary point $y$ has the following profile
\begin{equation}
  \sqrt{1-\frac{z^2}{\zh^2}}
  =\frac{\cosh\left(\frac{2x-y}{2\zh}\right)}
  {\cosh\left(\frac{y}{2\zh}\right)}.
  \label{eq:BTZ-bb-geodesic}
\end{equation}
The location of the boundary point $y$ can be determined in terms of the intersection point $P_m$ on the reference surface $\Sigma=\gamma_A$, by solving eqs.~\eqref{eq:BTZ-RT-profile} and \eqref{eq:BTZ-bb-geodesic} as follows
\begin{equation}
  \tanh\left(\frac{y}{2\zh}\right)
  =\coth\left(\frac{x_m}{\zh}\right)
  \left[1-\sech\left(\frac{b}{\zh}\right)\right].
  \label{eq:BTZ-bb-endpoint-map}
\end{equation}
Figure~\ref{fig:BTZ-class-I} displays this endpoint map: the solid magenta
geodesic starts at the fixed source, crosses the reference surface at
$P_m$, and returns to the (regulated) asymptotic boundary at
$y$.
Note that the sign of $y$ is determined by the sign of $x_m$.  This chamber occupies the two
outer portions of the RT geodesic,
\begin{equation}
  \zh\arctanh\left[1-\sech\left(\frac{b}{\zh}\right)\right]
  <|x_m|\leq b.
  \label{eq:BTZ-separatrix-location}
\end{equation}
At $|x_m|=b$ the partner is the adjacent endpoint $|y|=b$, whereas $|y|\to\infty$ as $|x_m|$ approaches the inner endpoint of this chamber. The divergence of the partner coordinate at the inner edge is the  description of the transition to the mixed channel.  It does not signal a
singularity of the local current: the limiting geodesic is smooth, while its global endpoint is being moved to the ideal boundary of the planar geometry.

Next, we consider the unit tangent to a boundary--boundary thread,
oriented toward increasing $x$,
\begin{equation}
  \tau_{\partial\partial}^{\mu}
  =\left(
  \frac{z^2}{\zh}\coth\left(\frac{y}{2\zh}\right),
  -z\frac{\sinh\left(\frac{2x-y}{\zh}\right)}
  {\sinh\left(\frac{y}{\zh}\right)}
  \right)\,.
  \label{eq:BTZ-bb-unit-tangent}
\end{equation}
We find it easier to switch variables from the crossing coordinate
$x_m$ to the endpoint $y$. On the reference RT surface, we then have the following relations
\begin{align}
  \left|n_{\gamma_A}\cdot\tau_{\partial\partial}\right|
  &=\sqrt{1-\tanh^2\left(\frac{b}{2\zh}\right)
  \coth^2\left(\frac{y}{2\zh}\right)},
  \label{eq:BTZ-bb-normal-tangent}\\
  \left|\frac{\dd\Sigma}{\dd y}\right|
  &=\frac{\sinh\left(\frac{b}{\zh}\right)}
  {2\zh\left[\cosh\left(\frac{y}{\zh}\right)
  -\cosh\left(\frac{b}{\zh}\right)\right]}.
  \label{eq:BTZ-bb-RT-measure-y}
\end{align}
The flux-matching condition \eqref{eq:flux-matching-general} now reads
\begin{equation}
  |V_O|\,
  \left|n_{\gamma_A}\cdot\tau_{\partial\partial}\right|
  \left|\frac{\dd\Sigma}{\dd y}\right|=\cI_{\del\del}(0,y)
  =\frac{1}{16G\zh^2}
  \csch^2\left(\frac{y}{2\zh}\right).
  \label{eq:BTZ-bb-flux-matching-explicit}
\end{equation}
The local meaning of this equation is the following: the PEE weight in the endpoint element
$\dd y$ equals the current crossing the corresponding RT element $\dd\Sigma$. In the above expression $|V_O|$ is the strength of the source-resolved current, the normal projection
accounts for the incidence angle, and $|\dd\Sigma/\dd y|$ converts between the endpoint and RT parametrizations.
Solving for the norm gives
\begin{equation}
  |V_O|
  =\frac{1}{8G\zh}
  \sqrt{\csch^2\left(\frac{b}{2\zh}\right)
  -\csch^2\left(\frac{y}{2\zh}\right)}.
  \label{eq:BTZ-bb-norm-on-RT}
\end{equation}
The vanishing of this expression at the adjacent endpoint $|y|=b$ and its growth toward the chamber boundary $|y|\to\infty$ describe how the fixed-source thread bundle is redistributed along $\gamma_A$.  The apparent growth near the separatrix  compensated by the endpoint-to-RT Jacobian in \eqref{eq:BTZ-bb-RT-measure-y}; the physical flux element remains the finite measure fixed by \eqref{eq:BTZ-bb-kernel}.

Eliminating the endpoint $y$ utilizing \eqref{eq:BTZ-bb-endpoint-map} gives the norm at the arbitrary bulk point $P_m$
\begin{equation}
	|V_{O}(x_m,z_m)|
	=\frac{1}{4G}\frac{z_m}{2\zh^2\left[
		\cosh\left(\frac{x_m}{\zh}\right)
		-\sqrt{1-\frac{z_m^2}{\zh^2}}\right]}.
	\label{eq:BTZ-elementary-norm}
\end{equation}
For a general source $x_0$, the partner coordinate associated with a bulk
point $P_m=(x_m,z_m)$ satisfies the full relation
\begin{equation}
  \tanh\left(\frac{y-x_0}{2\zh}\right)
  =\frac{\cosh\left(\frac{x_m-x_0}{\zh}\right)
  -\sqrt{1-\frac{z^2}{\zh^2}}}
  {\sinh\left(\frac{x_m-x_0}{\zh}\right)}.
  \label{eq:BTZ-bb-local-map}
\end{equation}
Upon eliminating $y$, the unit tangent to the thread reads
{\small
\begin{equation}
	\tau_{x_0}^{\mu}(x_m,z_m)
	=\frac{z_m}{\cosh\left(\frac{x_m-x_0}{\zh}\right)
		-\sqrt{1-\frac{z_m^2}{\zh^2}}}
	\begin{pmatrix}
		\dfrac{z_m}{\zh}\sinh\left(\dfrac{x_m-x_0}{\zh}\right)
		\\[4pt]
		\sqrt{1-\dfrac{z_m^2}{\zh^2}}
		\left[1-\sqrt{1-\dfrac{z_m^2}{\zh^2}}
		\cosh\left(\dfrac{x_m-x_0}{\zh}\right)\right]
	\end{pmatrix},
	\label{eq:BTZ-elementary-tangent}
\end{equation}
}
Equations \eqref{eq:BTZ-bb-unit-tangent} and \eqref{eq:BTZ-elementary-tangent},
together with the kernel \eqref{eq:BTZ-bb-kernel}, determine the PEE thread current, following the PEE-thread flow construction of \cite{Lin:2023rxc,Lin:2024fze}. It is straightforward to verify that the PEE thread flow is divergenceless. As a simple cross-check, we may take the zero temperature limit $z_h \to \infty$ and recover the Poincaré AdS$_3$ result reported in \cite{Lin:2023rxc},
	\begin{equation}
	V^\mu_{x_0}(x_m,z_m) = \frac{1}{4G} \frac{2 z_m^3}{\left[(x_m-x_0)^2 + z_m^2\right]^2} \left(x_m-x_0, \frac{z_m^2 - (x_m-x_0)^2}{2z_m} \right) \,.
	\label{eq:2.24}
\end{equation}

\subsubsection{Class II: boundary--horizon PEE thread flow}
\begin{figure}[t]
  \centering
  \includegraphics[width=0.82\textwidth]{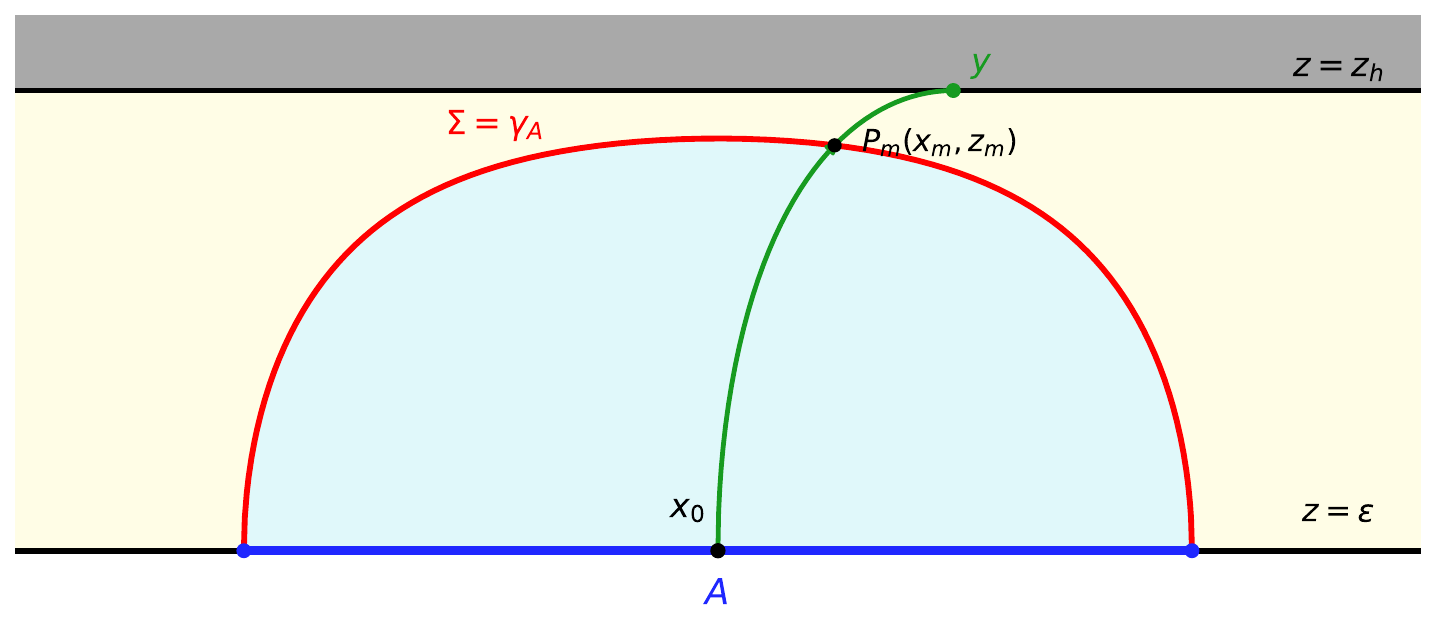}
  \caption{Schematic boundary--horizon (Class-II) PEE thread in the BTZ
  exterior (not to scale).  The solid green geodesic begins at $x_0=0$,
  crosses $\Sigma=\gamma_A$ at $P_m=(x_m,z_m)$, and terminates at the
  horizon point $(y,z_h)$.  The blue segment denotes $A$ and the cyan region
  its entanglement wedge.  These threads form the local endpoint sector
  carrying the extensive thermal contribution; the arrow indicates the
  chosen thread orientation.}
  \label{fig:BTZ-class-II}
\end{figure}

Next we consider the class of PEE threads reaching the event horizon at $x=y$.
A geodesic from the boundary source $x_0=0$ to the horizon point $y$ obeys (cf. \cref{eq:BTZ-bdy-horizon})
\begin{equation}
  \sqrt{1-\frac{z^2}{\zh^2}}
  =\frac{\sinh\left(\frac{y-x}{\zh}\right)}
  {\sinh\left(\frac{y}{\zh}\right)}.
  \label{eq:BTZ-bh-geodesic}
\end{equation}
Its intersection with $\gamma_A$ gives the horizon endpoint $y$ in terms of the crossing point $P_m$ as follows
\begin{equation}
  \tanh\left(\frac{y}{\zh}\right)
  =\frac{\tanh\left(\frac{x_m}{\zh}\right)}
  {1-\sech\left(\frac{b}{\zh}\right)}.
  \label{eq:BTZ-bh-endpoint-map}
\end{equation}
This geometry is displayed in figure~\ref{fig:BTZ-class-II}; the same source and reference surface are used, but the second endpoint now lies at $y$ on the horizon. In the one-sided exterior this endpoint is an allowed flux endpoint; in the complete TFD geometry it is the point at which the same geodesic continues into the second exterior. Accordingly, the word ``horizon-ending'' refers to the cut-open description and does not describe absorption by a physical membrane.
This covers the central portion
\begin{equation}
  |x_m|<\zh\arctanh\left[
  1-\sech\left(\frac{b}{\zh}\right)\right].
  \label{eq:BTZ-bh-chamber}
\end{equation}
The horizon endpoint moves from $y=0$ to $|y|\to\infty$ as $|x_m|$
approaches the separatrix. This is the mixed-sector counterpart of the boundary endpoint running to infinity in the outer chamber. Both limits approach the same measure-zero geodesic, which is why the endpoint type can change without producing a discontinuity in the bulk current.

The corresponding boundary--horizon unit tangent is
\begin{equation}
  \tau_{\partial h}^{\mu}
  =\left(
  \frac{z^2}{\zh}\tanh\left(\frac{y}{\zh}\right),
  -z\frac{\sinh\left(\frac{2(x-y)}{\zh}\right)}
  {\sinh\left(\frac{2y}{\zh}\right)}
  \right)\,,
  \label{eq:BTZ-bh-unit-tangent}
\end{equation}
and on the reference RT geodesic, we have the following relations
\begin{align}
  \left|n_{\gamma_A}\cdot\tau_{\partial h}\right|
  &=\sech\left(\frac{y}{\zh}\right)
  \sqrt{1+\sech^2\left(\frac{b}{2\zh}\right)
  \sinh^2\left(\frac{y}{\zh}\right)},
  \label{eq:BTZ-bh-normal-tangent}\\
  \left|\frac{\dd\Sigma}{\dd y}\right|
  &=\frac{\sinh\left(\frac{b}{\zh}\right)}
  {\zh\left[\cosh\left(\frac{b}{\zh}\right)
  +\cosh\left(\frac{2y}{\zh}\right)\right]}.
  \label{eq:BTZ-bh-RT-measure-y}
\end{align}
The flux-matching equation \eqref{eq:flux-matching-general} now becomes
\begin{equation}
  |V_O|\,
  \left|n_{\gamma_A}\cdot\tau_{\partial h}\right|
  \left|\frac{\dd\Sigma}{\dd y}\right|=\cI_{\del h}(0,y)
  =\frac{1}{8G\zh^2}
  \sech^2\left(\frac{y}{\zh}\right),
  \label{eq:BTZ-bh-flux-matching-explicit}
\end{equation}
The relative normalization of this equation compared to \eqref{eq:BTZ-bb-flux-matching-explicit} is not optional. The mismatch by a factor of two in the mixed kernel is the Jacobian of the TFD midpoint map derived in subsection~\ref{subsec:BTZ-TFD-map}. Its appearance here is what makes the mixed endpoint measure compatible with the same local flux normalization as the boundary--boundary sector.
Solving the matching condition gives the norm of the boundary-horizon PEE thread current as
\begin{equation}
  |V_O|
  =\frac{1}{8G\zh}
  \sqrt{\csch^2\left(\frac{b}{2\zh}\right)
  +\sech^2\left(\frac{y}{\zh}\right)}.
  \label{eq:BTZ-bh-norm-on-RT}
\end{equation}
For a general source $x_0$, the partner coordinate associated with a bulk
point $P_m=(x_m,z_m)$ satisfies the full relation
\begin{equation}
  \tanh\left(\frac{y-x_0}{\zh}\right)
  =\frac{\sinh\left(\frac{x_m-x_0}{\zh}\right)}
  {\cosh\left(\frac{x_m-x_0}{\zh}\right)
  -\sqrt{1-\frac{z_m^2}{\zh^2}}}.
  \label{eq:BTZ-bh-local-map}
\end{equation}
Substitution into \eqref{eq:flux-matching-general}, now with the mixed kernel
\eqref{eq:BTZ-bh-kernel}, gives exactly the same local norm as the
boundary--boundary calculation. Interestingly, upon eliminating $y$ using \eqref{eq:BTZ-bh-local-map} the tangent vector \eqref{eq:BTZ-bh-unit-tangent} takes the same form \eqref{eq:BTZ-elementary-tangent} as the boundary-boundary sector.
This agreement is a nontrivial
compatibility condition on the thermal endpoint measure and its geodesic
flow representative \cite{Caggioli:2024uza}.  Physically, it says that a
bulk observer cannot detect the change of endpoint component from the
local magnitude or direction of the current.  The distinction becomes
visible only after the geodesic is followed globally to the boundary of
the cut-open exterior. 

Recall that, the two families meet on the limiting geodesics already given in the limiting class of subsection~\ref{subsec:BTZ-two-point}. They form a
separatrix of the endpoint map but carry no independent PEE measure.
The important point is that the PEE thread flow is analytic across the separatrix \eqref{eq:BTZ-limiting-geodesic}.  Only the global endpoint of an integral curve changes.  In the outer chamber an integral curve reaches the asymptotic boundary again; in the central chamber it reaches the horizon.  Thus the two endpoint sectors are not two unrelated local flows.  They are two global completions of one PEE current.

\subsubsection{PEE-selected geodesic bit threads}
\label{subsec:BTZ-superposition}
We now superpose the physical PEE thread currents, including both of
their positive endpoint chambers.  For the subsystem $A=[-b,b]$ with RT surface $\gamma_A$, the superposition of all the PEE threads emanating from within $A$ leads to the coarse-grained current
\begin{align}
	v_A^\mu(x,z)&=\int_{-b}^{b}\dd x_0\,V_{x_0}^\mu(x,z)\notag\\
	&=
	\frac{1}{4G}
	\frac{z^2\sinh\left(\frac{b}{\zh}\right)}
	{\zh^2
		\left[\cosh\left(\frac{b-x}{\zh}\right)
		-\sqrt{1-\frac{z^2}{\zh^2}}\right]
		\left[\cosh\left(\frac{b+x}{\zh}\right)
		-\sqrt{1-\frac{z^2}{\zh^2}}\right]}
	\nonumber\\[-2pt]
	&\quad\times
	\left(
	z\sinh\left(\frac{x}{\zh}\right),
	\zh\sqrt{1-\frac{z^2}{\zh^2}}
	\left[\cosh\left(\frac{b}{\zh}\right)
	-\sqrt{1-\frac{z^2}{\zh^2}}
	\cosh\left(\frac{x}{\zh}\right)\right]
	\right).
	\label{eq:BTZ-bit-flow}
\end{align}
This is the analytic geodesic bit-thread field denoted by $\mathbf{V}_+$ in
\cite{Caggioli:2024uza}.  We call it the PEE-selected
geodesic representative; no uniqueness among unrestricted max flows is
implied. The adjective ``PEE-selected'' records the boundary calibration: the RT entropy fixes only the maximal flux, whereas the endpoint PEE kernels specify how that flux is resolved
among sources and boundary components. The norm of the bit-thread field reads
\begin{align}
	|v_A|=\frac{1}{4G}
	\frac{\frac{z}{\zh}\sinh\left(\frac{b}{\zh}\right)}
	{\sqrt{
			\left[\cosh\left(\frac{b-x}{\zh}\right)
			-\sqrt{1-\frac{z^2}{\zh^2}}\right]
			\left[\cosh\left(\frac{b+x}{\zh}\right)
			-\sqrt{1-\frac{z^2}{\zh^2}}\right]}}.
	\label{eq:BTZ-bit-norm}
\end{align}
The identity
\begin{align}
	&\left[\cosh\left(\frac{b-x}{\zh}\right)
	-\sqrt{1-\frac{z^2}{\zh^2}}\right]
	\left[\cosh\left(\frac{b+x}{\zh}\right)
	-\sqrt{1-\frac{z^2}{\zh^2}}\right]
	-\frac{z^2}{\zh^2}\sinh^2\left(\frac{b}{\zh}\right)
	\nonumber\\
	&\hspace{3cm}=
	\left[\cosh\left(\frac{x}{\zh}\right)
	-\sqrt{1-\frac{z^2}{\zh^2}}
	\cosh\left(\frac{b}{\zh}\right)\right]^2
	\label{eq:BTZ-norm-identity}
\end{align}
immediately implies $|v_A|\leq1/(4G)$. On the RT surface \eqref{eq:BTZ-RT-profile} the right-hand side vanishes, and the bound is saturated. Substituting the RT profile into \eqref{eq:BTZ-bit-flow} also shows that the flow is normal to $\gamma_A$. Together with $\nabla\cdot v_A=0$, these checks establish that \eqref{eq:BTZ-bit-flow} is a max flow.
This realizes the Riemannian flow-cut construction explicitly in the thermal
geometry \cite{Freedman:2016zud,Agon:2018lwq,Caggioli:2024uza}.
Pointwise saturation has a stronger meaning than agreement of the integrated
flux alone: every infinitesimal RT element is used at capacity, while the
strictly smaller norm away from $\gamma_A$ reflects the transverse expansion
of the thread bundle.  Appendix~\ref{appB} gives the intrinsic Fermi-coordinate
explanation of this expansion and proves directly that the final integral
curves split into boundary-returning and horizon-crossing families at the
correct macroscopic separatrix.

\paragraph{Thermal flux and the extended entanglement contour.}
\label{subsec:BTZ-flux-decomposition}
The detailed PEE integrals and their TFD interpretation were established
in subsection~\ref{subsec:BTZ-TFD-map}. Here, we collect the resulting densities in the form needed for a pointwise comparison with the macroscopic vector field.  Let
\begin{equation}
	A^c_\partial=(-\infty,-b)\cup(b,\infty),
	\qquad h=(-\infty,\infty)
\end{equation}
be the asymptotic complement and the planar horizon.  Thus
\begin{equation}
	S_A
	=\underbrace{\cI(A,A^c_\partial)}
	_{S_A-\frac{2b}{4G\zh}}
	+\underbrace{\cI(A,h)}
	_{\frac{2b}{4G\zh}}.
	\label{eq:BTZ-extended-normalization}
\end{equation}
The second term is precisely the thermal entropy
\begin{equation}
	S_A^{\rm th}=s_{\rm th}|A|,
	\qquad s_{\rm th}=\frac{1}{4G\zh}=\frac{\pi c}{3\beta}.
	\label{eq:BTZ-thermal-entropy}
\end{equation}
This equality holds for every $b/\beta$, not only in the
high-temperature or large-interval limit.  As emphasized in
subsection~\ref{subsec:BTZ-TFD-map}, however, this is a decomposition by
endpoint channel in the reflection-symmetric purification.  It is not a
purification-independent separation of $S_A$ into its vacuum value and a
thermal correction: the same-boundary term is itself temperature dependent.

The contour decomposition is local on $A$.  For $x\in(-b,b)$,
\begin{align}
	s_A^{\partial}(x)
	&=\int_{A^c_\partial}\dd y\,\cI_{\partial\partial}(x,y)=\frac{1}{8G\zh}\left[
	\coth\left(\frac{b-x}{2\zh}\right)
	+\coth\left(\frac{b+x}{2\zh}\right)-2\right],
	\label{eq:BTZ-boundary-contour}\\
	s_A^{h}(x)
	&=\int_h\dd y\,\cI_{\partial h}(x,y)
	=\frac{1}{4G\zh}.
	\label{eq:BTZ-horizon-contour}
\end{align}
Their sum is the ALC/PEE-selected thermal-CFT entanglement contour
\cite{Wen:2018whg,Wen:2020ech},
\begin{equation}
	s_A(x)=\frac{1}{8G\zh}\left[
	\coth\left(\frac{b-x}{2\zh}\right)
	+\coth\left(\frac{b+x}{2\zh}\right)\right].
	\label{eq:BTZ-total-contour}
\end{equation}
The constant term $s_A^h$ is the local thermal entropy density carried by
the TFD channel.  Its constancy follows from translation invariance of the
thermal state and should not be confused with a uniform distribution of
endpoints along the horizon.  The latter is the nontrivial function
$h_A(y)$ below.  The UV divergences near $x=\pm b$ reside entirely in
$s_A^\partial$, as expected because they arise from short-distance
correlations across the two entangling points rather than from the smooth
purifier channel.

Conversely, the horizon-resolved density sourced by the whole interval is
\begin{equation}
	h_A(y)=\int_{-b}^{b}\dd x\,\cI_{\partial h}(x,y)
	=\frac{1}{8G\zh}\left[
	\tanh\left(\frac{b-y}{\zh}\right)
	+\tanh\left(\frac{b+y}{\zh}\right)\right],
	\label{eq:BTZ-horizon-density}
\end{equation}
whose integral over the horizon equals
\eqref{eq:BTZ-horizon-PEE-direct}.  This is the endpoint-resolved origin of
the thermal horizon flux observed in the geodesic bit-thread construction
\cite{Caggioli:2024uza}.  In particular, a constant amount of horizon-channel
PEE is assigned to each source point $x\in A$, but the corresponding flux
arrives at the horizon with a profile concentrated beneath the interval.
Source resolution on $A$ and endpoint resolution on $h$ are therefore two
different marginals of the same mixed kernel.

The decisive check is that the macroscopic field reproduces these densities
pointwise, not merely after integration.  On the
regulated asymptotic boundary, with the inward unit normal pointing toward
larger $z$,
\begin{equation}
	\lim_{z\to0}
	\frac{v_A^z(x,z)}
	{z^2\sqrt{1-z^2/\zh^2}}
	=s_A(x),
	\qquad -b<x<b.
	\label{eq:BTZ-pointwise-boundary-flux}
\end{equation}
At the horizon, we use the signed proper-distance coordinate introduced in
\eqref{eq:BTZ-TFD-Fermi-metric}, for which
$v_A^\rho=-v_A^z/[z\sqrt{1-z^2/\zh^2}]$ and the outward normal of the
one-sided exterior is $-\partial_\rho$.  Since the horizon line element is
$\dd y/\zh$,
\begin{align}
	\left.\frac{-v_A^\rho}{\zh}\right|_{\rho=0}
	&=\frac{\sinh\left(\frac{b}{\zh}\right)
		\cosh\left(\frac{b}{\zh}\right)}
	{4G\zh
		\cosh\left(\frac{b-y}{\zh}\right)
		\cosh\left(\frac{b+y}{\zh}\right)}
	\nonumber\\
	&=h_A(y).
	\label{eq:BTZ-pointwise-horizon-flux}
\end{align}
Thus the endpoint kernels, their two marginal contour densities, and the
actual normal fluxes are mutually consistent point by point.  Together with
RT saturation, this closes the physical interpretation of the selected flow: the
boundary--boundary part accounts for correlations with the asymptotic
complement, while the horizon part accounts for the TFD purifier channel.

\subsubsection{Auxiliary threads}
\label{subsec:BTZ-auxiliary}
As advertized earlier, a second congruence is obtained by retaining the other half of a geodesic
continued through the horizon into the second BTZ exterior \cite{Caggioli:2024uza}.  In the signed proper-distance coordinate of subsection~\ref{subsec:BTZ-TFD-map}, this is
an ordinary continuation across $\rho=0$.  It becomes an ``auxiliary''
branch only when that second exterior is folded back onto the same
$0<z\leq\zh$ coordinate patch as the physical one-sided geometry; the
folding reverses the sign of
$\zeta(z)=\sqrt{1-z^2/\zh^2}$.  Thus this construction should be regarded first as
geometric continuation data, not as a third positive endpoint sector of the
one-sided PEE decomposition.
With the outward radial orientation on the folded second exterior, the
continued segment points from the horizon seam toward the auxiliary
asymptotic boundary.  In that limited sense it may be pictured as
horizon-originating.  This is an orientation of one half of a complete
cross-boundary chord, however, not a new physical source placed on the
one-sided horizon.

For a reference source $x_0=0$, the signed continuation may be represented
by
\begin{equation}
	\sqrt{1-\frac{z^2}{\zh^2}}
	=\frac{\sinh\left(\frac{x-y}{\zh}\right)}
	{\sinh\left(\frac{y}{\zh}\right)},
	\label{eq:BTZ-auxiliary-geodesic}
\end{equation}
where the sign of the square root is continued across the horizon.  If the
physical right segment runs from $x_0$ to the crossing point $y$, the folded
auxiliary segment runs from $y$ to the coordinate
$\widetilde x=2y-x_0$.  This is precisely the midpoint relation of the
complete TFD chord.  Consequently, $x_0$ labels the right endpoint of the
complete chord; it is not the asymptotic endpoint of the auxiliary segment
viewed in isolation.  At its
intersection $(x_m,z_m)$ with the reference RT geodesic, the horizon
coordinate satisfies
\begin{equation}
	\tanh\left(\frac{y}{\zh}\right)
	=\frac{\tanh\left(\frac{x_m}{\zh}\right)}
	{1+\sech\left(\frac{b}{\zh}\right)}.
	\label{eq:BTZ-auxiliary-endpoint-map}
\end{equation}
The corresponding tangent and the norm of the auxiliary PEE thread flow obtained by the same oriented flux-matching calculation are given by
\begin{align}
	\widehat\tau_{0}^{\mu}
	&=\left(
	\frac{z^2}{\zh}\tanh\left(\frac{y}{\zh}\right),
	-z\frac{\sinh\left(\frac{2(x-y)}{\zh}\right)}
	{\sinh\left(\frac{2y}{\zh}\right)}
	\right),
	\label{eq:BTZ-auxiliary-tangent}\\
	|\widehat V_O|
	&=\frac{1}{4G\zh}
	\sqrt{\sech^2\left(\frac{y}{\zh}\right)
		-\sech^2\left(\frac{b}{2\zh}\right)}.
	\label{eq:BTZ-auxiliary-elementary-norm}
\end{align}
After eliminating $y$ and translating the source to $x_0$, the elementary
auxiliary current is
\begin{align}
	\widehat V_{x_0}^{\mu}(x_m,z_m)
	&=\frac{1}{4G}
	\frac{z_m^2}
	{2\zh^2\left[\cosh\left(\frac{x_m-x_0}{\zh}\right)
		+\sqrt{1-\frac{z^2}{\zh^2}}\right]^2}
	\nonumber\\[-2pt]
	&\quad\times\left(
	\frac{z_m}
	{\zh}\sinh\left(\frac{x_m-x_0}{\zh}\right),
	\sqrt{1-\frac{z_m^2}{\zh^2}}
	\left[1-\sqrt{1-\frac{z_m^2}{\zh^2}}
	\cosh\left(\frac{x_m-x_0}{\zh}\right)\right]
	\right)\,.
	\label{eq:BTZ-auxiliary-elementary-current}
\end{align}
It is divergenceless and, unlike the physical fixed-source current, has no
point-source singularity at $x=x_0$ on the physical asymptotic boundary.
This is another indication that $x_0$ is an inherited TFD chord label rather
than a source of the folded auxiliary segment.  The current vanishes in the
zero-temperature limit $\zh\to\infty$, as expected for a branch whose
existence depends on the second exterior and its horizon seam. 

\paragraph{Auxiliary superposed flow.} Superposing the auxiliary currents emanating from within the physical interval $A$ leads to
\begin{align}
	\widehat v_A^\mu
	&=\frac{1}{4G}
	\frac{z^2\sinh\left(\frac{b}{\zh}\right)}
	{\zh^2
		\left[\cosh\left(\frac{b-x}{\zh}\right)
		+\sqrt{1-\frac{z^2}{\zh^2}}\right]
		\left[\cosh\left(\frac{b+x}{\zh}\right)
		+\sqrt{1-\frac{z^2}{\zh^2}}\right]}
	\nonumber\\[-2pt]
	&\quad\times\left(
	z\sinh\left(\frac{x}{\zh}\right),
	-\zh\sqrt{1-\frac{z^2}{\zh^2}}
	\left[\cosh\left(\frac{b}{\zh}\right)
	+\sqrt{1-\frac{z^2}{\zh^2}}
	\cosh\left(\frac{x}{\zh}\right)\right]
	\right),
	\label{eq:BTZ-auxiliary-flow}
\end{align}
which is identical to the auxiliary flow $\mathbf{V}_-$ obtained in \cite{Caggioli:2024uza}.
This field has a simple sheet interpretation.  If the physical flow below
is written as a function of the signed radial quantity $\zeta(z)$, then
\begin{equation*}
	\widehat v_A(x,z;\zeta)=v_A(x,z;-\zeta).
\end{equation*}
Hence $v_A$ and $\widehat v_A$ are the two exterior restrictions of one
analytic TFD flow.  They are to be glued across the horizon on the doubled
geometry, not added pointwise as two independent currents on the same
one-sided exterior.

At the seam their folded radial components are equal and oppositely
oriented.  Dividing by the horizon line element gives on both sides the
endpoint density $h_{A_R}(y)$ in
\eqref{eq:BTZ-TFD-horizon-endpoint-density}: the horizon sink of the right
segment is exactly the horizon source of its continued left segment.  This
matching is the local flux statement behind the complete TFD chord.
Its norm satisfies
\begin{align}
	|\widehat v_A|=\frac{1}{4G}
	\frac{\frac{z}{\zh}\sinh\left(\frac{b}{\zh}\right)}
	{\sqrt{
			\left[\cosh\left(\frac{b-x}{\zh}\right)
			+\sqrt{1-\frac{z^2}{\zh^2}}\right]
			\left[\cosh\left(\frac{b+x}{\zh}\right)
			+\sqrt{1-\frac{z^2}{\zh^2}}\right]}}
	<\frac{1}{4G}
	\label{eq:BTZ-auxiliary-norm}
\end{align}
throughout the physical exterior.  The strict inequality is consistent with
the folded interpretation: the physical RT surface is not a bottleneck for
this branch.  In particular, $\widehat v_A$ is neither normal to nor
saturating on $\gamma_A$.  With the same orientation convention used for
the physical $A$-flow,
\begin{equation}
	\widehat\Phi_A(\gamma_A)
	=\frac{1}{2G}\log\left[
	\sech\left(\frac{b}{\zh}\right)\right],
	\label{eq:BTZ-auxiliary-flux}
\end{equation}
which is negative and vanishes as $\zh\to\infty$.  The sign does not mean
that the norm or the underlying unoriented chord measure is negative.  It
means that the folded field crosses the physical RT surface opposite to the
normal chosen for the $A$-flow.  Equation
\eqref{eq:BTZ-auxiliary-flux} is therefore the signed flux of a non-maximal
field through a surface that is not its bottleneck.  It must not be added to
the two positive terms in \eqref{eq:BTZ-TFD-PEE-normalization}, and it does
not furnish an independent entropy channel of the one-sided thermal state.

\section{Finite-cutoff holography}
\label{sec:finite-cutoff}
In this section, we terminate the bulk on a timelike
Dirichlet surface $\Gamma_c$ and regard the theory living on that surface as
the finite-cutoff theory of
\cite{McGough:2016lol,Kraus:2018xrn,Hartman:2018tkw}, proposed as the holographic dual of $T\bar{T}$-deformed CFTs \cite{Zamolodchikov:2004ce,Cardy:2018sdv}.  The cutoff is therefore not merely a regulator to be removed at the end: it changes the proper
distance between endpoint degrees of freedom, rounds off the ultraviolet
singularities of the asymptotic two-point PEE density, and makes the RT surface a
finite geodesic segment whose endpoints lie directly on $\Gamma_c$.

This change also sharpens a distinction that is invisible at the conformal
boundary.  There are now three related, but logically different, objects:
(i) the exact endpoint-pair kernel obtained by differentiating the regulated
geodesic length; (ii) the macroscopic flow obtained by superposing the
corresponding endpoint-resolved geodesic currents; and (iii) an independent
max-flow representative constructed from the geodesics normal to the RT
surface.  The first two define what we call the \emph{PEE-selected}
representative.  The third solves the same max-flow/min-cut problem, but need
not reproduce the same local flux density on the cutoff boundary.  Keeping these
levels separate is essential: microscopic PEE chords may all be geodesic even
when the integral curves of their vector sum are not.

Throughout this section the two-point PEE kernel is built from the \emph{exact}
bulk distance, before any near-boundary expansion.  We first analyze the
Poincar\'e patch, where both max-flow representatives are available in closed
form and their difference can be isolated cleanly.  Subsection~\ref{subsec:cutoff-global}
then repeats the endpoint-resolved construction in global AdS$_3$.  The
finite-cutoff $T\bar T$ dictionary reviewed in
section~\ref{subsec:review-ttbar} supplies a useful interpretation of the
resulting scales.  Once the hard-cutoff interpretation specified in
section~\ref{subsec:review-ttbar} is assumed, we use the terms two-point PEE
kernel and PEE density for these positive endpoint measures.  The underlying
geometric identities do not depend on that field-theory interpretation.

\subsection{Poincar\'e \texorpdfstring{AdS$_3$}{AdS3}}
\label{subsec:cutoff-poincare}
We work on the truncated time slice of Poincar\'e AdS$_3$,
$\mathcal M_c=\{(x,z):z\geq\zc\}$, with metric
\begin{equation}
  \dd s^2=\frac{\dd x^2+\dd z^2}{z^2}\,,
  \label{eq:Poincare-cutoff-metric}
\end{equation}
and choose the interval
$A=[-b,b]\subset\Gamma_c$. The discussion proceeds in four steps. We first
derive the exact cutoff boundary two-point PEE kernel and its finite endpoint entanglement contour. We then lift the two-point PEEs to elementary geodesic currents and superpose their sources over $A$. We also construct the normal-geodesic representative in exact Fermi coordinates and compare its cutoff boundary calibration with the PEE-selected one.

\subsubsection{Two-point PEE and entanglement contour.}
\begin{figure}[ht]
	\centering
	\includegraphics[width=0.96\textwidth]{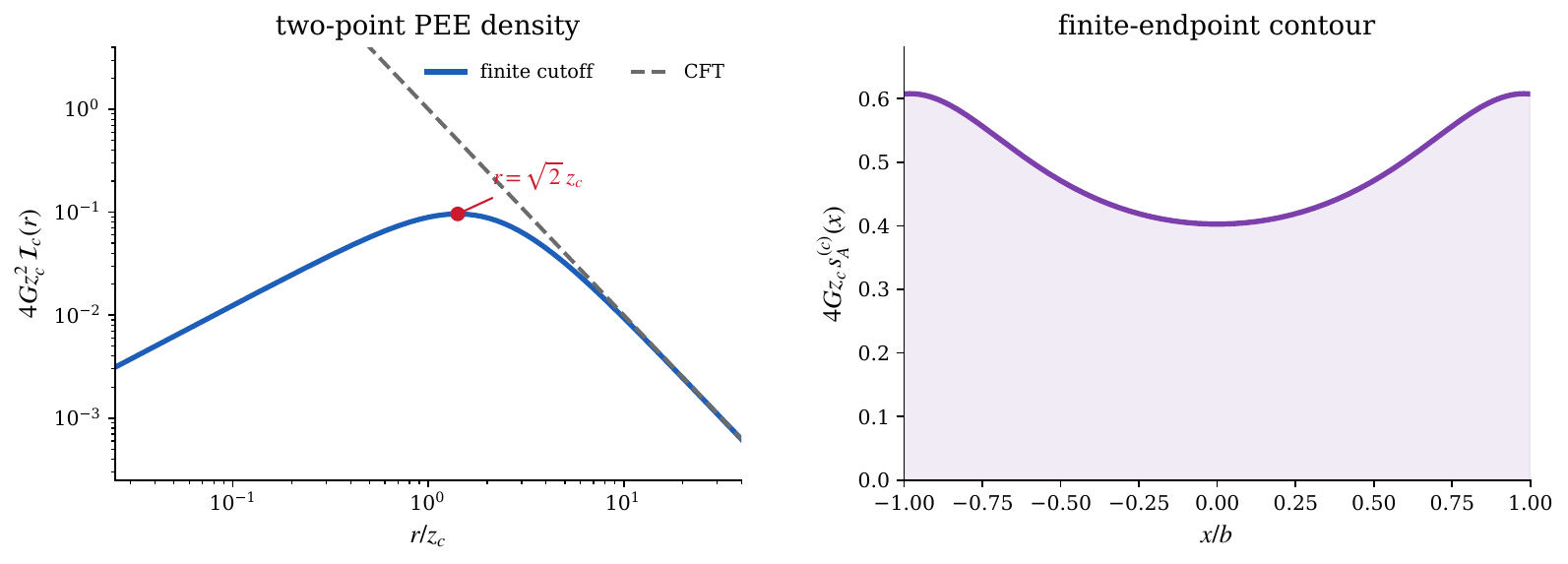}
	\caption{Finite-cutoff endpoint data for $b=1$ and $\zc=0.22$.
		Left: the dimensionless two-point PEE density
		$4G\zc^2\cI_c(r)$ (blue) and its CFT limit (gray dashed).  The exact
		density vanishes at coincidence, reaches its maximum at
		$r=\sqrt2\,\zc$, and approaches the CFT curve for $r\gg\zc$.
		Right: the dimensionless PEE contour
		$4G\zc s_A^{(c)}(x)$; the cutoff resolves the endpoint divergence into
		a finite enhancement.}
	\label{fig:pcut-kernel-contour}
\end{figure}

The RT surface homologous to $A$ is the portion of the circle
\begin{equation}
	\gamma_A:\qquad x^2+z^2=b^2+\zc^2.
	\label{eq:Poincare-cutoff-RT}
\end{equation}
The length of a geodesic joining two boundary points $(x_1,\zc)$ and $(x_2,\zc)$ is readily computed from \eqref{eq:length-Poincare} as
\begin{equation}
	\cL_c(x_1,x_2)=2\arcsinh\left(\frac{|x_1-x_2|}{2\zc}\right),
	\label{eq:Poincare-cutoff-length}
\end{equation}
so that the two-point PEE \eqref{eq:PEE-kernel-general} is given by
\begin{equation}
	\cI_c(x_1,x_2)=\frac{1}{4G}\frac{|x_1-x_2|}
	{\bigl[(x_1-x_2)^2+4\zc^2\bigr]^{3/2}}.
	\label{eq:Poincare-cutoff-kernel}
\end{equation}
Writing $r=|x_1-x_2|$, the kernel is symmetric, positive, and smooth for every
$r>0$.  Its two useful asymptotic regimes are
\begin{align}
	\cI_c(r)&=\frac{r}{32G\zc^3}
	\left[1-\frac{3r^2}{8\zc^2}
	+O\!\left(\frac{r^4}{\zc^4}\right)\right],
	&r\ll\zc,
	\label{eq:pcut-kernel-UV}\\
	\cI_c(r)&=\frac{1}{4G r^2}
	\left[1-\frac{6\zc^2}{r^2}+O\!\left(\frac{\zc^4}{r^4}\right)\right],
	&r\gg\zc.
	\label{eq:pcut-kernel-IR}
\end{align}
Thus the cutoff does not impose a hard minimum chord length.  Instead it
suppresses arbitrarily short chords continuously: $\cI_c(r)$ vanishes
linearly at coincidence, crosses over near the cutoff boundary scale, and approaches the
CFT density at large separation.  Its unique maximum occurs at
\begin{equation}
	r_{\rm peak}=\sqrt2\,\zc.
	\label{eq:pcut-kernel-peak}
\end{equation}
The peak therefore identifies a crossover in endpoint resolution, rather
than a sharp shortest thread.  Chords shorter than this scale are still
present, but their weight decreases because the finite cutoff prevents their
proper length from developing the asymptotic logarithmic singularity.
Figure~\ref{fig:pcut-kernel-contour} shows the same resolution mechanism in
both endpoint space and the interval contour.  The maximum of the kernel and
the finite endpoint enhancement of the contour are two projections of the
same fact: the cutoff boundary replaces arbitrarily short asymptotic bonds by a smooth
distribution over the cutoff scale.

The bipartite Crofton sum gives the exact RT entropy,
\begin{equation}
	S_A^{(c)}=\int_A\dd x\int_{A^c}\dd y\,\cI_c(x,y).
	\label{eq:pcut-PEE-sum}
\end{equation}
Integrating out the complementary endpoint defines the associated finite cutoff contour function,
\begin{align}
	s_A^{(c)}(x)
	&\equiv\int_{A^c}\dd y\,\cI_c(x,y)
	=\frac{1}{4G}\left[
	\frac{1}{\sqrt{(x+b)^2+4\zc^2}}
	+\frac{1}{\sqrt{(b-x)^2+4\zc^2}}
	\right],~~~-b<x<b.
	\label{eq:Poincare-cutoff-contour}
\end{align}
Direct integration gives \eqref{eq:pcut-TTbar-entropy}, consistently
with the finite-cutoff entanglement results of
\cite{Donnelly:2018bef,Chen:2018eqk,Jeong:2019ylz}.  We will show below that
this same function is the pointwise incoming flux of the PEE-selected max
flow.  Under the hard-cutoff interpretation stated at the beginning of this
section, it is therefore the finite-cutoff PEE contour.  In contrast with the
CFT contour, it remains finite at $x=\pm b$.  The endpoint enhancement has
not disappeared; it has been spread over a layer of proper width set by
$\zc$, while its integral continues to reproduce the complete RT entropy.

There is also a direct field-theory check that does not require interpreting
the full two-point PEE kernel microscopically.  Applying the translationally
invariant ALC formula \eqref{eq:review-ALC-translational} to the finite-cutoff
entropy \eqref{eq:pcut-TTbar-entropy} reproduces
\eqref{eq:Poincare-cutoff-contour} exactly.  Thus the one-point PEE contour
follows independently from the deformed interval entropy; the stronger
statement encoded in $\cI_c(x,y)$ is the endpoint-resolved pairing that
realizes the same contour.

\paragraph{$T\bar T$ interpretation.}
Using the convention and dictionary in
\eqref{eq:TTbar-flow}--\eqref{eq:TTbar-cutoff-dictionary}, the entropy of an
interval of length $\ell$ and the two-point PEE density at
separation $r$ are
\begin{align}
	S_\mu(\ell)
	&=\frac{c}{3}\arcsinh\left(\frac{\ell}{2\zc}\right)
	=\frac{c}{3}\arcsinh\left(
	\ell\sqrt{\frac{3\pi}{2c\mu}}\right),
	\label{eq:pcut-TTbar-entropy}\\
	\cI_\mu(r)
	&=\frac{c}{6}\frac{r}{(r^2+4\zc^2)^{3/2}}
	=\frac{c}{6}\frac{r}
	{\left(r^2+\frac{2c\mu}{3\pi}\right)^{3/2}}.
	\label{eq:pcut-TTbar-kernel}
\end{align}
For a general interval $A=[u,v]$, the corresponding contour is
\begin{equation}
	s_A^{(\mu)}(x)=\frac{c}{6}\left[
	\frac{1}{\sqrt{(x-u)^2+4\zc^2}}
	+\frac{1}{\sqrt{(v-x)^2+4\zc^2}}
	\right].
	\label{eq:pcut-TTbar-contour}
\end{equation}
The deformation dependence is summarized by
\begin{align}
	\left(2\mu\partial_\mu+\ell\partial_\ell\right)S_\mu(\ell)&=0,
	\label{eq:pcut-S-flow}\\
	\left(2\mu\partial_\mu+r\partial_r+2\right)\cI_\mu(r)&=0,
	\label{eq:pcut-I-flow}\\
	\partial_\mu\cI_\mu(r)
	&=-\frac{c^2r}{6\pi
		\left(r^2+\frac{2c\mu}{3\pi}\right)^{5/2}}<0.
	\label{eq:pcut-I-mu-derivative}
\end{align}
Thus the deformation suppresses arbitrarily short bonds.  The density is
maximal at
\begin{equation}
  r_{\rm peak}=\sqrt2\,\zc
  =\sqrt{\frac{c\mu}{3\pi}}.
  \label{eq:pcut-TTbar-resolution-scale}
\end{equation}
The two-point PEE kernel therefore exposes a real-space resolution scale
proportional to $\sqrt\mu$ in the present normalization.  This classical
hard-cutoff scale should be distinguished from, but compared with, the
nonperturbative $T\bar T$ entanglement scale discussed in
\cite{Lai:2025thy}: the latter arises from replica dynamics in the deformed
theory, whereas \eqref{eq:pcut-TTbar-resolution-scale} is read directly from
the positive classical cutoff PEE density.  Their common $\sqrt\mu$ scaling
is a useful point of contact rather than an identification of the two
mechanisms.

The same kernel gives a simple information-theoretic consistency check.  For
$r>0$,
\begin{equation}
  \cI_\mu(r)=-\frac12 S_\mu''(r)\geq0,
  \label{eq:pcut-kernel-entropy-concavity}
\end{equation}
so positivity of the static two-point PEE density is equivalent to concavity
of the equal-time interval entropy.  For three contiguous intervals
$A,B,C$ on the cutoff boundary,
\begin{equation}
  I(A:C\,|\,B)
  =S_{AB}+S_{BC}-S_B-S_{ABC}
  =2\int_A\dd x\int_C\dd y\,\cI_\mu(|x-y|)\geq0.
  \label{eq:pcut-CMI-PEE}
\end{equation}
Thus equal-time strong subadditivity is represented by positivity of a
bundle of endpoint-resolved PEE threads.  The contrast with boosted strong
subadditivity can be made explicit.  For a Lorentz-invariant interval entropy,
the infinitesimal boosted condition requires
\begin{equation}
  \ell^2 S''(\ell)+\ell S'(\ell)\leq0.
  \label{eq:pcut-boosted-SSA-condition}
\end{equation}
For the finite-cutoff entropy in \eqref{eq:pcut-TTbar-entropy}, however,
\begin{equation}
  \ell^2 S_\mu''(\ell)+\ell S_\mu'(\ell)
  =\frac{4c\zc^2\ell}{3(\ell^2+4\zc^2)^{3/2}}>0
  \qquad (\zc>0).
  \label{eq:pcut-boosted-SSA-violation}
\end{equation}
This is the boosted strong-subadditivity violation discussed in
\cite{Lewkowycz:2019xse}.  There is therefore no tension between the
positive static PEE kernel \eqref{eq:pcut-kernel-entropy-concavity} and the
known ultraviolet nonlocality of the deformed theory: our construction is
intrinsically static and does not imply a single Lorentz-covariant positive
pairing kernel for arbitrary boosted cuts.

A useful scale-dependent entropic response is
\begin{equation}
	3\ell\,\partial_\ell S_\mu
	=c\frac{\ell}{\sqrt{\ell^2+4\zc^2}}.
	\label{eq:pcut-entropic-response}
\end{equation}
This quantity interpolates from zero at distances far below the deformation scale to $c$
at distances far above it.  We do not call
\eqref{eq:pcut-entropic-response} a $c$-function: the $T\bar T$ deformation
is irrelevant, and this interpolation is not an application of a
two-dimensional RG monotonicity theorem.

The total two-point PEE weight emitted from an infinitesimal cutoff element is finite:
\begin{equation}
	\int_{-\infty}^{\infty}\dd y\,\cI_\mu(x,y)
	=\frac{1}{4G\zc}
	=\frac{c}{6\zc}
	=\sqrt{\frac{\pi c}{6\mu}}.
	\label{eq:pcut-PEE-capacity}
\end{equation}
The same quantity is the coefficient of the short-interval volume law,
\begin{equation}
	S_\mu(\ell)=\frac{\ell}{4G\zc}+O(\ell^3).
	\label{eq:pcut-short-volume-law}
\end{equation}
The cutoff therefore replaces the divergent local pairwise valence of a CFT
by a finite PEE capacity per unit cutoff-boundary length.

The finite capacity in \eqref{eq:pcut-PEE-capacity} is a property of the
positive real-cutoff boundary branch.  It should not be reinterpreted as a derivation of
Hagedorn growth: that phenomenon belongs to a different analytic branch of
the deformed spectrum \cite{Aharony:2018bad}.  Continuing
\eqref{eq:pcut-TTbar-kernel} to that branch destroys reality or positivity at
short separation, so it lies outside the Riemannian flow construction used
here.

\subsubsection{PEE thread flow}
We now determine the PEE thread current emanating from a generic source point
$P_0=(x_0,z_c)$ on $\Gamma_c$.  As earlier, we take the RT surface
$\gamma_A$ in \eqref{eq:Poincare-cutoff-RT} as the reference surface
$\Sigma$.  A generic PEE thread emitted from $P_0$, crossing $\Sigma$ and
ending at $(y,z_c)$ on $\Gamma_c$, has the circular profile,
\begin{figure}[ht]
	\centering
	\includegraphics[width=0.86\textwidth]{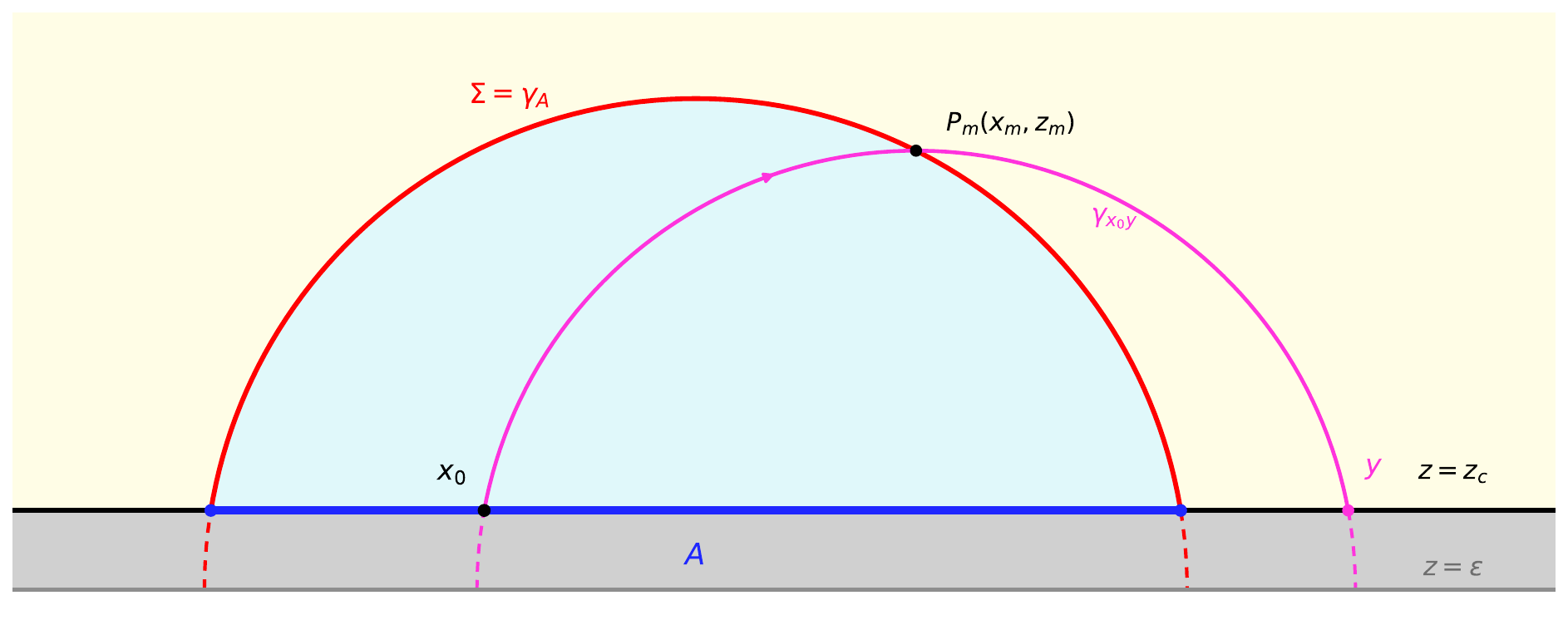}
	\caption{Finite-cutoff Poincar\'e geometry (schematic, not to scale).
		The black line is the cutoff boundary $z=\zc$, while the gray line
		$z=\epsilon$ indicates a regulated copy of the asymptotic boundary; the
		gray strip is the excised region between them.  The blue segment is
		$A$, the red circular geodesic is its RT surface
		$\Sigma=\gamma_A$, and the magenta circular geodesic
		$\gamma_{x_0y}$ is an elementary PEE thread joining the cutoff-boundary
		points $x_0$ and $y$ and crossing the RT surface at
		$P_m=(x_m,z_m)$.  Solid portions denote the physical geometry above the
		cutoff boundary, whereas the same-color dashed portions show the formal
		continuation of the circles into the excised region toward
		$z=\epsilon$.  The small nonzero $\epsilon$ is used only to display the
		regulated asymptotic line; the analytic formulas correspond to the
		usual $\epsilon\to0$ Poincar\'e boundary.  The chord flux across
		$\gamma_A$ is normalized by the exact finite-distance kernel
		\eqref{eq:Poincare-cutoff-kernel}.}
	\label{fig:Poincare-cutoff-threads}
\end{figure}
\begin{equation}
  \left(x-\frac{x_0+y}{2}\right)^2+z^2
  =\zc^2+\frac{(y-x_0)^2}{4}.
  \label{eq:Poincare-cutoff-geodesic}
\end{equation}
The
endpoint geometry and the role of the exact finite-distance normalization
are summarized in figure~\ref{fig:Poincare-cutoff-threads}.
Demanding that \eqref{eq:Poincare-cutoff-geodesic} pass through the point $P_m=(x_m,z_m)$ on $\Sigma=\gamma_A$ leads to the endpoint map
\begin{equation}
  y=x_m+\frac{z_m^2-\zc^2}{x_m-x_0}.
  \label{eq:Poincare-cutoff-endpoint-map}
\end{equation}
The unit normal covector to the reference surface is given by
\begin{align}
	n_{\Sigma,\mu}=\frac{1}{z\sqrt{x^2+z^2}}\left(x,z\right)
\end{align}
The oriented unit tangent to this PEE thread is obtained as
\begin{equation}
  \tau_{x_0}^{\mu}
  =\frac{2z^2}{D_c}\left(
  x-x_0,\frac{z^2-(x-x_0)^2-\zc^2}{2z}
  \right)\,.
  \label{eq:Poincare-cutoff-unit-tangent}
\end{equation}
where we have defined 
\begin{align}
	D_c=\sqrt{\left((x-x_0)^2+z^2-z_c^2\right)^2+4z_c^2(x-x_0)^2}
\end{align}
As earlier, the flux-matching condition \eqref{eq:flux-matching-general} leads to the norm of the PEE thread flow as
\begin{align}
	\left|V_{x_0}\right|=\frac{1}{4G}\frac{z}{D_c^2}\left((x-x_0)^2+z^2-z_c^2\right)\,.\label{eq:Poincare-cutoff-norm}
\end{align}
Away from its distributional cutoff boundary source, the current is tangent to
\eqref{eq:Poincare-cutoff-geodesic} and satisfies $\nabla\cdot V_{x_0}=0$
away from the source. The singularity at $P_0$ is a boundary distribution, while the cutoff boundary flux is precisely the endpoint-pair measure.
This is the sense in which $V_{x_0}$ resolves the contribution of one cutoff boundary
source.  It is important not to call $V_{x_0}$ by itself a max flow: an
elementary current need not satisfy the bit-thread norm bound.  The bound
emerges only after the sources in $A$ have been superposed.  Equations
\eqref{eq:Poincare-cutoff-unit-tangent} and \eqref{eq:Poincare-cutoff-norm} reduce smoothly to the asymptotic vacuum current obtained in \cite{Lin:2023rxc} as $\zc\to0$.

\subsubsection{PEE-selected bit threads}
The PEE-selected macroscopic field is the source superposition
\begin{equation}
	v_A^\mu(x,z)=\int_{-b}^{b}\dd x_0\,V_{x_0}^\mu(x,z).
	\label{eq:pcut-source-superposition}
\end{equation}
This integration is the coarse graining from endpoint space to a local bulk
current.  It preserves linear data such as divergence and calibrated flux,
but it does not preserve the trajectory of any particular microscopic
chord.  The streamlines of $v_A$ must therefore be determined only after the
vector sum has been performed.
Direct integration over $x_0\in[-b,b]$ gives
\begin{align}
  v_A^\mu&=\frac{z}{8G}\left(\frac{z^2-\zc^2-(x-b)^2}{R_-}
  -\frac{z^2-\zc^2-(x+b)^2}{R_+},2z\left(\frac{x+b}{R_+}-\frac{x-b}{R_-}\right)\right)\,,
  \label{eq:Poincare-cutoff-bit}
\end{align}
where we have defined
\begin{align}
	R_\pm&=\sqrt{\bigl[(x\pm b)^2+z^2-\zc^2\bigr]^2
		+4\zc^2(x\pm b)^2}\,.
\end{align}
We  may verify the norm bound directly from the expression \eqref{eq:Poincare-cutoff-bit}.  Introduce the vectors
\begin{align}
  \mathbf n_\pm=\frac{1}{R_\pm}\bigl((x\pm b)^2-z^2+\zc^2,2z(x\pm b)\bigr)\,,
  \label{eq:Poincare-unit-pairs}
\end{align}
with unit norm $\mathbf n_\pm^2=1$, where $A_\pm=z^2-\zc^2-(x\pm b)^2$.  In the orthonormal frame $e_{\hat x}=z\partial_x$, $e_{\hat z}=z\partial_z$, the bit-thread field \eqref{eq:Poincare-cutoff-bit} obeys the component identity
\begin{equation}
  \bigl(v_A^{\hat x},v_A^{\hat z}\bigr)
  =\frac{1}{8G}\left(\mathbf n_+-\mathbf n_-\right).
  \label{eq:Poincare-component-identity}
\end{equation}
Consequently,
\begin{equation}
  (4G|v_A|)^2
  =\frac12\left(1-\mathbf n_+\cdot\mathbf n_-\right)\leq1.
  \label{eq:Poincare-cutoff-norm-bound}
\end{equation}
The norm bound follows because the scalar product of two unit vectors lies in
$[-1,1]$. On the RT surface \eqref{eq:Poincare-cutoff-RT}, direct substitution gives $\mathbf n_+=-\mathbf n_-$.  The field saturates and its tangent component along the RT circle vanishes. More explicitly, the outward unit normal to the RT curve and the saturated
value of the PEE-selected flow are
\begin{equation}
	n_{\gamma_A}^{\mu}=\frac{z}{\sqrt{b^2+\zc^2}}(x,z),
	\qquad
	v_A^\mu\big|_{\gamma_A}=\frac{1}{4G}n_{\gamma_A}^{\mu}.
	\label{eq:pcut-RT-saturation}
\end{equation}
This proof is deliberately coordinate based: the unit vectors in \eqref{eq:Poincare-unit-pairs} are algebraic devices tied to the Poincar\'e chart. For the remainder of this subsection we use the single abbreviation $R_c=\sqrt{b^2+\zc^2}$ for the radius of the complete RT geodesic \eqref{eq:Poincare-cutoff-RT}.
\paragraph{Endpoint distance-difference representation.}
\label{subsec:pcut-DDP}
The explicit expression \eqref{eq:Poincare-cutoff-bit} conceals a simple
geometric structure.  Denote the two physical endpoints of $A$ by
$p_+=(b,\zc)$ and $p_-=(-b,\zc)$, and consider the geodesic distances
\begin{align}
  \cL_\pm(x,z)&=\cL\bigl((x,z),(\pm b,\zc)\bigr)
  =\arccosh\left[
  \frac{(x\mp b)^2+z^2+\zc^2}{2z\zc}\right]\,,
  \label{eq:pcut-endpoint-distances}
\end{align}
and define their signed half-difference by
\begin{equation}
  \Psi_A(x,z)=\frac12\left(\cL_+(x,z)-\cL_-(x,z)\right).
  \label{eq:pcut-stream-function}
\end{equation}
It is straightforward to
show that \eqref{eq:Poincare-cutoff-bit} is exactly
\begin{equation}
  v_A^x=\frac{z^2}{4G}\partial_z\Psi_A,
  \qquad
  v_A^z=-\frac{z^2}{4G}\partial_x\Psi_A.
  \label{eq:pcut-stream-field}
\end{equation}
Equivalently, with the orientation fixed by incoming flux on $A$, we may write the one-form identity
\begin{equation}
	v_A=\frac{1}{4G}\star\dd\Psi_A\,,
	\label{eq:pcut-form-flow}
\end{equation}
where $\star$ denotes the Hodge dual operation.
This compact answer is not an independent ansatz.  It follows from an exact
telescoping identity at the level of the elementary currents.  If $P(s)$
parametrizes the cutoff boundary $\Gamma_c$ and $\cL_s(X)=\cL(X,P(s))$ denotes the geodesic distance between an arbitrary bulk point $X$ and the cutoff boundary $P(s)$, then define the one-form field
\begin{equation}
	V_s=\frac{1}{8G}\star\dd\big(\partial_s\cL_s\big).
	\label{eq:pcut-elementary-potential}
\end{equation}
For $P(s)=(s,\zc)$, direct differentiation of
\eqref{eq:pcut-elementary-potential} gives precisely the PEE thread flow obtained in \cref{eq:Poincare-cutoff-norm,eq:Poincare-cutoff-unit-tangent}.  Hence, we have
\begin{equation}
	\int_{s_1}^{s_2}\dd s\,V_s
	=\frac{1}{8G}\star\dd\big(\cL_{s_2}-\cL_{s_1}\big).
	\label{eq:pcut-endpoint-telescoping}
\end{equation}
Taking $(s_1,s_2)=(-b,b)$ leads to the exact identity \eqref{eq:pcut-form-flow}.  Thus the continuum of microscopic boundary sources collapses, at the level of the coarse-grained
field, to the difference of the two endpoint distance functions. No source
has been placed on the RT surface; saturation there is a consequence of the
geometry of $\cL_\pm$.

Equation~\eqref{eq:pcut-endpoint-telescoping} is the two-finite-focus
specialization of the endpoint-reduction developed in Appendix A of
\cite{BasuWen:BCFTMinPurification}.  There the same distance-difference construction also covers
mixed finite--ideal configurations through Busemann functions and explains
why source sums over boundary, EOW-brane, horizon, or purifying RT-surface
pieces can reduce to the endpoints of the calibrated bottleneck.

Because $\star\dd\Psi_A$ is co-closed in two dimensions,
\eqref{eq:pcut-form-flow} immediately implies $\grad\cdot v_A=0$.  Moreover,
distance functions have unit hyperbolic gradient away from their foci, and
the triangle inequality gives
\begin{equation}
	\abs{v_A}
	=\frac{1}{4G}\abs{\grad\Psi_A}
	\leq\frac{1}{8G}
	\left(\abs{\grad \cL_+}+\abs{\grad \cL_-}\right)
	=\frac{1}{4G}.
	\label{eq:pcut-triangle-bound}
\end{equation}
On the segment $p_-p_+=\gamma_A$, the two unit distance gradients are
antiparallel.  The inequality therefore saturates and
\eqref{eq:pcut-RT-saturation} follows.  The component identity
\eqref{eq:Poincare-component-identity} and the distance-function proof
\eqref{eq:pcut-triangle-bound} are complementary descriptions of the same
PEE-selected field.

\paragraph{Integral curves, geodesicity and nesting.}
\label{sec:pcut-curves}
Because the bit-thread field is the Hodge dual of $\dd\Psi_A$, the stream function is a
first integral:
\begin{equation}
	\frac{\dd\Psi_A}{\dd s}
	=\partial_x\Psi_A\frac{\dd x}{\dd s}
	+\partial_z\Psi_A\frac{\dd z}{\dd s}=0.
	\label{eq:pcut-first-integral}
\end{equation}
Hence the corresponding coarse-grained integral curves are the hyperbolic Apollonius curves
\begin{equation}
	\cL_+(x,z)-\cL_-(x,z)=2\Psi_0.
	\label{eq:pcut-apollonius}
\end{equation}
An exact branch-sensitive representation of \eqref{eq:pcut-apollonius} is
\begin{align}
	\e^{2\Psi_0}
	&=\frac{
		(x-b)^2+z^2+\zc^2
		+\sqrt{\left[(x-b)^2+z^2+\zc^2\right]^2-4z^2\zc^2}}
	{(x+b)^2+z^2+\zc^2
		+\sqrt{\left[(x+b)^2+z^2+\zc^2\right]^2-4z^2\zc^2}}.
	\label{eq:pcut-apollonius-exp}
\end{align}
Eliminating the square roots gives the useful algebraic form
\begin{align}
	&\left[(x-b)^2+z^2+\zc^2\right]^2
	+\left[(x+b)^2+z^2+\zc^2\right]^2
	\nonumber\\
	&\quad-2\cosh(2\Psi_0)
	\left[(x-b)^2+z^2+\zc^2\right]
	\left[(x+b)^2+z^2+\zc^2\right]
	+4z^2\zc^2\sinh^2(2\Psi_0)=0,
	\label{eq:pcut-apollonius-quartic}
\end{align}
but this quartic must always be supplemented by the unsquared branch
condition \eqref{eq:pcut-apollonius-exp}; otherwise spurious components may
be introduced.  The central level $\Psi_0=0$ is $x=0$, the hyperbolic
perpendicular bisector of $p_-p_+$.  However, the non-central levels are generically
quartic rather than geodesic semicircles. 

The distance-difference form also makes the failure of macroscopic
geodesicity transparent.  Appendix~\ref{app:geodesic-curvature} derives the
universal curvature of a regular level set of $(L_+-L_-)/2$ in $\mathbb H^2$.
Applied here, it shows that the reflection-symmetric central streamline is
geodesic, whereas every regular noncentral streamline has nonzero curvature
at finite $\zc$.  The leading bending is $O(\zc^2)$, or equivalently $O(\mu)$
under \eqref{eq:TTbar-cutoff-dictionary},
\begin{equation}
	\abs{k_g}
	=16\zc^2\frac{b^2\abs{x}z^3(b^2+x^2+z^2)}
	{\left[((x-b)^2+z^2)((x+b)^2+z^2)\right]^{5/2}}
	+O(\zc^4).
	\label{eq:pcut-curvature-small-zc}
\end{equation}
This is the precise physical distinction between microscopic and
macroscopic geodesicity.  Each source-resolved current transports its weight
along a geodesic chord, but the superposed vector at a bulk point is the
density-weighted sum of many chord tangents.  Following that resultant
direction is a coarse-graining operation and need not reproduce any one of
the microscopic chords.  The curvature therefore diagnoses the
PEE-selected representative, not the entropy or the max-flow problem itself;
the normal construction below carries the same bottleneck flux with exactly
geodesic streamlines.

\paragraph{One-to-one endpoint map.}

The value of the stream function at a cutoff point $x_0\in(-b,b)$ is
\begin{equation}
  \Psi_c(x_0)
  =\arcsinh\left(\frac{b-x_0}{2\zc}\right)
  -\arcsinh\left(\frac{b+x_0}{2\zc}\right).
  \label{eq:pcut-cutoff-stream-label}
\end{equation}
It is strictly monotone,
\begin{equation}
  \Psi_c'(x_0)
  =-\frac{1}{\sqrt{(b-x_0)^2+4\zc^2}}
  -\frac{1}{\sqrt{(b+x_0)^2+4\zc^2}}<0,
  \label{eq:pcut-stream-monotone}
\end{equation}
and maps $(-b,b)$ onto $(-s_b,s_b)$ with reversed orientation, where
\begin{equation}
  s_b=\arcsinh\left(\frac{b}{\zc}\right).
  \label{eq:pcut-sb}
\end{equation}
We now parametrize the RT curve by
\begin{equation}
  x=R_c\tanh\eta,
  \qquad
  z=R_c\sech\eta,
  \qquad
  -s_b\leq\eta\leq s_b.
  \label{eq:pcut-RT-eta}
\end{equation}
The induced line element is $\dd s=\dd\eta$.  On $\gamma_A$, we may write 
\begin{equation}
  \cL_+=s_b-\eta~~,~~ \cL_-=s_b+\eta~~,~~ \Psi_A=-\eta.
  \label{eq:pcut-Psi-on-RT}
\end{equation}
Therefore the streamline emitted from $x_0$ reaches the RT curve at
\begin{equation}
  \eta_\star=-\Psi_c(x_0),
  \qquad
  (x_\star,z_\star)
  =\left(-R_c\tanh\Psi_c(x_0),
  R_c\sech\Psi_c(x_0)\right).
  \label{eq:pcut-endpoint-map-RT}
\end{equation}
The strict monotonicity in \eqref{eq:pcut-stream-monotone} proves that the ordering
of the threads is preserved from the cutoff interval to the bottleneck.

\paragraph{Regular non-crossing foliation.}
The function $\Psi_A$ is smooth and single-valued in the open entanglement
wedge $W_A^\circ$; its only
singularities are the foci $p_\pm$, which lie on the boundary.  A critical
point would require $\grad \cL_+=\grad \cL_-$.
The two sides are the unit tangents, at $X$, to the geodesics emitted from
$p_+$ and $p_-$.  Equality implies, by uniqueness of hyperbolic geodesics,
that $X$ lies on the complete geodesic through $p_-$ and $p_+$.  On the
segment $p_-p_+$ the two gradients are antiparallel, not equal.  Equality can
occur only on the two exterior rays.  In the present coordinates those rays
are the parts of $x^2+z^2=R_c^2$ with $\abs{x}>b$ and $z<\zc$, so they lie
outside the truncated wedge.  Hence
\begin{equation}
	\grad\Psi_A\neq0
	\qquad\text{throughout }W_A^\circ.
	\label{eq:pcut-no-critical-points}
\end{equation}
By the implicit-function theorem, every level set is a smooth
one-dimensional curve.  Two distinct level sets cannot intersect because
$\Psi_A$ is single-valued.  Equivalently, uniqueness of solutions to the
first-order flow equations forbids two distinct streamlines from crossing.
Thus the integral curves form a regular, ordered foliation of the open wedge.

This is \emph{streamline ordering within one max-flow representative}.  It
should not be conflated with simultaneous locking of an arbitrary nested
family of boundary regions, which is a stronger global statement about one
flow shared by several boundary regions.

\subsubsection{Geodesic bit threads}
\label{sec:pcut-geodesic-flow}
We have built the bit-thread field from the PEE threads, but the max-flow problem itself
does not select a unique representative \cite{Freedman:2016zud,Agon:2018lwq,Caggioli:2024uza}.  To explore the non-uniqueness, we
now apply the normal-geodesic prescription of
\cite{Agon:2018lwq}.  Starting at the bottleneck, the construction shoots
the unique hyperbolic geodesic normal to $\gamma_A$ at each point and then
fixes the magnitude along the congruence by flux conservation.  It remains
exact at finite cutoff and saturates the same RT surface, but its cutoff boundary
calibration will differ from the PEE contour.  This provides a
controlled comparison in which the global entropy is held fixed while the
local boundary-to-bulk assignment changes.

\paragraph{Normal congruence and Fermi coordinates.}

Use the unit-speed parametrization of the RT surface \eqref{eq:pcut-RT-eta}. The hyperbolic
geodesic normal to $\gamma_A$ at the point labelled by \(\eta\neq0\) is
\begin{equation}
	\left(x-R_c\coth\eta\right)^2+z^2
	=R_c^2\csch^2\eta .
	\label{eq:pcut-normal-geodesic-circle}
\end{equation}
For \(\eta=0\), the limiting geodesic is the vertical line \(x=0\).
The circles \eqref{eq:pcut-normal-geodesic-circle} are mutually disjoint and
form the standard normal foliation of \(\Htwo\) around the complete geodesic
$x^2+z^2=R_c^2$.  Restricting this foliation to the active RT segment gives
the thread bundle relevant to $A$.

Let \(\lambda\) be signed proper distance along the normal curves, with
\(\lambda=0\) on the complete RT geodesic.  We orient it so that $\lambda$
increases from the cutoff boundary interval, through $\gamma_A$, toward the complementary
side.  Exact Fermi normal coordinates follow from the construction reviewed in appendix~\ref{appB},
\begin{align}
	x(\eta,\lambda)
	&=R_c\frac{\cosh\lambda\,\sinh\eta}
	{\cosh\lambda\cosh\eta-\sinh\lambda},
	\label{eq:pcut-Fermi-x}\\
	z(\eta,\lambda)
	&=\frac{R_c}
	{\cosh\lambda\cosh\eta-\sinh\lambda}.
	\label{eq:pcut-Fermi-z}
\end{align}
They put the metric in the form
\begin{equation}
	\dd s^2=\dd\lambda^2+\cosh^2\lambda\,\dd\eta^2 .
	\label{eq:pcut-Fermi-metric}
\end{equation}
Thus \(\partial_\lambda\) is the hyperbolic unit tangent to the normal
geodesics and $\cosh\lambda$ is the transverse Jacobi factor.  This is the
only input needed to determine the norm from flux conservation.  The same
Jacobi factor appears in the BTZ construction of
appendix~\ref{appB}, because both constant-time slices
are locally $\Htwo$.  Their different endpoint sectors arise from the global
placement of the cutoff boundary and horizon, not from a different local dilution law.

\paragraph{Flux conservation and the exact vector field.}
As in appendix~\ref{appB}, we write a flow tangent to the congruence as
\begin{equation}
	v_{\rm geo}=\frac{1}{4G}f(\lambda)\,\partial_\lambda .
	\label{eq:pcut-geo-ansatz}
\end{equation}
The divergencelessness condition gives
\begin{equation}
	0=\grad_\mu v_{\rm geo}^{\mu}
	=\frac{1}{4G\cosh\lambda}
	\partial_\lambda\!\left(\cosh\lambda\,f\right).
	\label{eq:pcut-geo-divergence-Fermi}
\end{equation}
The conservation equation fixes $\cosh\lambda\,f(\lambda)$ to be constant.
Pointwise saturation on \(\gamma_A\) supplies the initial condition
\(f(0)=1\), and hence
\begin{equation}
v_{\rm geo}
		=\frac{1}{4G}\sech\lambda\,\partial_\lambda,
	\label{eq:pcut-geo-flow-Fermi}
\end{equation}
which is the same local dilution law as
\eqref{eq:app-BTZ-Fermi-flow}.  This agreement is expected: both spatial
geometries are locally $\Htwo$, so their normal congruences have the same
transverse Jacobi factor even though their global endpoint sectors differ.
It follows immediately that
\begin{equation}
	\grad\cdot v_{\rm geo}=0,
	\qquad
	\abs{v_{\rm geo}}
	=\frac{1}{4G}\sech\lambda
	\leq\frac{1}{4G},
	\label{eq:pcut-geo-properties-Fermi}
\end{equation}
with equality only at the RT bottleneck $\lambda=0$.  In other words, the same Jacobi field
that measures the transverse spreading of the normal congruence fixes the decay of its thread density.

To compare this representative bit-thread field directly with the PEE-selected bit threads in \eqref{eq:Poincare-cutoff-bit}, we return to Poincar\'e coordinates. The inverse Fermi relations are
\begin{equation}
	\sinh\lambda=\frac{x^2+z^2-R_c^2}{2R_cz}~~,~~
	\tanh\eta=\frac{2R_cx}{R_c^2+x^2+z^2}.
	\label{eq:pcut-inverse-Fermi}
\end{equation}
The unit tangent $\partial_\lambda$ is
\begin{equation}
	\del_\lambda=\tau_{\rm geo}^{\mu}
	=\left(
	\frac{2xz^2}{Q_c},
	\frac{z(R_c^2-x^2+z^2)}{Q_c}
	\right).
	\label{eq:pcut-geo-unit-tangent}
\end{equation}
where we have defined the positive function
\begin{equation}
	Q_c^2(x,z)
	=\left(R_c^2+x^2+z^2\right)^2-4R_c^2x^2\,.
	\label{eq:pcut-Qgeo}
\end{equation}
Substitution in \eqref{eq:pcut-geo-flow-Fermi} gives the geodesic bit-thread field
\begin{align}
	v_{\rm geo}^{\mu}=\frac{1}{4G}\frac{2R_c z^2}{Q_c^2}\left(2xz,R_c^2-x^2+z^2\right)\,,
	\label{eq:pcut-geo-v}
\end{align}
identical to the geodesic bit threads obtained in \cite{Agon:2018lwq,Caggioli:2024uza} with the replacement $R\to R_c$.
On $\gamma_A$, equation \eqref{eq:pcut-geo-v} reduces to
\begin{equation}
	v_{\rm geo}^{\mu}\big|_{\gamma_A}
	=\frac{1}{4G}\left(\frac{xz}{R_c},\frac{z^2}{R_c}\right)
	=\frac{1}{4G}n_{\gamma_A}^{\mu}.
	\label{eq:pcut-geo-RT-saturation}
\end{equation}
Thus \(v_{\rm geo}\) and \(v_A\) agree pointwise on the bottleneck, not only
in their integrated flux.  Away from $\gamma_A$, however, their directions
and magnitudes need not agree.

The corresponding stream function is the difference of the two geodesic lengths associated with the ideal endpoints $x=\pm R_c$ of the complete RT
geodesic ending on the asymptotic boundary $z=\epsilon$:
\begin{equation}
	\Psi_{\rm geo}(x,z)
	=-\eta(x,z)
	=\frac12\log
	\frac{(x-R_c)^2+z^2}{(x+R_c)^2+z^2}.
	\label{eq:pcut-geo-stream-function}
\end{equation}
Indeed,
\begin{equation}
	v_{\rm geo}^{x}
	=\frac{z^2}{4G}\partial_z\Psi_{\rm geo},
	\qquad
	v_{\rm geo}^{z}
	=-\frac{z^2}{4G}\partial_x\Psi_{\rm geo}.
	\label{eq:pcut-geo-stream-field}
\end{equation}
Its level sets are precisely \eqref{eq:pcut-normal-geodesic-circle}.  Hence,
in sharp contrast with the noncentral PEE-selected curves, we have $k_g^{(\rm geo)}=0$
for every streamline.

\paragraph{Cutoff endpoint map and boundary density.}

The cutoff boundary calibration follows from the same congruence \eqref{eq:pcut-normal-geodesic-circle}.  A normal geodesic
that intersects $\Gamma_c$ at $x$ has its two boundary intersections as the
roots of \eqref{eq:pcut-normal-geodesic-circle} evaluated at $z=z_c$. 
The product of these roots immediately gives the integral-curve partner map
\begin{equation}
	x\in A\quad\longleftrightarrow\quad
		y=\frac{b^2+2\zc^2}{x}\in A^c\,
	\label{eq:pcut-geo-partner-map}
\end{equation}
For $x\neq0$, while the central line reaches the ideal endpoint at infinity.
For $0<x\leq b$, one has
\begin{equation}
	y\geq\frac{b^2+2\zc^2}{b}=b+\frac{2\zc^2}{b},
	\label{eq:pcut-geo-gap}
\end{equation}
and similarly on the left.  The active normal-geodesic bundle therefore does
not reach the immediate complementary strips
\(b<\abs{y}<b+2\zc^2/b\).  The full smooth Fermi flow contains additional
\(A^c\)-to-\(A^c\) curves, which do not contribute to the flux of $A$.
Alternatively, one may set the field to zero outside
\(\abs{\eta}\leq s_b\).  Because the two support boundaries are themselves
streamlines, this truncation introduces no normal flux across the jump and
remains divergenceless in the weak sense. The gap is therefore not missing entropy.  It records a particular integral-curve pairing chosen by the normal congruence, whereas the PEE-selected microscopic ensemble continues to assign positive weight to all pairs across the entangling cut.

The cutoff-boundary point $x\in(-b,b)$ reaches the RT curve at
\begin{equation}
	\eta_{\rm geo}(x)
	=\arctanh\left(\frac{2R_cx}{x^2+b^2+2\zc^2}\right).
	\label{eq:pcut-geo-landing-map}
\end{equation}
Differentiating this monotone map gives the incoming cutoff boundary contour function,
\begin{align}
	s_{\rm geo}(x)
	&=\left.\frac{v_{\rm geo}^{z}}{\zc^2}\right|_{z=\zc}
	=\frac{1}{4G}\frac{\dd\eta_{\rm geo}}{\dd x}
	=\frac{1}{4G}
	\frac{2R_c(b^2+2\zc^2-x^2)}
	{(x^2+b^2+2\zc^2)^2-4R_c^2x^2}.
	\label{eq:pcut-geo-boundary-density}
\end{align}
Since $\eta_{\rm geo}(\pm b)=\pm s_b$, we recover the normalization
\begin{equation}
	\int_{-b}^{b}\dd x\,s_{\rm geo}(x)
	=\frac{2s_b}{4G}
	=S_A^{(c)}.
	\label{eq:pcut-geo-total-flux}
\end{equation}

\paragraph{Comparison with the PEE-selected max flow.}
\label{subsec:pcut-flow-comparison}

\begin{figure}[t]
	\centering
	\includegraphics[width=0.96\textwidth]{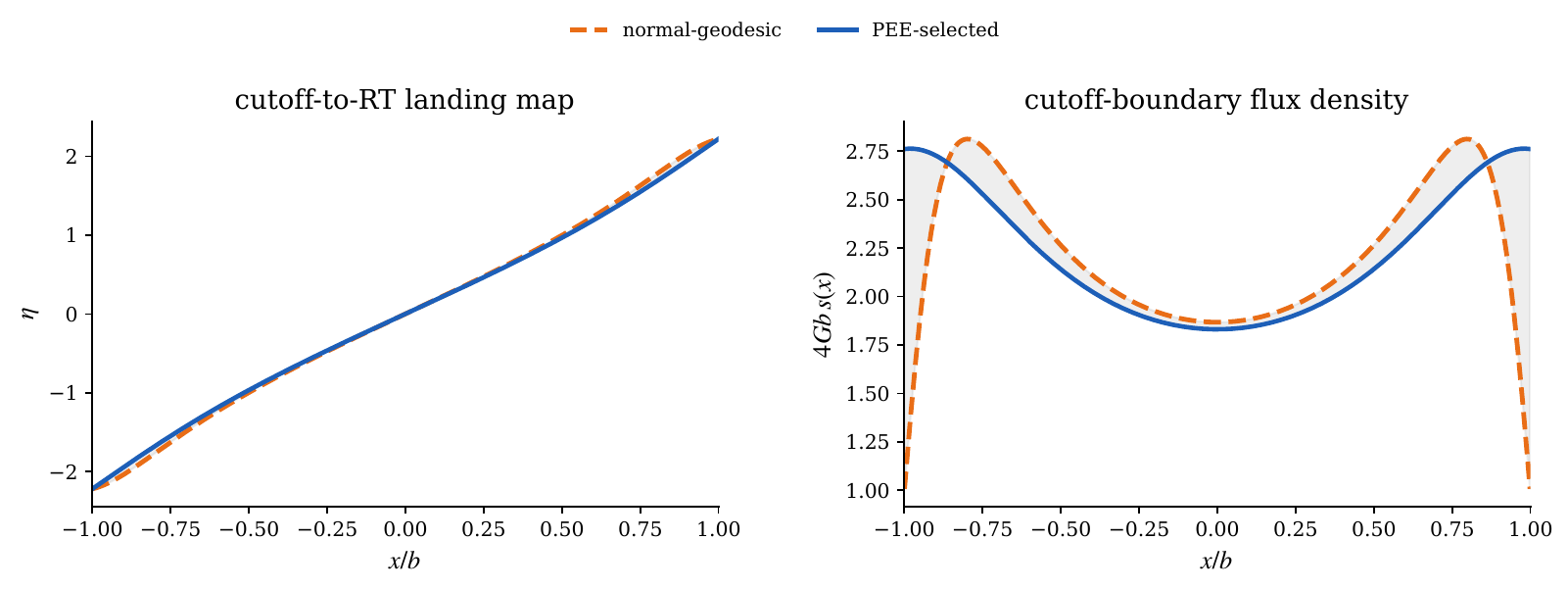}
	\caption{Exact cutoff boundary calibration for $b=1$ and $\zc=0.22$.  Left: the
		cutoff--to--RT maps $\eta_{\rm geo}(x)$ (orange dashed) and
		$\eta_{\rm PEE}(x)$ (blue).  They share the same endpoint range but differ
		in the interior.  Right: their derivatives, equivalently the dimensionless
		incoming cutoff boundary densities.  The finite difference between the curves is the
		local information that is invisible to the integrated RT flux.}
	\label{fig:pcut-wall-calibration}
\end{figure}

For the PEE-selected representative, the boundary-to-RT map obtained in
\eqref{eq:pcut-endpoint-map-RT} can be written as
\begin{equation}
	\eta_{\rm PEE}(x)
	=-\Psi_c(x)
	=\arcsinh\left(\frac{b+x}{2\zc}\right)
	-\arcsinh\left(\frac{b-x}{2\zc}\right),
	\label{eq:pcut-PEE-landing-map}
\end{equation}
and
\begin{equation}
	s_{\rm PEE}(x)
	=\frac{1}{4G}\frac{\dd\eta_{\rm PEE}}{\dd x}
	=s_A^{(c)}(x).
	\label{eq:pcut-PEE-density-comparison}
\end{equation}
Although both maps take $(-b,b)$ onto the same RT parameter interval
$(-s_b,s_b)$, they are not equal at finite cutoff.  Their derivatives, and
hence their pointwise cutoff boundary fluxes, differ as well.  For example, the
one-sided endpoint values are
\begin{equation}
	s_{\rm geo}(b^-)=\frac{1}{4G R_c},
	\qquad
	s_{\rm PEE}(b^-)
	=\frac{1}{8G}\left(\frac{1}{R_c}+\frac{1}{\zc}\right).
	\label{eq:pcut-endpoint-density-comparison}
\end{equation}
The common endpoint values $\eta(\pm b)=\pm s_b$ explain how distinct local
densities can integrate to the same maximal flux.  Figure~\ref{fig:pcut-wall-calibration} displays this distinction directly: the two
landing maps agree at the ends of the RT segment, while the finite area
between their derivatives records the redistribution of flux along the
cutoff boundary.

\begin{figure}[t]
	\centering
	\includegraphics[width=0.98\textwidth]{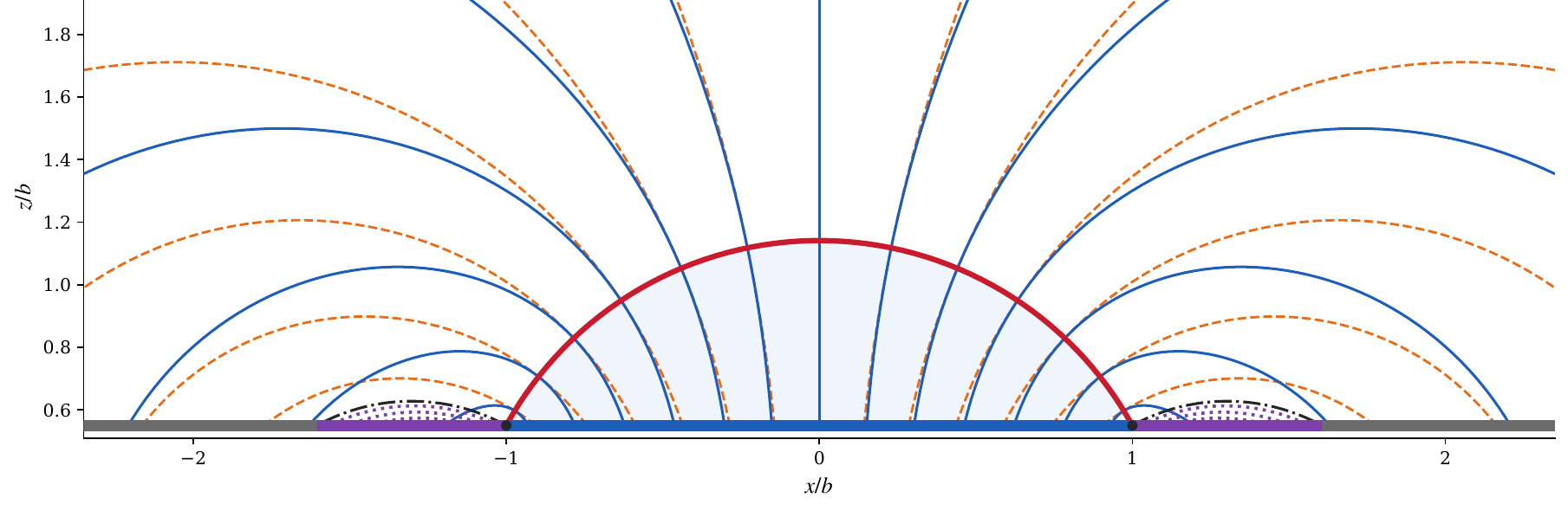}
	\par{\normalsize
		\begin{tabular}{@{}c@{\qquad}c@{\qquad}c@{}}
			\flowkey{floworange,dashed}{1.1pt}{active normal geodesics}
			&
			\flowkey{flowblue}{1.1pt}{PEE-selected streamlines}
			&
			\flowkey{flowpurple,dotted}{1.3pt}{normal spectators}
			\\[0.45em]
			\flowkey{flowgray,dash dot}{1.1pt}{support boundary}
			&
			\flowkey{flowred}{2.0pt}{RT surface}
			&
			{}
		\end{tabular}
	}\par
	\caption{Complete finite-cutoff Poincar\'e streamlines for $b=1$ and
		$\zc=0.55$, drawn on a single set of axes.  The larger illustrative
		cutoff makes the spectator cutoff-boundary strip visible at manuscript scale.
		Orange dashed curves are the
		active normal geodesics and blue solid curves are the PEE-selected
		distance-difference levels with the same RT labels.  All active curves are
		extended beyond the shaded entanglement wedge and through the red RT
		bottleneck to their second cutoff-boundary intersection; the central member reaches
		the ideal point at infinity.  Purple dotted curves are the smooth
		normal-congruence spectators with both endpoints in $A^c$, and the
		dash-dotted support curves pass through the RT endpoints.  The spectators
		carry no flux from $A$ and may be removed by the weakly divergenceless
		support truncation described in the text.}
	\label{fig:pcut-flow-comparison}
\end{figure}

The PEE-selected \emph{microscopic} thread bundle remains distributed over all
endpoint pairs through $\cI_c(x,y)$.  Its \emph{macroscopic} integral curve
through $x\in A$ nevertheless has a definite second boundary intersection: it
is the other solution $y\in A^c$ of
$\Psi_A(y,\zc)=\Psi_A(x,\zc)$.  The normal-geodesic partner map
\eqref{eq:pcut-geo-partner-map}, the PEE chord pairing, and the second boundary
intersection of a coarse-grained PEE streamline are therefore three
different notions and should not be identified.  Figure~\ref{fig:pcut-flow-comparison} consequently overlays the two macroscopic
representatives using the same RT labels and follows every displayed curve
past the red bottleneck into the complementary side of the cutoff geometry.
The continuation is essential: inside the entanglement wedge the two
families can look like alternative ways of reaching the same cut, whereas
their different cutoff-boundary endpoints become visible only after the complete
streamlines are shown.

The purple curves in figure~\ref{fig:pcut-flow-comparison} also resolve the
apparent missing-cutoff-boundary region in \eqref{eq:pcut-geo-gap}.  They are the
$|\eta|>s_b$ members of the same smooth normal Fermi congruence and join two
points of the immediate complement.  No spectator crosses the physical RT
segment, so adding or deleting them cannot change the entropy.  By contrast,
the PEE-selected curves bend so that the active bundle itself reaches the
adjacent complement.  Appendix~\ref{app:cutoff-fermi-spectators} gives the
exact spectator label range and shows why this difference disappears when
the cutoff is removed.

\begin{figure}[t]
	\centering
	\includegraphics[width=0.75\textwidth]{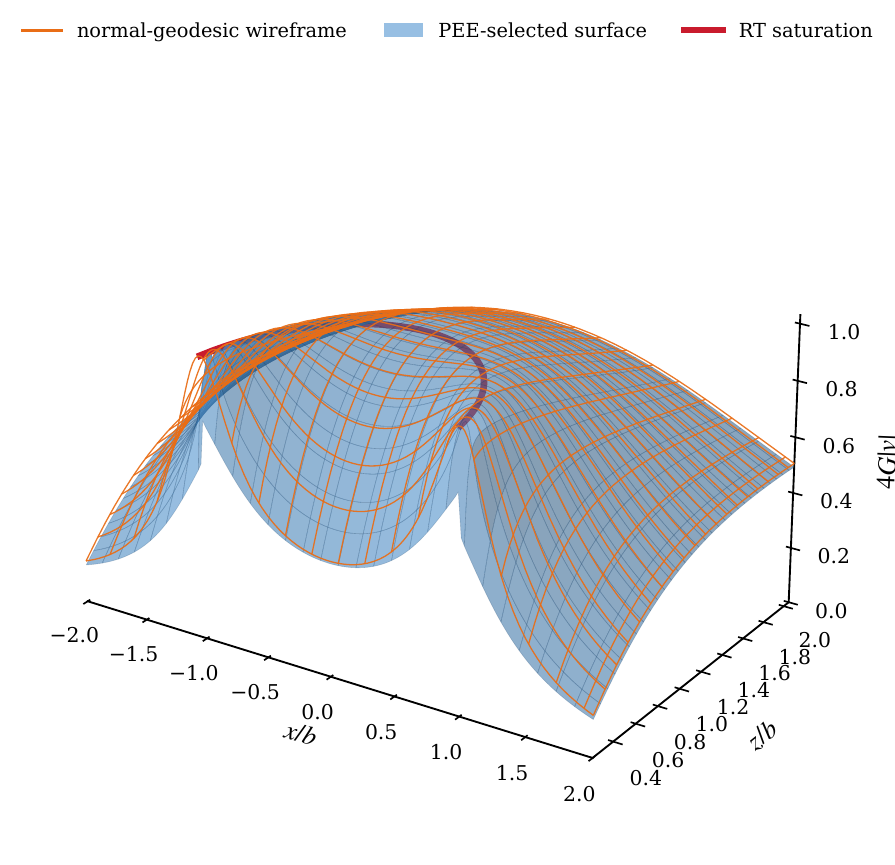}
	\caption{Invariant norms $4G|v|$ for the same parameters and Poincar\'e
		domain, superposed in one plot.  The translucent blue surface is the
		PEE-selected norm, the orange wireframe is the normal-geodesic norm, and
		the red unit ridge is their common RT saturation locus.  Both remain below
		one away from the bottleneck although their off-RT profiles differ.}
	\label{fig:pcut-norm-comparison}
\end{figure}

Figure~\ref{fig:pcut-norm-comparison} compares the corresponding capacities
on one three-dimensional plot.  The PEE-selected norm is shown as a
translucent surface and the normal-geodesic norm as an orange wireframe; their
common red ridge is the RT surface.  The separation away from that ridge is
the norm-space counterpart of the different landing maps, not a failure of
either field to maximize the flux.

In the asymptotic limit \(\zc\to0\), for every fixed \(x\) in the open
interval,
\begin{equation}
	\eta_{\rm geo}(x),\,\eta_{\rm PEE}(x)
	\longrightarrow
	\log\frac{b+x}{b-x},
	\qquad
	s_{\rm geo}(x),\,s_{\rm PEE}(x)
	\longrightarrow
	\frac{b}{2G(b^2-x^2)}.
	\label{eq:pcut-CFT-flow-limit}
\end{equation}
Likewise \(\Psi_{\rm geo}\to\Psi_A\) and
\(v_{\rm geo}\to v_A\) at fixed bulk points away from the endpoint foci.
The qualification is important: the convergence of the cutoff boundary densities is
not uniform at $x=\pm b$.  A cutoff-sized endpoint layer retains the local
distinction even while every fixed point in the open interval approaches the
CFT answer, as is already clear from
\eqref{eq:pcut-endpoint-density-comparison}.

\subsection{BTZ black brane at finite cutoff}
\label{subsec:cutoff-BTZ}
In this section, we place a finite Dirichlet wall in the BTZ black brane and refer to it below as the cutoff boundary. We first derive and interpret the two-point PEEs, then superpose the PEE thread currents and check the max-flow
conditions, and finally compare its PEE-selected endpoint map with an
independent normal-geodesic max flow.  The comparison is especially useful
here because the total RT flux is common to both representatives while the
amount assigned to horizon-ending threads is not.
\subsubsection{PEE threads and endpoint sectors}

We retain the finite surface $z=\zc$ in intrinsic BTZ coordinates
\begin{align}
  \left.\dd s^2\right|_{t=\mathrm{const.}}
  &=\frac{1}{z^2}\left(
  \dd x^2+\frac{\dd z^2}{f(z)}\right),
  \qquad \zc\leq z\leq\zh.
  \label{eq:pcut-cutoff-BTZ-slice}
\end{align}
After multiplying the induced cutoff boundary metric by the conventional Weyl factor
$\zc^2$, one obtains
\begin{equation}
  \dd s_{\Gamma_c}^2=-f(\zc)\dd t^2+\dd x^2.
  \label{eq:pcut-BTZ-wall-metric}
\end{equation}
Thus $x$ and $b$ below are physical cutoff boundary spatial coordinates in this
normalization (the unrescaled proper length is $|\dd x|/\zc$), whereas
normalized cutoff boundary time is $\tau_c=\sqrt{f(\zc)}\,t$.  If the Euclidean
coordinate period is $\beta_\infty=2\pi\zh$, the physical cutoff boundary period is
\begin{equation}
  \beta_c=\sqrt{f(\zc)}\,\beta_\infty
  =2\pi\sqrt{\zh^2-\zc^2}\implies 
  2\pi\zh=\sqrt{\beta_c^2+4\pi^2\zc^2}.
  \label{eq:pcut-BTZ-temperatures}
\end{equation}
For the interval $A=[-b,b]$ on the cutoff surface, the RT geodesic is
\begin{equation}
  \sqrt{1-\frac{z^2}{\zh^2}}
  =\sqrt{1-\frac{\zc^2}{\zh^2}}
  \frac{\cosh\left(\frac{x}{\zh}\right)}
  {\cosh\left(\frac{b}{\zh}\right)}.
  \label{eq:BTZ-cutoff-RT}
\end{equation}
The exact same-cutoff geodesic length and two-point PEE density are the thermal
finite-cutoff boundary counterparts of the interval observables studied in
\cite{Jeong:2019ylz,Hartman:2018tkw,He:2023xnb,Chang:2024voo},
\begin{align}
  \cL_{cc}(x_1,x_2)
  &=2\arcsinh\left[
  \frac{\zh}{\zc}\sinh\left(\frac{|x_1-x_2|}{2\zh}\right)\right],
  \label{eq:BTZ-cutoff-bb-length}\\
  \cI_{cc}(x_1,x_2)
  &=\frac{\zh^2-\zc^2}{16G\zh}
  \frac{\sinh\left(\frac{|x_1-x_2|}{2\zh}\right)}
  {\left[\zc^2+\zh^2
  \sinh^2\left(\frac{x_1-x_2}{2\zh}\right)\right]^{3/2}}.
  \label{eq:BTZ-cutoff-bb-kernel}
\end{align}
The cutoff-to-horizon distance supplies the endpoint sector required by the
thermal geodesic flow \cite{Caggioli:2024uza},
\begin{equation}
  \cL_{ch}(x,y)=\arccosh\left[
  \frac{\zh}{\zc}\cosh\left(\frac{x-y}{\zh}\right)\right],
  \label{eq:BTZ-cutoff-bh-length}
\end{equation}
with mixed Crofton PEE density
\begin{equation}
  \cI_{ch}(x,y)
  =\frac{\zh^2-\zc^2}{8G\zh}
  \frac{\cosh\left(\frac{x-y}{\zh}\right)}
  {\left[\zh^2\cosh^2\left(\frac{x-y}{\zh}\right)
  -\zc^2\right]^{3/2}}.
  \label{eq:BTZ-cutoff-bh-kernel}
\end{equation}
Both expressions reduce to \eqref{eq:BTZ-bb-kernel} and
\eqref{eq:BTZ-bh-kernel} when $\zc\to0$.

\begin{figure}[t]
	\centering
	\includegraphics[width=0.98\textwidth]
	{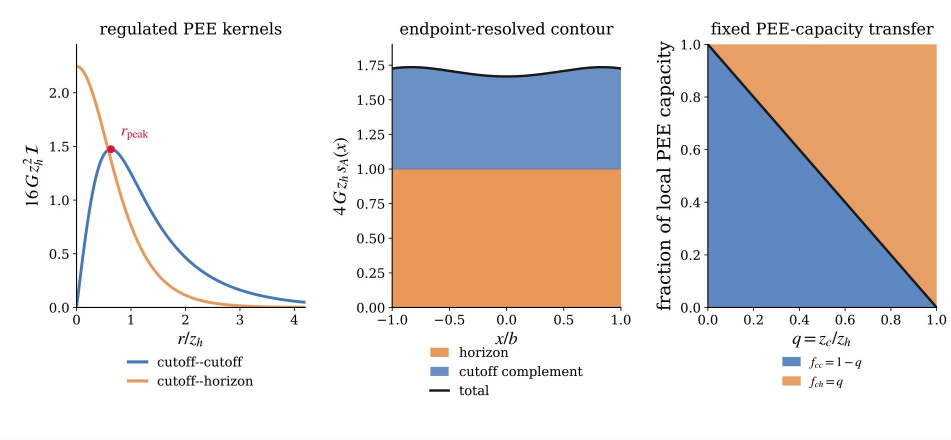}
	\caption{Finite-cutoff BTZ endpoint data.  Left: the regulated
		boundary--boundary and boundary--horizon kernels, with the boundary--boundary maximum at \eqref{eq:BTZ-cutoff-kernel-peak}.  Middle: the boundary--boundary and
		horizon-ending contributions to the interval contour.  Right: the exact
		transfer of the fixed local PEE capacity between the two endpoint
		sectors as $q=\zc/\zh$ is varied.}
	\label{fig:BTZ-cutoff-kernels-transfer}
\end{figure}

For finite $\zc$, the cutoff boundary--boundary kernel is not largest at coincident
endpoints.  It vanishes there and reaches its maximum at
\begin{equation}
	r_{\rm peak}
	=2\zh\arcsinh\left(\frac{\zc}{\sqrt2\,\zh}\right),
	\qquad r=|x_1-x_2|.
	\label{eq:BTZ-cutoff-kernel-peak}
\end{equation}
This is the thermal analogue of the finite-resolution peak in
Poincar\'e AdS: the cutoff boundary converts the ultraviolet singularity into chords of
finite preferred separation.  By contrast, the boundary--horizon kernel is
centered at $x=y$, because a horizon endpoint already supplies a distinct
geometric component and no second boundary point must be resolved.  Figure~\ref{fig:BTZ-cutoff-kernels-transfer} places these two kernels beside their
contributions to the interval contour and makes the fixed local-capacity
transfer visible as the cutoff boundary approaches the horizon.

\paragraph{Finite-cutoff thermofield-double map.}

The horizon endpoint sector again follows from the standard two-sided
purification rather than from assigning an autonomous Hilbert space to the
horizon; finite-cutoff thermofield-double states were constructed in
\cite{Coleman:2022tfd}.  At equal normalized cutoff boundary time, the geodesic joining
$(x,\zc)$ in one exterior to $(\tilde x,\zc)$ in the other has length
\begin{equation}
  \cL_{c\tilde c}(x,\tilde x)
  =2\arccosh\left[
  \frac{\zh}{\zc}
  \cosh\left(\frac{x-\tilde x}{2\zh}\right)\right].
  \label{eq:BTZ-cutoff-cross-length}
\end{equation}
Consequently the two-point PEE may be obtained from the Crofton kernel \eqref{eq:PEE-kernel-general},
\begin{equation}
  \cI_{c\tilde c}(x,\tilde x)
  =\frac{\zh^2-\zc^2}{16G\zh}
  \frac{\cosh\left(\frac{x-\tilde x}{2\zh}\right)}
  {\left[
  \zh^2\cosh^2\left(\frac{x-\tilde x}{2\zh}\right)-\zc^2
  \right]^{3/2}}.
  \label{eq:BTZ-cutoff-cross-kernel}
\end{equation}
The reflection-symmetric complete geodesic crosses the bifurcation line at
\begin{equation}
  y=\frac{x+\tilde x}{2},
  \qquad
  \cL_{c\tilde c}(x,\tilde x)=2\cL_{ch}(x,y).
  \label{eq:BTZ-cutoff-cross-midpoint}
\end{equation}
Since $\tilde x=2y-x$ and $\dd\tilde x=2\dd y$, the measures obey
\begin{equation}
  \cI_{c\tilde c}(x,\tilde x)\,
  \dd x\,\dd\tilde x
  =\cI_{ch}(x,y)\,\dd x\,\dd y.
  \label{eq:BTZ-cutoff-TFD-pushforward}
\end{equation}
Thus the mixed kernel is exactly the push-forward of the two-sided
finite-cutoff boundary measure in this purification; the factor of two is its
Jacobian.
\paragraph{PEE thread flow.}
The finite-cutoff current is most transparent when derived directly from the
source-endpoint distance.  Let
\begin{equation}
  \cL_{x_0}(X)=\cL\bigl(X;(x_0,\zc)\bigr).
  \label{eq:BTZ-cutoff-source-distance}
\end{equation}
The oriented source-resolved PEE thread current can be written intrinsically
as
\begin{equation}
  V_{x_0}^{\flat}
  =\frac{1}{8G}\star\dd\!\left(\partial_{x_0}\cL_{x_0}\right).
  \label{eq:BTZ-cutoff-elementary-distance-current}
\end{equation}
This form makes divergence conservation in the open bulk immediate and shows
that the component formulas below follow from the exact finite-endpoint
distance rather than from an independent vector-field ansatz.  Its normal
flux on the cutoff boundary gives the cutoff--cutoff PEE density
\eqref{eq:BTZ-cutoff-bb-kernel}; continuing the same analytic geodesic family
into the cutoff--horizon chamber gives the mixed PEE density
\eqref{eq:BTZ-cutoff-bh-kernel}.  The two endpoint sectors therefore differ
in their global termination and support, not in the local bulk current
emitted by a fixed cutoff source.

For a source $x_0$ on the cutoff surface, the exact distance from
$(x_0,\zc)$ to a bulk point $(x,z)$ is encoded by the following functions
\begin{align}
  \mathcal W_{x_0}(x,z)
  &=\frac{\zh^2}{zz_c}\left[\cosh\left(\frac{x-x_0}{\zh}\right)-\sqrt{\left(1-\frac{z^2}{z_h^2}\right)\left(1-\frac{z_c^2}{z_h^2}\right)}\right],
  \nonumber\\
  \mathcal S_{x_0}(x,z)
  &=\sqrt{\mathcal W_{x_0}(x,z)^2-1},
  \nonumber\\
  \mathcal K_{x_0}(x,z)
  &=\frac{\zh^2}{z\zc}\left[
  \sqrt{1-\frac{z^2}{\zh^2}}\cosh\left(\frac{x-x_0}{\zh}\right)
  -\sqrt{1-\frac{\zc^2}{\zh^2}}\right].
  \label{eq:BTZ-cutoff-source-functions}
\end{align}
Here $0<\zc<\zh$, $\zc\leq z\leq\zh$, and
$\mathcal W_{x_0}=\cosh \cL((x,z),(x_0,\zc))\geq1$.
We take the principal root $\mathcal S_{x_0}>0$ away from the source, where
$\mathcal S_{x_0}=0$.  The quantity $\mathcal K_{x_0}$ is signed and obeys
\begin{equation}
  \mathcal S_{x_0}^2
  =\mathcal K_{x_0}^2
  +\frac{\zh^2}{\zc^2}
  \sinh^2\left(\frac{x-x_0}{\zh}\right).
  \label{eq:BTZ-cutoff-SK-domain}
\end{equation}
These statements fix all square-root branches used below.  Differentiating
\eqref{eq:BTZ-cutoff-elementary-distance-current} gives the elementary PEE
thread flow current below.  Direct flux matching in the cutoff--cutoff and
cutoff--horizon chambers, as in section~\ref{sec:BTZ-threads}, provides an
independent check of its normalization:
\begin{align}
  V_{x_0}^{\mu}(x,z)
  &=\frac{z_h}
  {8G\zc\,\mathcal S_{x_0}(x,z)^3}
  \Bigg(\sinh\left(\frac{x-x_0}{\zh}\right)\Big[
  \sqrt{1-\frac{z^2}{z_h^2}}\mathcal S_{x_0}(x,z)^2
  -\mathcal K_{x_0}(x,z)\mathcal W_{x_0}(x,z)
  \Big],\notag\\
  &\frac{z\sqrt{1-\frac{z^2}{z_h^2}}}{z_c}  \Big[
  \cosh\left(\frac{x-x_0}{\zh}\right)
  \mathcal S_{x_0}(x,z)^2
  -\frac{\zh^2}{z\zc}\mathcal W_{x_0}(x,z)
  \sinh^2\left(\frac{x-x_0}{\zh}\right)
  \Big]\Bigg)
  \label{eq:BTZ-cutoff-elementary}
\end{align}
The two endpoint chambers are therefore globally distinct but locally
compatible at finite cutoff, just as in the asymptotic thermal geometry.

\subsubsection{Bit threads, max-flow tests, and fluxes}
Direct integration of the finite-cutoff elementary currents \eqref{eq:BTZ-cutoff-elementary} over $x_0\in[-b,b]$ gives the coarse-grained vector field
\begin{align}
  v_A^\mu(x,z)&=\frac{z}{8G}\left(\frac{\mathcal K_{-b}(x,z)}{\mathcal S_{-b}(x,z)}
  -\frac{\mathcal K_{+b}(x,z)}{\mathcal S_{+b}(x,z)},\frac{z_h}{z_c}\sqrt{1-\frac{z^2}{\zh^2}} \left[
  \frac{\sinh\left(\frac{x+b}{\zh}\right)}
  {\mathcal S_{-b}(x,z)}
  -\frac{\sinh\left(\frac{x-b}{\zh}\right)}
  {\mathcal S_{+b}(x,z)}
  \right]\right)
  \label{eq:BTZ-cutoff-flow}
\end{align}
This superposition has the intrinsic endpoint-distance representation
\begin{align}
  \cL_\pm(x,z)&=\arccosh\mathcal W_{\pm b}(x,z),
  \qquad
  \Psi_A(x,z)=\frac12\left[\cL_+(x,z)-\cL_-(x,z)\right],
  \nonumber\\
  v_A&=\frac{1}{4G}\star\dd\Psi_A,
  \label{eq:BTZ-cutoff-distance-potential}
\end{align}
with orientation chosen toward the interior of $A$ on the cutoff boundary.  It follows
at once that the field is divergenceless away from its boundary sources and
that its streamlines are regular level sets of $\Psi_A$ wherever
$\dd\Psi_A\neq0$.  Because the BTZ slice is locally $\mathbb H^2$, the
geodesic-curvature identity \eqref{eq:app-curvature-exact} applies with
these two endpoint distances.  The reflection-symmetric central streamline
is geodesic; away from it, regular streamlines are generically non-geodesic
at finite cutoff.  As in Poincar\'e AdS$_3$, this does not conflict with the
geodesicity of each elementary chord.

We may also check the norm bound directly from \eqref{eq:BTZ-cutoff-flow}, by introducing the auxiliary unit vectors
\begin{align}
  \mathbf q_{\pm b}
  &=\frac{1}{\mathcal S_{\pm b}(x,z)}\left(
  \mathcal K_{\pm b}(x,z),
  -\frac{\zh}{\zc}\sinh\left(\frac{x\mp b}{\zh}\right)
  \right)\,.
  \label{eq:BTZ-cutoff-unit-pairs}
\end{align}
In the orthonormal frame
\begin{equation}
  e_{\hat x}=z\partial_x,
  \qquad
  e_{\hat\rho}=-z\sqrt{1-\frac{z^2}{\zh^2}}\,\partial_z,
\end{equation}
the explicit field satisfies
\begin{equation}
  4G\bigl(v_A^{\hat x},v_A^{\hat\rho}\bigr)
  =\frac12\left(\mathbf q_{-b}-\mathbf q_{+b}\right).
  \label{eq:BTZ-cutoff-component-identity}
\end{equation}
Thus $|v_A|\leq1/(4G)$.  On the RT surface \eqref{eq:BTZ-cutoff-RT},
$\mathbf q_{-b}=-\mathbf q_{+b}$, and a direct contraction with the RT tangent
shows normality.  The entanglement entropy is reproduced as the flux through the RT bottleneck,
\begin{equation}
  S_A^{(c)}=\frac{1}{2G}\arcsinh\left[
  \frac{\zh}{\zc}\sinh\left(\frac{b}{\zh}\right)\right].
  \label{eq:BTZ-cutoff-entropy}
\end{equation}
The extended Crofton normalization also survives at finite cutoff.  Since
\begin{equation}
  \int_{-\infty}^{\infty}\dd y\,\cI_{ch}(x,y)
  =\frac{1}{4G\zh},
  \label{eq:BTZ-cutoff-horizon-per-site}
\end{equation}
we obtain
\begin{align}
  \cI_{cc}(A,A^c_\Gamma)
  &=S_A^{(c)}-\frac{b}{2G\zh},
  \nonumber\\
  \cI_{ch}(A,h)&=\frac{b}{2G\zh},
  \label{eq:BTZ-cutoff-extended-decomp}
\end{align}
where $A^c_\Gamma$ is the complement on the cutoff line.  Therefore, the horizon contribution remains $2b/(4G\zh)$ and is independent of the cutoff boundary position; the cutoff boundary sector contains all explicit $\zc$ dependence.  This is not the same as varying
$\zc$ at fixed physical cutoff boundary temperature: equation
\eqref{eq:pcut-BTZ-temperatures} then makes $\zh$ cutoff dependent.

The local contour decomposition, again written without dimensionless
coordinates, is
\begin{align}
  s_A^{\Gamma}(x)
  &=\frac{1}{8G}\Bigg[
  \frac{\cosh\left(\frac{b-x}{2\zh}\right)}
  {\sqrt{\zc^2+\zh^2\sinh^2\left(\frac{b-x}{2\zh}\right)}}
  +\frac{\cosh\left(\frac{b+x}{2\zh}\right)}
  {\sqrt{\zc^2+\zh^2\sinh^2\left(\frac{b+x}{2\zh}\right)}}
  -\frac{2}{\zh}\Bigg],
  \nonumber\\
  s_A^{h}(x)&=\frac{1}{4G\zh},
  \nonumber\\
  s_A(x)
  &=\frac{1}{8G}\Bigg[
  \frac{\cosh\left(\frac{b-x}{2\zh}\right)}
  {\sqrt{\zc^2+\zh^2\sinh^2\left(\frac{b-x}{2\zh}\right)}}
  +\frac{\cosh\left(\frac{b+x}{2\zh}\right)}
  {\sqrt{\zc^2+\zh^2\sinh^2\left(\frac{b+x}{2\zh}\right)}}
  \Bigg].
  \label{eq:BTZ-cutoff-contours}
\end{align}

The total contour in \eqref{eq:BTZ-cutoff-contours} also follows directly
from the ALC prescription.  Rewriting \eqref{eq:BTZ-cutoff-entropy} as a
function of the physical interval length $\ell=2b$ and applying
\eqref{eq:review-ALC-translational} reproduces $s_A(x)$ exactly.  This fixes
the one-point PEE contour independently of the endpoint decomposition.  The
separate pieces $s_A^{\Gamma}$ and $s_A^h$ contain the additional
endpoint-resolved information supplied by the cutoff-boundary and TFD
purifier sectors.

These densities may also be realized by the actual fluxes of the bit-thread field \eqref{eq:BTZ-cutoff-flow}.  On $\Gamma_c$, the inward unit normal is
$n_{\Gamma_c}=\zc\sqrt{1-\zc^2/\zh^2}\,\partial_z$ and
$\dd\Sigma=\dd x/\zc$.  Directly evaluating
\eqref{eq:BTZ-cutoff-flow} gives
\begin{equation}
  \left.
  \frac{v_A^z}
  {\zc^2\sqrt{1-\zc^2/\zh^2}}
  \right|_{z=\zc}
  =s_A(x),
  \qquad -b<x<b.
  \label{eq:BTZ-cutoff-pointwise-cutoff-flux}
\end{equation}
For the horizon, setting $z=\zh\sech\rho$, so that
$v_A^\rho=-v_A^z/[z\sqrt{1-z^2/\zh^2}]$.  The outward normal of the
one-sided exterior is $-\partial_\rho$ and the line element at $\rho=0$ is
$\dd y/\zh$.  The endpoint density is therefore
\begin{align}
  h_A^{(c)}(y)
  &\equiv\int_{-b}^{b}\dd x\,\cI_{ch}(x,y) \nonumber\\
  &=\frac{1}{8G}\left[
  \frac{\sinh\left(\frac{b-y}{\zh}\right)}
  {\sqrt{\zh^2\cosh^2\left(\frac{b-y}{\zh}\right)-\zc^2}}
  +\frac{\sinh\left(\frac{b+y}{\zh}\right)}
  {\sqrt{\zh^2\cosh^2\left(\frac{b+y}{\zh}\right)-\zc^2}}
  \right]=\left.\frac{-v_A^\rho}{\zh}\right|_{\rho=0}.
  \label{eq:BTZ-cutoff-pointwise-horizon-flux}
\end{align}

The endpoint-distance potential also permits a direct comparison with the
normal-geodesic representative.  In the Fermi chart of appendix
\ref{app:cutoff-fermi-nongeodesic}, the physical RT segment is
$|\eta|\leq\sigma_b$, with
\begin{equation}
  \sigma_b=\arcsinh\left[
  \frac{\zh}{\zc}\sinh\left(\frac b\zh\right)\right].
  \label{eq:BTZ-cutoff-Fermi-half-length-main}
\end{equation}
The PEE-selected levels are non-geodesic and divide at
$|\eta_\star|=b/\zh$ into horizon-ending and cutoff-boundary-returning sectors.  The
normal congruence instead uses constant-$\eta$ geodesics and has the distinct
horizon threshold \eqref{eq:cutoff-fermi-normal-threshold}.  The different
thresholds show that even the division of the common RT flux among endpoint
components is representative dependent; only the total flux is fixed by
maximality.

\begin{figure}[t]
  \centering
  \includegraphics[width=0.98\textwidth]{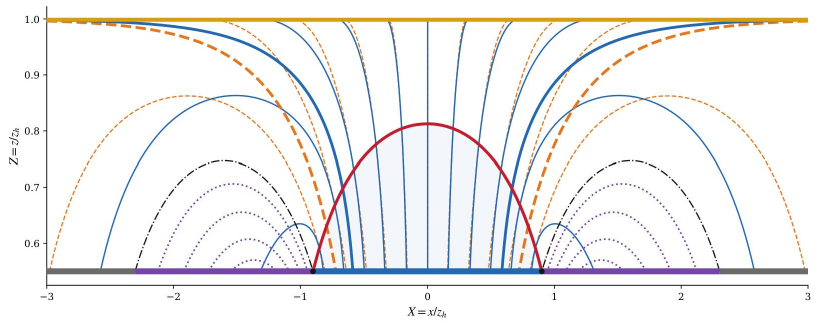}\vspace{0.25em}
  {\small
  	\begin{center}
  		\flowkey{floworange,dashed}{1.0pt}{active normal geodesics}
  		\hspace{1.2em}
  		\flowkey{flowblue}{1.0pt}{PEE-selected streamlines}
  		\hspace{1.2em}
  		\flowkey{flowpurple,dotted}{1.2pt}{normal spectators}
  		
  		\vspace{0.35em}
  		
  		\flowkey{floworange,dashed}{2.0pt}{normal separatrix}
  		\hspace{1.2em}
  		\flowkey{flowblue}{2.0pt}{PEE separatrix}
  		\hspace{1.2em}
  		\flowkey{flowred}{2.0pt}{RT surface}
  		\hspace{1.2em}
  		\flowkey{flowgold}{2.2pt}{horizon}
  	\end{center}
  }
  \caption{Complete finite-cutoff BTZ flow comparison for
  $q=\zc/\zh=0.55$ and $B=b/\zh=0.90$.  PEE-selected streamlines are blue
  solid curves and active normal geodesics are orange dashed curves; both
  are extended from the cutoff boundary, through the red RT segment, to their
  cutoff-boundary or horizon endpoints.  Their thicker members are the respective
  endpoint-sector separatrices.  Purple dotted normal geodesics lie outside
  $|\eta|\leq\sigma_b$ and connect two points of $A^c$; they are the spectator
  completion.  The shaded region is the entanglement wedge, the gold upper
  boundary is the horizon, and the colored cutoff-boundary bars distinguish $A$, the
  unused adjacent complement and the remaining complement.}
  \label{fig:BTZ-cutoff-flow-comparison}
\end{figure}

Figure~\ref{fig:BTZ-cutoff-flow-comparison} superposes the two families over
the full one-sided exterior.  It also includes the normal curves with
$|\eta|>\sigma_b$.  These spectators begin and end on the immediate
complementary cutoff-boundary strips, never cross the physical RT segment and never
reach the horizon.  They are therefore neither an additional thermal sector
nor a correction to \eqref{eq:BTZ-cutoff-extended-decomp}.  Their only role
is to continue the normal Fermi field smoothly beyond the active support;
the exact cutoff-boundary partner map, tangency point and label range are given in
\eqref{eq:cutoff-fermi-BTZ-normal-partner}--\eqref{eq:cutoff-fermi-BTZ-spectator-band}.

\begin{figure}[t]
  \centering
  \includegraphics[width=0.75\textwidth]{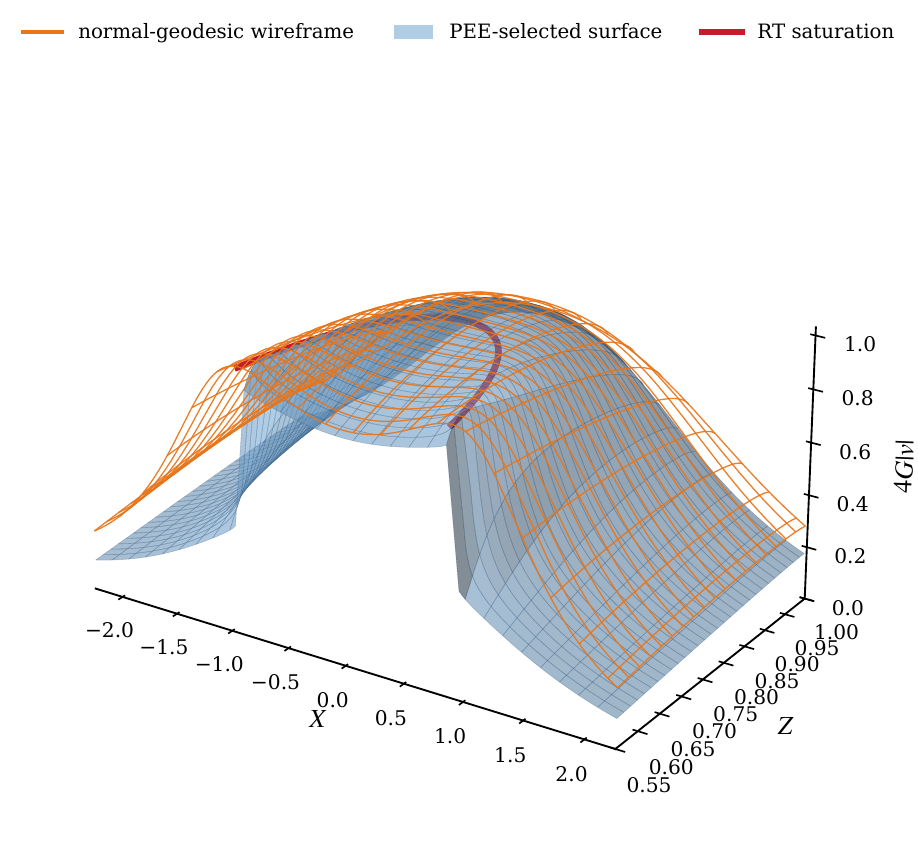}
  \caption{Finite-cutoff BTZ invariant norms on one set of axes.  The
  translucent blue surface is $4G|v_A|$ for the PEE-selected flow, the
  orange wireframe is $4G|v_{\rm geo}|$, and the red unit ridge is their
  common saturation locus on the RT segment.  The visual separation away
  from the ridge measures max-flow degeneracy rather than a difference in
  entropy.}
  \label{fig:BTZ-cutoff-norm-comparison}
\end{figure}

The superposed norm plot in figure~\ref{fig:BTZ-cutoff-norm-comparison}
provides an independent max-flow check.  Both representatives reach the same
unit ridge on the RT geodesic and remain below the bound in the entire
exterior, while their different off-bottleneck profiles encode the
tangential redistribution present only in the PEE-selected field.

The two integrations underlying these statements can be performed in closed
form.  Because the kernels are even functions of $r=x-y$, set
$u=r/(2\zh)$ for the cutoff--cutoff sector.  Then
\begin{align}
  \int_{-\infty}^{\infty}\dd r\,\cI_{cc}(r)
  &=\frac{\zh^2-\zc^2}{4G}
  \int_0^\infty\dd u\,
  \frac{\sinh u}{[\zc^2+\zh^2\sinh^2u]^{3/2}}
  =\frac{1}{4G}\left(\frac{1}{\zc}-\frac{1}{\zh}\right),
  \label{eq:pcut-thermal-bb-integral-check}
\end{align}
where $t=\cosh u$.  For the cutoff--horizon sector, set
$u=r/\zh$ and then $t=\sinh u$:
\begin{align}
  \int_{-\infty}^{\infty}\dd r\,\cI_{ch}(r)
  &=\frac{\zh^2-\zc^2}{4G}
  \int_0^\infty\dd u\,
  \frac{\cosh u}{[\zh^2\cosh^2u-\zc^2]^{3/2}}
  =\frac{1}{4G\zh}.
  \label{eq:pcut-thermal-bH-integral-check}
\end{align}
These integrals imply the exact local PEE identity
\begin{equation}
  \int_{-\infty}^{\infty}\dd y\,\cI_{cc}(x,y)
  +\int_{-\infty}^{\infty}\dd y\,\cI_{ch}(x,y)
  =\frac{1}{4G\zc}.
  \label{eq:pcut-thermal-Crofton-budget}
\end{equation}
The right-hand side is independent of temperature at fixed $\zc$.  Its
coordinate density is best read as the invariant PEE-capacity measure
\begin{equation}
  \frac{\dd x}{4G\zc}=\frac{\dd s_{\Gamma_c}}{4G},
  \label{eq:pcut-local-capacity-measure}
\end{equation}
where $\dd s_{\Gamma_c}=\dd x/\zc$ is the unrescaled proper line element on
the cutoff boundary.  Thus the local statement is reparametrization
independent: an infinitesimal cutoff-boundary segment carries a fixed
endpoint-resolved PEE capacity, while temperature partitions that capacity
between the cutoff--cutoff and cutoff--horizon sectors.  Equivalently,
integrating the source-resolved conserved current over a thin pillbox based
on the same cutoff-boundary segment gives the same identity by Gauss' law:
the outgoing source flux equals the sum of its two positive PEE endpoint
channels.  This interpretation uses the hard-cutoff state and the
thermofield-double purification specified above.

The near-horizon cutoff limit is also sharp.  At fixed $\zh$,
\begin{equation}
  \cI_{cc}(r)\longrightarrow0,
  \qquad
  \int_{-\infty}^{\infty}\dd r\,\cI_{cc}(r)
  \longrightarrow0
  \qquad (\zc\to\zh^-).
  \label{eq:BTZ-cutoff-near-horizon-cc}
\end{equation}
For the mixed sector set
\begin{equation}
  a^2=\zh^2-\zc^2,
  \qquad
  t=\zh\sinh\!\left(\frac r\zh\right).
  \label{eq:BTZ-cutoff-delta-variable}
\end{equation}
Then the endpoint measure is exactly
\begin{equation}
  \cI_{ch}(r)\,\dd r
  =\frac{1}{8G\zh}
  \frac{a^2\,\dd t}{(a^2+t^2)^{3/2}},
  \qquad
  \int_{-\infty}^{\infty}
  \frac{a^2\,\dd t}{(a^2+t^2)^{3/2}}=2.
  \label{eq:BTZ-cutoff-delta-sequence}
\end{equation}
The measure is non-negative and has fixed total mass $1/(4G\zh)$.  Moreover,
for any $\delta>0$, with
$T_\delta=\zh\sinh(\delta/\zh)$,
\begin{equation}
  \int_{|r|>\delta}\dd r\,\cI_{ch}(r)
  =\frac{1}{4G\zh}
  \left(1-\frac{T_\delta}{\sqrt{T_\delta^2+a^2}}\right)
  \xrightarrow[a\to0]{}0.
  \label{eq:BTZ-cutoff-delta-tail}
\end{equation}
Hence all of the fixed mass concentrates at $r=0$, and therefore
\begin{equation}
  \cI_{ch}(r)
  \ \rightharpoonup\
  \frac{1}{4G\zh}\,\delta(r),
  \qquad
  \lim_{\zc\to\zh^-}\int_{-\infty}^{\infty}\dd r\,
  \cI_{ch}(r)\,\varphi(r)
  =\frac{\varphi(0)}{4G\zh}
  \label{eq:BTZ-cutoff-near-horizon-delta}
\end{equation}
for every smooth compactly supported test function $\varphi$.  The fixed
local PEE capacity therefore becomes ultralocal in the horizon channel.  At
fixed $\beta_c$, by contrast, $\zh$ varies with $\zc$, and this limiting
statement must be re-expressed using
\eqref{eq:pcut-BTZ-temperatures}.

\paragraph{$T\bar T$ temperature and thermal transfer.}

Using the physical cutoff boundary period $\beta_c$ in
\eqref{eq:pcut-BTZ-temperatures}, rather than the coordinate period
$\beta_\infty$, removes an otherwise common normalization ambiguity.
For an interval of physical length $\ell=2b$, the entropy and the thermal
entropy density are therefore
\begin{align}
  S_{\mu,\beta_c}(\ell)
  &=\frac{c}{3}\arcsinh\left[
  \frac{\sqrt{\beta_c^2+4\pi^2\zc^2}}{2\pi\zc}
  \sinh\left(\frac{\pi\ell}
  {\sqrt{\beta_c^2+4\pi^2\zc^2}}\right)\right],
  \label{eq:pcut-BTZ-entropy-mu}\\
  s_{\rm th}(\mu,\beta_c)
  &=\frac{1}{4G\zh}
  =\frac{\pi c}{3\sqrt{\beta_c^2+4\pi^2\zc^2}}.
  \label{eq:pcut-real-wall-thermal}
\end{align}
The fraction of the fixed local PEE capacity
\eqref{eq:pcut-thermal-Crofton-budget} carried by horizon-ending threads is
\begin{equation}
  \frac{\zc}{\zh}
  =\frac{2\pi\zc}{\sqrt{\beta_c^2+4\pi^2\zc^2}}.
  \label{eq:pcut-horizon-fraction}
\end{equation}
At fixed $\zc$, it vanishes as $\beta_c\to\infty$ and tends to one as
$\beta_c\to0$.  Raising the physical cutoff boundary temperature therefore transfers
the fixed PEE capacity continuously from cutoff--cutoff threads to
cutoff--horizon chords.  The same deformation scale appears in exact
thermal spectra and partition functions
\cite{Smirnov:2016lqw,Datta:2018thy,Aharony:2018bad}.

\section{Summary and discussion}
\label{sec:summary}

\subsection*{Exact geometric results}

This paper develops endpoint-resolved chord currents and their
source-superposed max flows on static, constant-curvature slices.  The
following statements are exact within classical Einstein gravity.

For the asymptotic planar BTZ exterior, the boundary--boundary and
boundary--horizon PEE kernels are the two positive endpoint sectors of one
local current.  In the reflection-symmetric thermofield double, the mixed
horizon measure is the push-forward of the cross-boundary measure under
$y=(x+\tilde x)/2$; its relative factor of two is therefore a Jacobian.
The source superposition is the geodesic BTZ max-flow representative.  Its
intrinsic Fermi coordinates supply a global stream function and locate the
separatrices at $|\eta|=b/\zh$, or equivalently at
\begin{equation}
  |x_\beta|=\frac{\zh}{2}\log\cosh\left(\frac{2b}{\zh}\right)
  \qquad\text{on }\gamma_A.
  \label{eq:summary-BTZ-separatrix}
\end{equation}
Consequently, the horizon-ending streamlines intersect precisely the RT
segment whose length is $2b/\zh$.  This establishes the thermal-flux
assignment from the actual congruence, rather than inferring it only from an
entropy decomposition.  The asymptotic-boundary and horizon contour densities also agree
pointwise with the normal fluxes of this selected field.

For finite-cutoff Poincar\'e AdS$_3$, global AdS$_3$, and planar BTZ, the
mixed second derivatives of exact finite-endpoint distances determine
positive PEE kernels.  Flux matching fixes elementary divergenceless
currents; direct source integration then gives explicit macroscopic fields.
In all three geometries, independent two-unit-vector identities prove the
norm bound and pointwise RT saturation.  The cutoff boundary fluxes reproduce the
endpoint densities pointwise, including the periodic principal branch in
global AdS$_3$.

Finite-cutoff Poincar\'e AdS$_3$ makes the microscopic/macroscopic
distinction especially sharp.  The endpoint-calibrated field is the Hodge
dual of a distance difference.  Its integral curves are nested hyperbolic
Apollonius curves and every regular noncentral curve has nonzero geodesic
curvature.  The independent normal-geodesic construction is another max
flow with the same bottleneck flux but a different cutoff boundary density, RT landing
map, and deterministic boundary pairing.  Under the smoothness,
single-crossing, geodesicity, and pointwise-saturation assumptions stated in
section~\ref{sec:pcut-geodesic-flow}, flux conservation uniquely fixes that
representative within the chosen RT-normal congruence; no uniqueness among
all max flows is implied.  Thus geodesicity of the elementary chords does not imply
geodesicity of the source-superposed field.

The common Fermi chart makes the origin of this degeneracy especially
transparent.  Finite RT endpoints act as two hyperbolic foci: their
distance-difference potential produces a tangential drift relative to the
RT-normal foliation, while the drift vanishes when the foci are pushed to
ideal endpoints.  The same chart also identifies the
complement-to-complement spectator bands of the unrestricted normal flow in
Poincar\'e and BTZ.  Since their support boundaries are streamlines, the
spectators may either be retained as a smooth completion or removed by a
weakly divergenceless truncation without changing the entropy flux.

At finite cutoff in BTZ, the same endpoint-distance structure generically
produces non-geodesic macroscopic streamlines.  The cutoff--cutoff and
cutoff--horizon sectors satisfy the exact local budget
\eqref{eq:pcut-thermal-Crofton-budget}.  At fixed $\zh$, moving the cutoff boundary
toward the horizon sends the boundary--boundary measure to zero and the mixed kernel
to $\delta(r)/(4G\zh)$ in the distributional sense.  At fixed physical
cutoff boundary period $\beta_c$, $\zh$ instead varies according to
\eqref{eq:pcut-BTZ-temperatures}; these are different limiting procedures.

\subsection*{Interpretation and relation to earlier work}

The logical status of the construction may be summarized as follows.  A Crofton density is geometric data.  It becomes a PEE only after specifying a state, an endpoint support, and the
assumptions that relate that support to degrees of freedom.  In particular,
the horizon in the one-sided exterior is a geometric flux boundary.  In the
standard thermofield double it encodes correlations with the second
asymptotic system through the explicit push-forward proved in
sections~\ref{subsec:BTZ-TFD-map} and~\ref{subsec:cutoff-BTZ}; it is not
claimed to be a literal, unique Hilbert-space factor or a
purification-independent endpoint.

The relation to previous thread constructions is correspondingly precise.
Reference~\cite{Lin:2023rxc} established endpoint-resolved PEE currents in
vacuum AdS$_3$, while ref.~\cite{Caggioli:2024uza} constructed geodesic
black-brane max flows and identified their thermal horizon flux.
Reference~\cite{Lin:2025btz} subsequently developed two-sided planar-BTZ
entanglement threads.  We do not claim those kernels or the existence of
the thermal max flow as new.  The new BTZ ingredients here are the
one-sided push-forward of the two-sided measure, the chamberwise analytic
current in intrinsic BTZ coordinates, the exact separatrix proof for the
superposed field, and the pointwise flux audit.

For finite cutoff boundaries, earlier work established the hard-cutoff and
mixed-boundary dictionaries, finite-cutoff interval entropies, their RT and
replica interpretations, and finite-cutoff thermofield-double states
\cite{McGough:2016lol,Donnelly:2018bef,Murdia:2019fax,Jeong:2019ylz,Lewkowycz:2019xse,He:2023xnb,Chang:2024voo,Coleman:2022tfd}.
The new results are the endpoint-differentiated two-point PEE kernels and elementary
currents for all three geometries, their explicit max-flow superpositions,
the distinction between PEE-selected and normal-geodesic representatives,
the global-cylinder branch analysis, and the finite-BTZ local and
distributional PEE identities.  Subregion kinematic space
\cite{Basu:2026hbg} explains why finite geometric endpoints are natural,
but it does not by itself supply their microscopic field-theory
interpretation.

In \cite{BasuWen:BCFTMinPurification}, we develop the complementary extended-boundary
realization of the same PEE-thread program.  We enlarge the endpoint
support to include end-of-the-world brane segments in AdS/BCFT
\cite{Takayanagi:2011zk,Fujita:2011fp} and
RT-surface segments in the surface/state description of a gravitational
subregion.  The latter gives a particularly useful minimal-purification
picture: the RT surface bounding the entanglement wedge is promoted from a
passive homology boundary to an auxiliary purifying surface, split into
pieces associated with the purified subsystems.  PEE-thread currents sourced
from the original boundary region and from the appropriate purifying RT
segment then superpose into a divergenceless, norm-bounded flow that
saturates the entanglement-wedge cross section in the examples analyzed
there.  In this way the construction supplies a microscopic thread
realization of the usual holographic entanglement-of-purification geometry
within the surface/state class of purifications.  Appendix A of \cite{BasuWen:BCFTMinPurification}
also isolates the geometric reason why these apparently multi-component
source integrals simplify: the final stream function depends only on the
oriented endpoints of the relevant bottleneck and is expressed as a
difference of finite-point distances, or of distance and Busemann functions
when an endpoint lies at the conformal boundary.  The finite-cutoff
Poincar\'e potential used in the present paper is precisely the two-finite-focus
member of that general construction.


\subsection*{Limitations and outlook}

The analysis is restricted to static slices, classical Einstein gravity,
connected interval phases, and the positive real-cutoff boundary branch.  With generic
bulk matter, the simple Dirichlet-wall description is not a universal
definition of the deformed theory \cite{Guica:2019nzm}.  Max flows are also
nonunique: PEE data select one representative, but the RT entropy alone
does not.  For disconnected intervals, a direct source sum must be replaced
by an intersection-weighted construction \cite{Lin:2023rxc}.

The compact BTZ black hole is qualitatively more subtle than the planar
black brane studied here.  Its spatial exterior is an annulus, so the flow
may carry a nontrivial period around the horizon, while geodesic lengths and
endpoint kernels require minimizing over winding images of the quotient.
A vector field constructed in a cut-open fundamental domain must in addition
satisfy the gluing condition on the identified faces.  Consequently, a
simple periodization of the planar boundary--horizon construction need not
respect the homology and winding data of all sectors simultaneously.  One
possible route is to work with closed one-forms on the annulus, or
quasi-periodic/equivariant potentials on the universal cover, and to divide
endpoint space into image-switching chambers whose periods are fixed by the
horizon flux.  A locking-multiflow formulation may then provide the natural
bookkeeping for simultaneous boundary, winding, and horizon sectors.  We
leave the detailed compact-BTZ construction to future work; related
obstructions to a single global geodesic congruence were emphasized in
\cite{Caggioli:2024uza}.

Beyond these questions, higher-dimensional regions would replace endpoint
integrals by higher-dimensional integral geometry.
Finally, Hagedorn behavior belongs to an opposite analytic branch and is not
derived by the positive cutoff boundary kernels.  A geometric thread account of that
physics should instead begin in a real single-trace $T\bar T$ or
linear-dilaton background \cite{Giveon:2017nie}.

\acknowledgments

The authors are supported by the NSFC Grant No.~12447108 and the Shing-Tung Yau Center of Southeast University.

\appendix

\section{Geodesic lengths in AdS geometries}
\label{AppA}

In this appendix, we provide the embedding-coordinate derivation of the
geodesic lengths used throughout the main text; see, for example,
\cite{Balasubramanian:2013lsa,Headrick:2014eia,Wen:2018mev}.  The AdS$_3$ covering
space is the hyperboloid in $\mathbb R^{2,2}$ with metric
\begin{equation}
  \dd s^2=\eta_{AB}\dd X^A\dd X^B,
  \qquad
  \eta_{AB}=\operatorname{diag}(-1,-1,1,1),
  \label{eq:AdS-embedding-metric}
\end{equation}
subject to
\begin{equation}
  \eta_{AB}X^AX^B=-1.
  \label{eq:AdS-embedding-constraint}
\end{equation}
For two spacelike-separated points $i$ and $j$, their geodesic distance is
\begin{equation}
  \cL_{ij}=\arccosh\left(-\eta_{AB}X_i^AX_j^B\right).
  \label{eq:length-AdS}
\end{equation}
For a locally AdS$_3$ quotient, these coordinates describe a lift to the
covering space.  A quotient distance requires minimizing over the relevant
images.  Throughout the planar-BTZ analysis we use the noncompact covering
branch, so no angular image sum is implied.

\subsection*{BTZ black brane}

The embedding coordinates for the BTZ black brane are
\begin{align}
  X_0&=\frac{\sqrt{\zh^2-z^2}}{z}
  \sinh\left(\frac{t}{\zh}\right),
  &
  X_1&=\frac{\zh}{z}\cosh\left(\frac{x}{\zh}\right),
  \nonumber\\
  X_2&=\frac{\sqrt{\zh^2-z^2}}{z}
  \cosh\left(\frac{t}{\zh}\right),
  &
  X_3&=\frac{\zh}{z}\sinh\left(\frac{x}{\zh}\right).
  \label{eq:BTZ-embedding-coordinates}
\end{align}
Consequently, the geodesic distance between two arbitrary bulk points
$(x_i,t_i,z_i)$ and $(x_j,t_j,z_j)$ is
\begin{align}
  \cL_{ij}
  =\arccosh\Bigg[
  &\frac{\zh^2}{z_i z_j}
  \cosh\left(\frac{x_i-x_j}{\zh}\right)-\frac{\sqrt{(\zh^2-z_i^2)(\zh^2-z_j^2)}}{z_i z_j}
  \cosh\left(\frac{t_i-t_j}{\zh}\right)
  \Bigg].
  \label{eq:length-BTZ}
\end{align}
For two points at $z_i=z_j=\eps$ on the same time slice,
\begin{align}
  \cL(x_i,x_j)
  &=\arccosh\left[
  1+\frac{\zh^2}{\eps^2}
  \left(\cosh\left(\frac{x_i-x_j}{\zh}\right)-1\right)
  \right]
  \nonumber\\
  &=2\log\left[
  \frac{2\zh}{\eps}
  \sinh\left(\frac{|x_i-x_j|}{2\zh}\right)\right]
  +O(\eps^2).
  \label{eq:length-BTZ-boundary}
\end{align}

\subsection*{Global AdS$_3$}

For the global AdS$_3$ metric used in
\eqref{eq:global-cutoff-metric}, the embedding coordinates are
\begin{align}
  X_0&=\frac{\sqrt{1+z^2}}{z}\cos t,
  &
  X_2&=\frac{1}{z}\cos\theta,
  \nonumber\\
  X_1&=\frac{\sqrt{1+z^2}}{z}\sin t,
  &
  X_3&=\frac{1}{z}\sin\theta.
  \label{eq:global-embedding-coordinates}
\end{align}
The geodesic distance between two arbitrary bulk points
$(\theta_i,t_i,z_i)$ and $(\theta_j,t_j,z_j)$ is therefore
\begin{equation}
  \cL_{ij}=\arccosh\left[
  \frac{\sqrt{(1+z_i^2)(1+z_j^2)}
  \cos(t_i-t_j)-\cos(\theta_i-\theta_j)}
  {z_i z_j}\right].
  \label{eq:length-global}
\end{equation}

\subsection*{Poincar\'e AdS$_3$}

The Poincar\'e AdS$_3$ metric is
\begin{equation}
  \dd s^2=\frac{-\dd t^2+\dd x^2+\dd z^2}{z^2},
  \label{eq:Poincare-full-metric}
\end{equation}
and the embedding coordinates are
\begin{align}
  X_0&=\frac{-t^2+x^2+z^2+1}{2z},
  &
  X_1&=\frac{t}{z},
  \nonumber\\
  X_2&=-\frac{-t^2+x^2+z^2-1}{2z},
  &
  X_3&=\frac{x}{z}.
  \label{eq:Poincare-embedding-coordinates}
\end{align}
For two spacelike-separated bulk points, \eqref{eq:length-AdS} gives
\begin{equation}
  \cL_{ij}=\arccosh\left[
  \frac{(x_i-x_j)^2-(t_i-t_j)^2+z_i^2+z_j^2}
  {2z_i z_j}\right].
  \label{eq:length-Poincare}
\end{equation}
\subsection*{Geodesics on a fixed time-slice}
The geodesics on the constant time slice of the BTZ black brane  \cref{eq:BTZ-slice} may be found by solving the Euler-Lagrange’s equations for the induced area functional. Writing the profile of the geodesic as $z=z(x)$, a first integral of motion reads
\begin{equation}
		z\sqrt{\frac{1+(z')^2}{1-z^2/z_h^2}}=\text{const.}
	\label{eq:2.8}
\end{equation}
where the prime denotes differentiation with respect to $x$.
The general solutions may be written in terms of the integration constants $(x_0,c_0)$ as 
\begin{equation}
	\sqrt{1 - \frac{z^2}{z_h^2}} = \frac{1}{2 z_h} \left( e^{(x - x_0)/z_h} + (z_h^2 - c_0^2) e^{-(x + x_0)/z_h} \right)\,.
	\label{eq:BTZ-generic-geodesic}
\end{equation}
For example, for a geodesic anchored at two boundary points $x_1$ and $x_2$ (cf. \cref{eq:class-I-geodesic}), we may find
\begin{align}
	c_0 = z_h \tanh \left( \frac{x_2 - x_1}{z_h} \right)~~,~~
	x_0= z_h \log \left( \frac{e^{x_1/z_h} + e^{x_2/z_h}}{2 z_h} \right)\,.
	\label{eq:BTZ-bdy-bdy}
\end{align}
On the other hand, for the geodesics belonging to class-II (cf. \cref{eq:class-II-geodesic}) with endpoints $(x_1,\epsilon)$ and $(x_2,z_h)$, we may obtain
\begin{align}
	c_0 = z_h \coth \left( \frac{x_1 - x_2}{z_h} \right)~~,~~
	x_0 = x_2 + z_h \log \left[ \frac{\sinh \left( \frac{x_1 - x_2}{z_h} \right)}{z_h} \right] \,.
	\label{eq:BTZ-bdy-horizon}
\end{align}

\section{Geodesic curvature of distance-difference flows}
\label{app:geodesic-curvature}

This appendix collects the curvature calculation used for the
endpoint-superposed flows in section~\ref{sec:finite-cutoff}.  The derivation is
intrinsic to unit hyperbolic space and therefore applies equally to the
Poincar\'e, global, and BTZ coordinate systems. 

Let $p_\pm$ be two fixed points of $\mathbb H^2$, let
$\cL_\pm=\cL(\,\cdot\,,p_\pm)$, and define
\begin{equation}
	\Psi=\frac12(\cL_+-\cL_-).
	\label{eq:app-curvature-streamfunction}
\end{equation}
Away from the endpoints and the critical set, a streamline of
$\star\dd\Psi$ is a regular level set of $\Psi$.  Denote by
$\vartheta$ the angle between the two unit distance gradients,
\begin{equation}
	\cos\vartheta
	=g\!\left(\nabla \cL_+,\nabla\cL_-\right).
	\label{eq:app-curvature-angle}
\end{equation}
The Hessian of a distance function on unit \(\mathbb H^2\) obeys
\begin{equation}
	\nabla^2\cL=\coth \cL\left(g-\dd \cL\otimes\dd \cL\right).
	\label{eq:app-distance-Hessian}
\end{equation}

Let $T$ be a unit tangent to a regular level set.  Since
$T\cdot(\nabla \cL_+-\nabla \cL_-)=0$ and both distance gradients have unit
norm,
\begin{equation}
	(T\cdot\nabla \cL_+)^2=(T\cdot\nabla \cL_-)^2
	=\frac{1+\cos\vartheta}{2}.
	\label{eq:app-tangent-projections}
\end{equation}
Applying \eqref{eq:app-distance-Hessian} to the two distances gives
\begin{equation}
	\nabla^2\Psi(T,T)
	=\frac{1-\cos\vartheta}{4}
	\left(\coth \cL_+-\coth \cL_-\right),
	\qquad
	|\nabla\Psi|=\sqrt{\frac{1-\cos\vartheta}{2}}.
	\label{eq:app-Hess-Psi}
\end{equation}
The geodesic curvature of a regular level set is defined through
$|k_g|=|\nabla^2\Psi(T,T)|/|\nabla\Psi|$.  Therefore
\begin{equation}
		|k_g|
		=\frac{\sqrt{2(1-\cos\vartheta)}}{4}
		\left|\coth \cL_+-\coth \cL_-\right|.
	\label{eq:app-curvature-exact}
\end{equation}
The angular factor vanishes only where the two gradients coincide, while
the second factor vanishes on the reflection locus $\cL_+=\cL_-$.  In the
regular wedges considered in the main text, the latter is the central
streamline.  Thus reflection symmetry protects one geodesic integral curve;
it does not make the entire source superposition geodesic.

\section{Geodesic bit threads in the BTZ black brane revisited}\label{appB}
In this appendix, we revisit the geodesic bit threads in the BTZ black brane \eqref{eq:BTZ-slice} discussed in \cite{Agon:2018lwq}, in the Fermi normal coordinates \eqref{eq:BTZ-TFD-Fermi-metric}. The proper radial coordinate $\rho$ brings the constant time BTZ metric \eqref{eq:BTZ-slice} to \eqref{eq:BTZ-TFD-Fermi-metric}.  Thus the exterior slice is one Rindler wedge of the unit hyperbolic plane
$\Htwo$.  We embed the covering hyperbolic plane in $\mathbb R^{1,2}$ with
metric $\eta=\operatorname{diag}(-1,1,1)$, as
\begin{equation}
	Y(\rho,x)
	=\left(\cosh\rho\cosh \left(\frac{x}{z_h}\right),\,
	\cosh\rho\sinh\left(\frac{x}{z_h}\right),\,
	\sinh\rho\right)~~,~~Y^2=-1.
	\label{eq:app-BTZ-H2-embedding}
\end{equation}
Pulling back the ambient metric immediately reproduces \eqref{eq:BTZ-TFD-Fermi-metric}. The RT surface \eqref{eq:BTZ-RT-profile} is given by the intersection of the
hyperboloid with a plane $N\cdot Y=0$, where
\begin{equation}
	N=\left(\csch \left(\frac{b}{z_h}\right),0,\coth \left(\frac{b}{z_h}\right)\right),
	\qquad
	N\cdot N=1.
	\label{eq:app-BTZ-plane-normal}
\end{equation}
A unit-speed parametrization of this geodesic is
\begin{equation}
	P(\eta)
	=\left(\coth \left(\frac{b}{z_h}\right)\cosh\eta,\,
	\sinh\eta,\,
	\csch \left(\frac{b}{z_h}\right)\cosh\eta\right).
	\label{eq:app-BTZ-RT-parametrization}
\end{equation}
where $\eta$ parametrizes the proper length along it. Comparing the first two components of \cref{eq:app-BTZ-H2-embedding,eq:app-BTZ-RT-parametrization}, we find the following useful identity valid on the RT surface
\begin{align}
	\tanh \left(\frac{x}{z_h}\right)=\tanh \left(\frac{b}{z_h}\right)\,\tanh\eta
	\qquad \text{on }~\gamma_A\,.\label{eq:app-BTZ-RT-X-eta-map}
\end{align}
\subsection*{Oriented normal congruence and bit-thread flow}
The normal $N$ in \eqref{eq:app-BTZ-plane-normal} points from the RT
geodesic toward the interval $A$.  We orient the bit-thread flow in the
opposite direction, from $A$ through the RT surface toward its complement.
Accordingly, the signed Fermi distance $\lambda$ is defined by
\begin{equation}
	Y(\eta,\lambda)
	=\cosh\lambda\,P(\eta)-\sinh\lambda\,N.
	\label{eq:app-BTZ-oriented-Fermi-map}
\end{equation}
Thus $\lambda=0$ is the RT surface, $\lambda<0$ is its $A$ side, and
$\lambda>0$ is its complementary side.  The coordinate tangent vectors are
\begin{align}
	\partial_\lambda Y
	&=\sinh\lambda\,P-\cosh\lambda\,N\,,
	\nonumber\\
	\partial_\eta Y
	&=\cosh\lambda\,P'(\eta)\,,
	\label{eq:app-BTZ-Fermi-tangents}
\end{align}
leading to the following metric adapted to the normal congruence
\begin{align}
	\dd s^2=\dd\lambda^2+\cosh^2\lambda\,\dd\eta^2
\end{align}
Therefore, the curves
$\eta=\textrm{constant}$ are unit-speed geodesics orthogonal to $\gamma_A$. 

The geodesic bit thread flow constructed in \cite{Agon:2018lwq} is a divergenceless vector field tangent to this normal congruence with unit norm (in Planck units) on $\gamma_A$. Consider a candidate vector field 
\begin{align}
	v_A=\frac{1}{4G}f(\eta,\lambda)\,\del_\lambda\,.
\end{align}
Its divergence is
\begin{align}
	\nabla_\mu v_A^\mu
	&=\frac{1}{4G\cosh\lambda}
	\partial_\lambda\!\left(\cosh\lambda\,f(\lambda,\eta)\right).
\end{align}
Consequently, $\nabla\cdot v_A=0$ implies
\begin{equation}
	\cosh\lambda\,f(\lambda,\eta)=C(\eta).
	\label{eq:app-BTZ-Fermi-conservation-law}
\end{equation}
Pointwise saturation at the RT bottleneck requires
$f(0,\eta)=1$.  Hence,  we have $C(\eta)=1$ and the bit threads field is obtained as
\begin{align}
	v_A=\frac{1}{4G}\sech\lambda\,\del_\lambda\,.\label{eq:app-BTZ-Fermi-flow}
\end{align}
The norm bound follows immediately from $\sech\lambda\leq1$, with equality
only at $\lambda=0$.  Geometrically, a transverse strip has width
$\cosh\lambda\,\dd\eta$, so its flux is
\begin{equation}
	\dd\Phi
	=|v_A|\cosh\lambda\,\dd\eta
	=\frac{\dd\eta}{4G},
	\label{eq:app-BTZ-strip-flux}
\end{equation}
which is independent of $\lambda$.  Equation
\eqref{eq:app-BTZ-Fermi-flow} is therefore the unique divergenceless field
within this fixed normal-geodesic congruence that saturates the RT surface.

We now transform the bit threads vector field in the ADM coordinates $(x,z)$. To express $(\eta,\lambda)$ in terms of $(x,z)$, note the relation $N\cdot P(\eta)=0$, which leads to 
\begin{align}
		N\cdot Y=-\sinh\lambda.
		\label{eq:app-BTZ-N-dot-Y}
\end{align}
Using \eqref{eq:app-BTZ-H2-embedding}, this gives
\begin{equation}
		\sinh\lambda
		=\frac{\zh}{z\sinh \left(\frac{b}{z_h}\right)}
		\left[\cosh \left(\frac{x}{z_h}\right)-\sqrt{1-\frac{z^2}{z_h^2}}\cosh \left(\frac{b}{z_h}\right)\right].
	\label{eq:app-BTZ-sinh-lambda}
\end{equation}
It is useful to define the shorthand notations
\begin{align}
	\Delta_\pm:=\cosh\left(\frac{b\pm x}{z_h}\right)-\sqrt{1-\frac{z^2}{z_h^2}}~~,~~
	\mathcal D:=\Delta_-\Delta_+.
	\label{eq:app-BTZ-Delta-definitions}
\end{align}
Combining these with \eqref{eq:app-BTZ-sinh-lambda} gives the norm of the bit-htreads field
\begin{align}
	\left|v_A\right|=\frac{1}{4G}\sech\lambda=\frac{1}{4G}\frac{z}{\zh\sqrt{\mathcal D}}\sinh \left(\frac{b}{z_h}\right)\,.\label{eq:app-norm}
\end{align}
which is identical to \eqref{eq:BTZ-bit-norm}. Because $\lambda$ is a unit-distance coordinate, its coordinate vector is
the metric gradient of the corresponding scalar function,
\begin{equation}
	(\partial_\lambda)^\mu=g^{\mu\nu}\partial_\nu\lambda.
	\label{eq:app-BTZ-lambda-gradient}
\end{equation}
Consequently, a short calculation leads to
\begin{equation}
	\partial_\lambda
	=\frac{z^2\sinh \left(\frac{x}{z_h}\right)}{\zh\sqrt{\mathcal D}}\,\partial_x
	+\frac{z}{\sqrt{\mathcal D}}\sqrt{1-\frac{z^2}{z_h^2}}\left[\cosh \left(\frac{b}{z_h}\right)-\sqrt{1-\frac{z^2}{z_h^2}}\cosh \left(\frac{x}{z_h}\right)\right]
	\,\partial_z.
	\label{eq:app-BTZ-lambda-coordinate-vector}
\end{equation}
Multiplying the norm \eqref{eq:app-norm}, this precisely leads to the PEE-selected geodesic max flow
\eqref{eq:BTZ-bit-flow}.

For completeness, we note that the other Fermi coordinate follows from comparing the $Y_1$ component of \cref{eq:app-BTZ-H2-embedding,eq:app-BTZ-oriented-Fermi-map},
\begin{align}
	\sinh\eta=\frac{\sinh B\sinh X}{\sqrt{\mathcal D}}\,.\label{eq:app-BTZ-sinh-cosh-eta}
\end{align}
It follows that
\begin{equation}
	\e^{\eta}=\frac{\Delta_+}{\sqrt{\mathcal D}},
	\qquad
	\e^{-\eta}=\frac{\Delta_-}{\sqrt{\mathcal D}},
	\label{eq:app-BTZ-exp-eta}
\end{equation}
and therefore
\begin{equation}
		\eta(x,z)=\frac12\log\left(\frac{\Delta_+}{\Delta_-}\right)
		=\frac12\log\left[
		\frac{\cosh\!\left(\frac{b+x}{\zh}\right)
			-\sqrt{1-\frac{z^2}{\zh^2}}}
		{\cosh\!\left(\frac{b-x}{\zh}\right)
			-\sqrt{1-\frac{z^2}{\zh^2}}}
		\right]\,.
	\label{eq:app-BTZ-eta-stream-function}
\end{equation}
Thus $\eta$ is simultaneously the proper length on the RT surface and a global
label for the normal-geodesic streamlines.

\paragraph{Endpoint sectors and the separatrix.}
For fixed $\eta$, the two ideal endpoints of the complete normal geodesic
are the null rays represented by $P(\eta)\pm N$.  With the
orientation in \eqref{eq:app-BTZ-oriented-Fermi-map}, the endpoint reached
as $\lambda\to+\infty$ is $P(\eta)-N$.  Its third embedding component is
\begin{equation}
	\bigl[P(\eta)-N\bigr]_2
	=\frac{\cosh\eta-\cosh \left(\frac{b}{z_h}\right)}{\sinh \left(\frac{b}{z_h}\right)}.
	\label{eq:app-BTZ-endpoint-component}
\end{equation}
It lies on the opposite asymptotic boundary when $|\eta|<\frac{b}{z_h}$ and on the
same component when $|\eta|>\frac{b}{z_h}$.  Hence the corresponding one-sided
streamline ends on the horizon in the first case and returns to the
asymptotic boundary in the second.  The measure-zero cases $|\eta|=\frac{b}{z_h}$ are
the two separatrices.  Using \eqref{eq:app-BTZ-RT-X-eta-map} at
$\eta=\pm \frac{b}{z_h}$ gives
\begin{align}
	\tanh \left(\frac{x_\beta}{z_h}\right)
	&=\tanh^2\left(\frac{b}{z_h}\right)\implies 
	x_{\beta}
	=\zh\arctanh\!\left(\tanh^2\!\frac{b}{\zh}\right)
	=\frac{\zh}{2}\log\!\left[
	\cosh\!\left(\frac{2b}{\zh}\right)\right].
	\label{eq:app-BTZ-separatrix}
\end{align}
The RT segment pierced by horizon-ending flow lines is therefore
$-\frac{b}{z_h}<\eta<\frac{b}{z_h}$.  Since $\eta$ is proper length on $\gamma_A$, its flux is
\begin{equation}
	\Phi_{A\to h}
	=\frac{1}{4G}\int_{-\frac{b}{z_h}}^{\frac{b}{z_h}}\dd\eta
	=\frac{b}{2G\zh}.
	\label{eq:app-BTZ-thermal-flux}
\end{equation}
This is exactly the extensive thermal contribution for an interval of
length $2b$. 

\section{Finite-cutoff global AdS$_3$}
\label{subsec:cutoff-global}

The compact geometry provides a useful finite-size counterpart of the
Poincar\'e analysis.  We follow the same logical order: first the exact chord
kernel and contour, then the elementary current and its source
superposition, followed by the norm and flux checks, and finally a comparison
with the independent normal-geodesic representative.  The only new global
ingredient is the principal angular branch; it changes the endpoint topology
but not the local hyperbolic identities.

On a constant-time slice of global AdS$_3$,
\begin{equation}
	\dd s^2=\frac{\dd z^2}{z^2(1+z^2)}+\frac{\dd\theta^2}{z^2},
	\qquad \Gamma:\ z=\zc.
	\label{eq:global-cutoff-metric}
\end{equation}
The angular coordinate obeys $\theta\sim\theta+2\pi$.  We use the principal
circle separation
\begin{equation}
	d_{S^1}(\theta_1,\theta_2)
	=\min_{n\in\mathbb Z}|\theta_1-\theta_2+2\pi n|
	\in[0,\pi].
	\label{eq:global-principal-separation}
\end{equation}
We take $A=[-b,b]$ in the direct connected phase, $0<b\leq\pi/2$.
For larger arcs the RT prescription uses the complementary geodesic.  The RT
profile is
\begin{equation}
	\sqrt{1+z^2}=\sqrt{1+\zc^2}\,\frac{\cos\theta}{\cos b}.
	\label{eq:global-cutoff-RT}
\end{equation}
Consequently, from \eqref{eq:length-global}, we have 
\begin{align}
	L_c(\theta_1,\theta_2)
	&=2\arcsinh\left[
	\frac{\sin(d_{S^1}(\theta_1,\theta_2)/2)}{\zc}\right],
	\label{eq:global-cutoff-length}\\
	\cI_c(\theta_1,\theta_2)
	&=\frac{(1+\zc^2)\sin(d_{S^1}(\theta_1,\theta_2)/2)}
	{16G\bigl[\zc^2+\sin^2(d_{S^1}(\theta_1,\theta_2)/2)\bigr]^{3/2}}.
	\label{eq:global-cutoff-kernel}
\end{align}

The cutoff resolves the coincident-endpoint singularity into a finite peak.
For \(0<\zc\leq\sqrt2\), the maximum occurs at
\begin{equation}
	\delta_{\rm peak}
	=2\arcsin\left(\frac{\zc}{\sqrt2}\right),
	\qquad \delta=d_{S^1}(\theta_1,\theta_2).
	\label{eq:global-kernel-peak}
\end{equation}
Thus the kernel suppresses angular separations below the cutoff boundary resolution
scale instead of concentrating arbitrarily strongly at coincidence.  Its
integral over the complete cutoff circle nevertheless obeys the exact local
capacity identity
\begin{equation}
	\int_0^{2\pi}\dd\theta_2\,
	\cI_c(\theta_1,\theta_2)
	=\frac{1}{4G\zc}.
	\label{eq:global-circle-capacity}
\end{equation}
The same capacity appears on the noncompact Poincar\'e cutoff boundary.  Compactness
redistributes that fixed amount around the circle rather than changing its
local normalization.

\begin{figure}[t]
	\centering
	\includegraphics[width=0.96\textwidth]
	{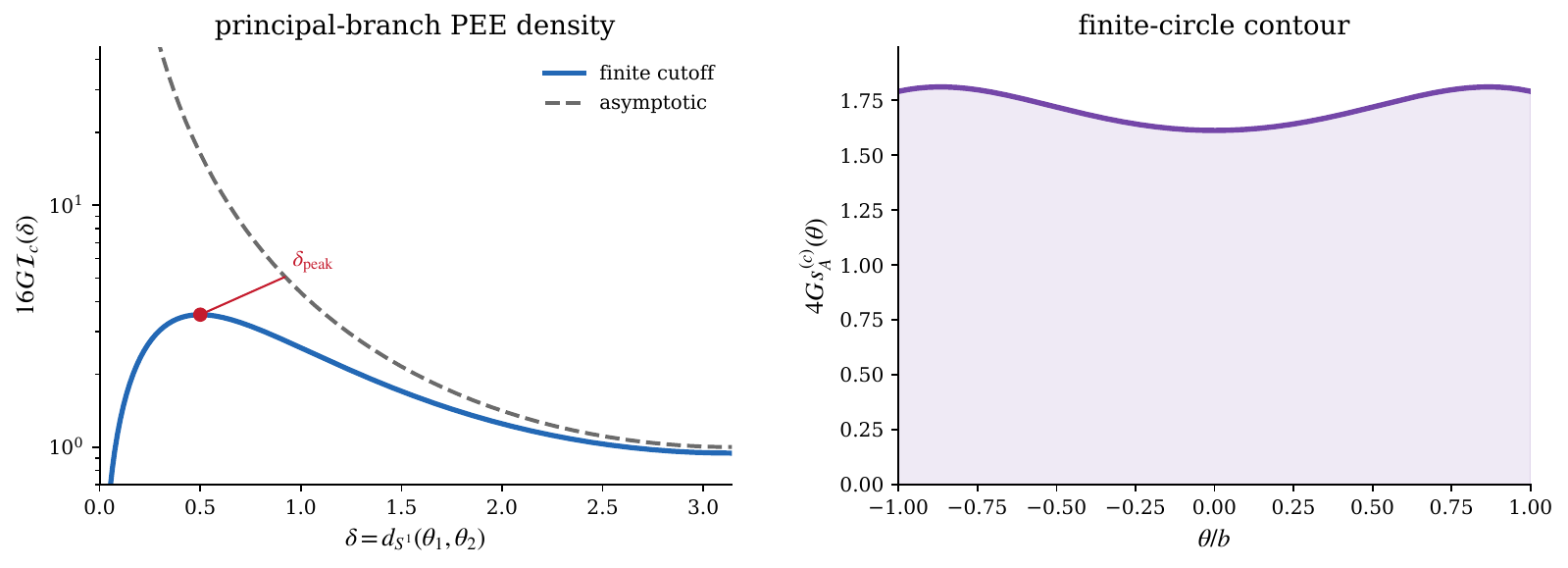}
	\caption{Finite-cutoff global-AdS endpoint data on the principal branch.
		Left: the regulated PEE density and its peak
		\eqref{eq:global-kernel-peak}.  Right: the interval contour for the same
		parameters.  The finite peak is a resolution effect, whereas the area
		under the full-circle kernel is fixed by
		\eqref{eq:global-circle-capacity}.}
	\label{fig:global-kernel-contour}
\end{figure}

Integrating the principal-branch density over $A^c$ gives the cutoff boundary contour
\begin{align}
	s_A^{(c)}(\theta)
	&=\frac{1}{8G}\left[
	\frac{\cos\left(\frac{b-\theta}{2}\right)}
	{\sqrt{\zc^2+\sin^2\left(\frac{b-\theta}{2}\right)}}
	+\frac{\cos\left(\frac{b+\theta}{2}\right)}
	{\sqrt{\zc^2+\sin^2\left(\frac{b+\theta}{2}\right)}}
	\right],
	\quad -b<\theta<b,
	\label{eq:global-cutoff-contour}
\end{align}
and hence
\begin{equation}
	\int_{-b}^{b}\dd\theta\,s_A^{(c)}(\theta)
	=\frac{1}{2G}\arcsinh\left(\frac{\sin b}{\zc}\right).
	\label{eq:global-contour-integral}
\end{equation}
The density is exact geometric data; its PEE interpretation carries
the finite-cutoff assumptions stated in section~\ref{subsec:review-ttbar}.
It is useful to integrate it only up to a variable cutoff boundary point.  With
\begin{equation}
	s_b=\arcsinh\left(\frac{\sin b}{\zc}\right),
\end{equation}
the PEE-selected boundary-to-RT landing coordinate is
\begin{equation}
	\eta_{\rm PEE}(\theta)
	=\arcsinh\left[
	\frac{\sin\left(\frac{b+\theta}{2}\right)}{\zc}\right]
	-\arcsinh\left[
	\frac{\sin\left(\frac{b-\theta}{2}\right)}{\zc}\right],
	\qquad
	\eta_{\rm PEE}(\pm b)=\pm s_b .
	\label{eq:global-PEE-wall-map}
\end{equation}
Indeed \(s_A^{(c)}=(4G)^{-1}\dd\eta_{\rm PEE}/\dd\theta\).
The map is monotone, so the source superposition preserves endpoint order
even though its integral curves will not generally be geodesics.  Figure~\ref{fig:global-kernel-contour} shows how compactness modifies both the
regulated kernel and the contour: the resolution peak is local, whereas the
integrated full-circle capacity remains fixed.

\paragraph{PEE thread flow.}
The geodesic joining $(\theta_0,\zc)$ and $(y,\zc)$ has profile
\begin{equation}
	\sqrt{1+z^2}
	=\sqrt{1+\zc^2}\,
	\frac{\cos\left(\theta-\frac{\theta_0+y}{2}\right)}
	{\cos\left(\frac{y-\theta_0}{2}\right)}.
	\label{eq:global-cutoff-geodesic}
\end{equation}
For $\theta_0=0$, the second endpoint of the geodesic through $Q=(\theta,z)$ is determined by
\begin{equation}
	\tan\frac{y}{2}
	=\frac{\sqrt{1+z^2}-\sqrt{1+\zc^2}\cos\theta}
	{\sqrt{1+\zc^2}\sin\theta}.
	\label{eq:global-cutoff-endpoint-map}
\end{equation}
This relation selects the connected chord inside the direct interval phase.  The complementary branch must be used once the interval is represented by its shorter complementary arc.

For any angle $\phi$ define
\begin{equation}
	\Delta(\phi)=\bigl(\sqrt{1+z^2}-\sqrt{1+z_c^2}\cos\phi\bigr)^2+\zc^2\sin^2\phi~~,~~
	F(\phi)=\frac{\sqrt{1+z^2}\cos\phi-\sqrt{1+z_c^2}}{\sqrt{\Delta(\phi)}}.
	\label{eq:global-cutoff-functions}
\end{equation}
For a source at $\theta_0$, let $\phi=\theta-\theta_0$.  Flux matching with \eqref{eq:global-cutoff-kernel} leads to the PEE thread current
\begin{align}
	V_{\theta_0}^{\mu}&=\frac{1}{8G}\frac{z^3\bigl(\sqrt{1+z^2}-\sqrt{1+z_c^2}\cos\phi\bigr)}{\Delta(\phi)^{3/2}}\left(\sin\phi,\frac{\sqrt{1+z^2}(\sqrt{1+z^2}\cos\phi-\sqrt{1+z_c^2})}{z}\right)\,,\label{eq:global-cutoff-elementary}\\
	|V_{\theta_0}|
	&=\frac{1}{8G}
	\frac{z\bigl(\sqrt{1+z^2}-\sqrt{1+z_c^2}\cos\phi\bigr)}{\Delta(\phi)}\,.
	\label{eq:global-cutoff-elementary-norm}
\end{align}
Direct differentiation verifies $\nabla_\mu V_{\theta_0}^{\mu}=0$.  The formulas reduce smoothly to the global-vacuum current when $\zc\to0$.
\paragraph{PEE-selected bit threads.}
Integrating the PEE thread flow \eqref{eq:global-cutoff-elementary} over $\theta_0\in[-b,b]$ gives the coarse-grained current
\begin{align}
	v_A^\mu&=\frac{z}{8G}\left(F(\theta-b)-F(\theta+b)\,,\,z\sqrt{1+z^2}\left[
	\frac{\sin(\theta+b)}{\sqrt{\Delta(\theta+b)}}
	-\frac{\sin(\theta-b)}{\sqrt{\Delta(\theta-b)}}\right]\right)\,.
	\label{eq:global-cutoff-bit}
\end{align}
For the global flow \eqref{eq:global-cutoff-bit}, define the unit vector
\begin{equation}
	\mathbf m(\phi)=\frac{1}{\sqrt{\Delta(\phi)}}
	\left(\sqrt{1+z^2}\cos\phi-\sqrt{1+z_c^2},-z\sin\phi\right),
	\qquad \mathbf m(\phi)^2=1.
	\label{eq:global-unit-pair}
\end{equation}
In the orthonormal frame $e_{\hat\theta}=z\partial_\theta$ and $e_{\hat z}=z\sqrt{1+z^2}\partial_z$, equations \eqref{eq:global-cutoff-bit} gives
\begin{equation}
	4G\bigl(v_A^{\hat\theta},v_A^{\hat z}\bigr)
	=\frac12\left[\mathbf m(\phi_R)-\mathbf m(\phi_L)\right].
	\label{eq:global-component-identity}
\end{equation}
from which, it follows that $|v_A|\leq1/(4G)$.  On the RT profile \eqref{eq:global-cutoff-RT}, $\mathbf m(\phi_R)=-\mathbf m(\phi_L)$; the field saturates and is normal.  The RT flux is therefore
\begin{equation}
	\Phi_A=\frac{1}{2G}\arcsinh\left(\frac{\sin b}{\zc}\right)=S_A^{(c)}\,,
\end{equation}
which is the finite-radius global-AdS entropy familiar from the cutoff construction. Its small-$\zc$ expansion is
\begin{equation}
	S_A^{(c)}=\frac c3\left[
	\log\frac{2\sin b}{\zc}
	+\frac{\zc^2}{4\sin^2b}+O(\zc^4)\right].
	\label{eq:global-cutoff-expansion}
\end{equation}
The leading logarithm is the usual cylinder result with the cutoff replacing
the ultraviolet regulator.  The positive \(O(\zc^2)\) term records the
finite radial position and is enhanced as the interval approaches the
antipodal transition, where the direct and complementary branches compete.
It should therefore be interpreted together with the branch prescription,
not as a noncompact planar correction.

The inward normal to the cutoff boundary is
$n_{\Gamma_c}=\zc\sqrt{1+\zc^2}\,\partial_z$ and
$\dd\Sigma=\dd\theta/\zc$.  Therefore the cutoff boundary flux agrees with the contour
pointwise,
\begin{equation}
	\left.\frac{v_A^z}{\zc^2\sqrt{1+\zc^2}}\right|_{z=\zc}
	=s_A^{(c)}(\theta),
	\qquad -b<\theta<b.
	\label{eq:global-pointwise-wall-flux}
\end{equation}

\paragraph{Geodesic bit threads.}
To compare the PEE-selected bit-thread flow with a geodesic max flow, consider the coordinate transformation $z=\csch\rho$.  The spatial metric becomes
\begin{equation}
	\dd s^2=\dd\rho^2+\sinh^2\rho\,\dd\theta^2,
	\qquad
	\csch\rho_c=\zc.
	\label{eq:global-polar-metric}
\end{equation}
In the hyperboloid embedding
\(X=(\cosh\rho,\sinh\rho\cos\theta,
\sinh\rho\sin\theta)\), set
\begin{equation}
	\begin{aligned}
		D&=\sqrt{\zc^2+\sin^2b},
		&E_0&=\frac{(\sqrt{1+\zc^2},\cos b,0)}{D},\\
		E_1&=(0,0,1),
		&N&=\frac{(\cos b,\sqrt{1+\zc^2},0)}{D}.
	\end{aligned}
	\label{eq:global-Fermi-frame}
\end{equation}
Here \(E_0^2=-1\), \(E_1^2=N^2=1\), and the complete geodesic
containing \(\gamma_A\) is \(N\cdot X=0\).  Exact Fermi coordinates are
defined by
\begin{equation}
	X(\eta,\lambda)
	=\cosh\lambda\left(\cosh\eta\,E_0+
	\sinh\eta\,E_1\right)+\sinh\lambda\,N,
	\qquad
	\dd s^2=\dd\lambda^2+\cosh^2\lambda\,\dd\eta^2.
	\label{eq:global-Fermi-map}
\end{equation}
The physical RT segment is \(\lambda=0\), \(|\eta|\leq s_b\).
Flux conservation along the normal congruence therefore gives
\begin{equation}
	v_{\rm geo}
	=\frac{1}{4G\cosh\lambda}\,\partial_\lambda,
	\qquad
	4G|v_{\rm geo}|=\sech\lambda .
	\label{eq:global-normal-flow}
\end{equation}
It is divergenceless, saturates on and is normal to \(\gamma_A\), and obeys
the norm bound pointwise.

Intersecting a fixed-\(\eta\) normal with the cutoff circle yields the exact
boundary-to-RT map
\begin{equation}
	\eta_{\rm geo}(\theta)
	=\arctanh\left[
	\frac{D\sin\theta}
	{1+\zc^2-\cos b\cos\theta}\right],
	\qquad
	\eta_{\rm geo}(\pm b)=\pm s_b .
	\label{eq:global-normal-wall-map}
\end{equation}
The corresponding cutoff boundary density is
\begin{equation}
	s_{\rm geo}(\theta)
	=\frac{1}{4G}\frac{\dd\eta_{\rm geo}}{\dd\theta}
	=\frac{D\left[(1+\zc^2)\cos\theta-\cos b\right]}
	{4G\left[
		(1+\zc^2-\cos b\cos\theta)^2
		-D^2\sin^2\theta\right]} .
	\label{eq:global-normal-density}
\end{equation}
Equations \eqref{eq:global-PEE-wall-map} and
\eqref{eq:global-normal-wall-map} have the same endpoints and hence the same
integrated RT flux, but differ pointwise for finite \(\zc\).  This is the
global-cylinder version of the Poincar\'e calibration ambiguity: entropy
fixes the bottleneck measure, not how cutoff boundary points are assigned to it.

\begin{figure}[t]
	\centering
	\includegraphics[width=0.98\textwidth]
	{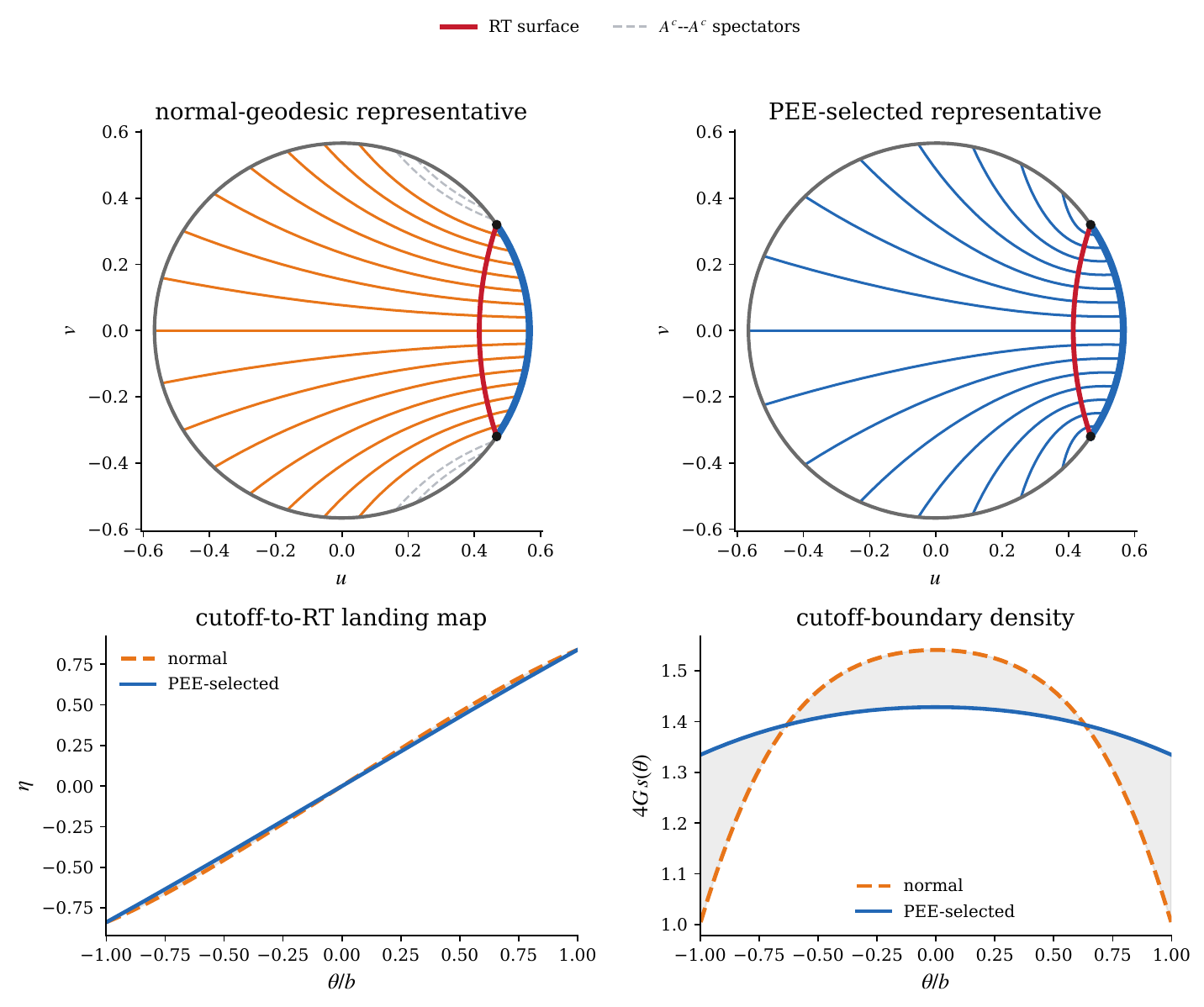}
	\caption{Two max-flow representatives in the cutoff global-AdS disk.
		Active curves cross the physical RT segment; light gray normal curves are
		the spectator continuation with both endpoints in \(A^c\).  The lower
		panels compare the exact landing maps and cutoff boundary densities.  Agreement of
		the endpoint values fixes the common entropy, while the separation in the
		interior measures the off-bottleneck max-flow degeneracy.}
	\label{fig:global-flow-comparison}
\end{figure}

A normal geodesic entering from \(\theta\in A\) re-enters the cutoff boundary at
\(\theta'\in A^c\), with the branch chosen continuously on the circle,
\begin{equation}
	\tan\frac{\theta'}2
	=\frac{1+\zc^2-\cos b}{1+\zc^2+\cos b}
	\cot\frac\theta2 .
	\label{eq:global-normal-partner}
\end{equation}
At finite cutoff the image of \(A\) does not begin immediately at
\(\pm b\); the unused adjacent complement arcs are bounded by the partner
of the endpoints.  Extending the smooth Fermi field beyond
\(|\eta|=s_b\) then produces complement-to-complement spectator curves.
They do not cross the physical RT segment and carry no flux out of \(A\).
Their role is to complete a smooth divergenceless representative, not to add
a new PEE sector.  In the asymptotic limit the active PEE and normal maps
coincide, the unused arcs collapse, and the spectators are pushed to the
ideal endpoints.  Figure~\ref{fig:global-flow-comparison} keeps the normal
and PEE-selected disk plots separate in this compact geometry so that the
spectator arcs and their cutoff-boundary endpoints remain legible; the lower plots then
compare the two calibrations using a common scale.

\begin{figure}[t]
	\centering
	\includegraphics[width=0.75\textwidth]
	{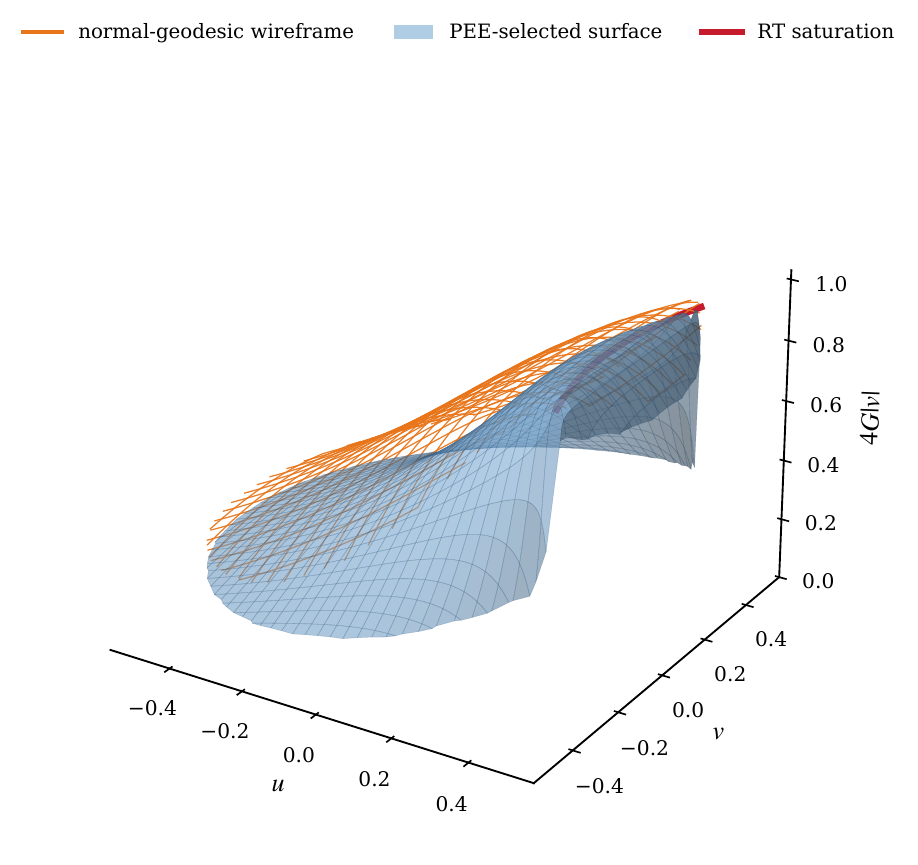}
	\caption{Invariant norms \(4G|v|\) of the cutoff global-AdS flows,
		superposed in one plot.  The translucent blue surface is the PEE-selected
		norm, the orange wireframe is the normal-geodesic norm, and the red unit
		ridge is their common RT saturation locus.  Both remain below the bound
		away from the bottleneck.}
	\label{fig:global-norm-comparison}
\end{figure}

The common RT saturation is easier to compare after superposing the norm
representatives.  Figure~\ref{fig:global-norm-comparison} displays the
PEE-selected surface and the normal wireframe on the same axes.  Their
coincident red ridge fixes the entropy, while the separation elsewhere is
the invariant imprint of their different streamline geometries.

The endpoint-telescoping argument of
\eqref{eq:pcut-elementary-potential}--\eqref{eq:pcut-endpoint-telescoping} is
intrinsic to $\mathbb H^2$ and applies here on the direct branch.  Thus the
coarse-grained streamlines are level sets of the distance difference to the
finite endpoints $(-b,\zc)$ and $(b,\zc)$.  The curvature formula
\eqref{eq:app-curvature-exact} consequently applies without change: the
reflection-symmetric central streamline is geodesic, while regular
noncentral streamlines are generically not.  This is a statement about the
PEE-selected representative, not about every global-AdS max flow.

At $d_{S^1}=\pi$ the two angular representatives of the principal branch
meet.  This measure-zero locus corresponds to the unique diameter as an
unoriented bulk chord.  The chord divides the disk, and the homology choice
selects whether it represents $A$ or its complement; their geodesic lengths
agree at the transition.  No additional endpoint sector is generated.

For completeness, the Lorentzian cutoff boundary metric obtained after the standard
Weyl normalization is
\begin{equation}
	\dd s_{\Gamma_c}^2=-(1+\zc^2)\dd t^2+\dd\theta^2.
	\label{eq:global-wall-metric}
\end{equation}
Hence normalized cutoff boundary time is
$\tau=\sqrt{1+\zc^2}\,t$.  The result is naturally a finite-size cylinder
observable in the hard-cutoff sector; it should not be identified without
qualification with the noncompact planar $T\bar T$ theory.

\section{Universal Fermi-coordinate form of the finite-cutoff PEE-selected flow}
\label{app:cutoff-fermi-nongeodesic}
In this appendix we express the PEE-selected, generically non-geodesic bit-threads flow directly in the same Fermi chart used for the normal geodesic constructions.  The result is universal on a unit-curvature hyperbolic slice.  The Poincar\'e, global-AdS and planar-BTZ
cutoff geometries differ only in the proper half-length of the RT segment and in the way the cutoff boundary and horizon truncate the Fermi chart.

\subsection*{Universal endpoint-distance potential}
\label{app:cutoff-universal-fermi}
Let $(\eta,\lambda)$ be Fermi coordinates about the complete geodesic
containing the RT surface,
\begin{equation}
	\mathrm ds^2=\mathrm d\lambda^2+
	\cosh^2\!\lambda\,\mathrm d\eta^2\,.
	\label{eq:cutoff-fermi-metric}
\end{equation}
We orient $\lambda$ so that it increases from the interval, through the RT
surface at $\lambda=0$, toward the complementary side.  Write the physical
RT segment as $|\eta|\leq a$, with its finite endpoints at
\begin{equation}
	p_\pm:(\eta,\lambda)=(\pm a,0).
	\label{eq:cutoff-fermi-foci}
\end{equation}
The hyperbolic distances $\cL_\pm=\cL(X,p_\pm)$ obey the right-triangle identity
\begin{equation}
	\cosh \cL_\pm
	=\cosh\lambda\cosh(\eta\mp a).
	\label{eq:cutoff-fermi-distances}
\end{equation}
Consequently, the endpoint-distance potential selected by the PEE
superposition is given by
\begin{equation}
	\Psi_A(\eta,\lambda)
	=\frac12\left[
	\operatorname{arccosh}\left(\cosh\lambda\cosh(\eta-a)\right)
	-\operatorname{arccosh}\left(\cosh\lambda\cosh(\eta+a)\right)
	\right]\,.
	\label{eq:cutoff-fermi-PEE-potential}
\end{equation}
For $\lambda=0$ and $|\eta|<a$, we have
\begin{equation}
	\cL_+=a-\eta~~,~~
	\cL_-=a+\eta~~,~~
	\Psi_A(\eta,0)=-\eta.
	\label{eq:cutoff-fermi-potential-RT}
\end{equation}
This last identity is responsible for pointwise saturation and normality on
the RT surface.

\subsection*{Exact vector field and max-flow properties}
\label{app:cutoff-fermi-vector}

With orientation $\mathrm d\lambda\wedge\mathrm d\eta$, the
distance-difference bit-threads field is given by the one-form identity
\begin{equation}
	v_A=\frac{1}{4G}\star\mathrm d\Psi_A.
	\label{eq:cutoff-fermi-form}
\end{equation}
The Hodge star associated with \eqref{eq:cutoff-fermi-metric} satisfies
\begin{equation}
	\star\mathrm d\lambda=\cosh\lambda\,\mathrm d\eta~~,~~
	\star\mathrm d\eta=-\operatorname{sech}\lambda\,\mathrm d\lambda\,.
	\label{eq:cutoff-fermi-Hodge}
\end{equation}
Raising the index in \eqref{eq:cutoff-fermi-form} therefore yields the compact
expression
\begin{equation}
	v_A=\frac{\operatorname{sech}\lambda}{4G}
	\left(-\partial_\eta\Psi_A\,\partial_\lambda
	+\partial_\lambda\Psi_A\,\partial_\eta\right)\,.
	\label{eq:cutoff-fermi-vector-compact}
\end{equation}
The coordinate components are
\begin{align}
	v_A^\mu
	=\frac{1}{8G}\left(\frac{\sinh(\eta+a)}{\Delta_-}
	-\frac{\sinh(\eta-a)}{\Delta_+}\,,\,\tanh\lambda
	\left[
	\frac{\cosh(\eta-a)}{\Delta_+}
	-\frac{\cosh(\eta+a)}{\Delta_-}
	\right]\right)\,.
	\label{eq:cutoff-fermi-v}
\end{align}
where we have defined
\begin{equation}
	\Delta_\pm(\eta,\lambda)
	=\sqrt{\cosh^2\!\lambda\cosh^2(\eta\mp a)-1}
	=\sinh \cL_\pm .
	\label{eq:cutoff-fermi-deltas}
\end{equation}
The second component in \eqref{eq:cutoff-fermi-v} is the feature that is absent from the normal-geodesic
ansatz.  It describes a tangential redistribution of the threads relative to
the RT-normal foliation.

On the RT segment, equations \eqref{eq:cutoff-fermi-potential-RT} and
\eqref{eq:cutoff-fermi-vector-compact} give
\begin{equation}
	v_A\big|_{\lambda=0}
	=\frac{1}{4G}\partial_\lambda,
	\qquad
	|v_A|_{\lambda=0}=\frac{1}{4G}.
	\label{eq:cutoff-fermi-saturation}
\end{equation}
Thus the PEE-selected and normal-geodesic flows agree pointwise at the
bottleneck.  Their difference is entirely off the RT surface.

For comparison, the normal representative is
\begin{equation}
	v_{\rm geo}
	=\frac{1}{4G\cosh\lambda}\partial_\lambda,
	\qquad
	\Psi_{\rm geo}=-\eta.
	\label{eq:cutoff-fermi-normal-comparator}
\end{equation}
It is important that one cannot obtain the PEE-selected field merely by
replacing the scalar coefficient of $\partial_\lambda$ in
\eqref{eq:cutoff-fermi-normal-comparator}: doing so changes the density but
leaves every integral curve geodesic.  The tangential component in
\eqref{eq:cutoff-fermi-v} is essential.

\subsection*{Exact integral curves and their curvature}
\label{app:cutoff-fermi-curves}

Because $v_A\propto\star\mathrm d\Psi_A$, the stream function is
constant along every integral curve.  Let $\eta_\star$ denote the point at
which a curve crosses the RT surface.  Equation
\eqref{eq:cutoff-fermi-potential-RT} fixes its level to
\begin{equation}
	\Psi_A(\eta,\lambda)=-\eta_\star.
	\label{eq:cutoff-fermi-level}
\end{equation}
The implicit distance-difference equation can be solved exactly.  On the
branch that passes through $(\eta_\star,0)$ one finds the following integral curves
\begin{equation}
	\eta(\lambda;\eta_\star)
	=\operatorname{arcsinh}\!\left[
	\frac{\sinh\eta_\star}{\cosh\lambda}
	\sqrt{
		\frac{\cosh^2\!\lambda\cosh^2a-\cosh^2\eta_\star}
		{\sinh^2a-\sinh^2\eta_\star}}
	\right],
	\qquad |\eta_\star|<a\,.
	\label{eq:cutoff-fermi-exact-curve}
\end{equation}
This formula is even in $\lambda$, so the curve meets the RT geodesic
orthogonally.  It is nevertheless not a constant-$\eta$ line unless
$\eta_\star=0$.  The unique bulk geodesic with the same initial point and
normal tangent is $\eta=\eta_\star$; hence
\eqref{eq:cutoff-fermi-exact-curve} gives a particularly direct diagnosis of
the macroscopic non-geodesicity.

Expanding about the bottleneck gives
\begin{equation}
	\eta(\lambda;\eta_\star)
	=\eta_\star+
	\frac{\sinh(2\eta_\star)}
	{4(\sinh^2a-\sinh^2\eta_\star)}\lambda^2
	+O(\lambda^4).
	\label{eq:cutoff-fermi-curve-near-RT}
\end{equation}
At $\lambda=0$ the Fermi connection coefficients that could contribute to
the transverse acceleration vanish.  The magnitude of the geodesic
curvature at the RT crossing is therefore
\begin{equation}
		|k_g|_{\lambda=0}
		=\frac{|\sinh(2\eta_\star)|}
		{2(\sinh^2a-\sinh^2\eta_\star)}\,.
	\label{eq:cutoff-fermi-curvature-RT}
\end{equation}
This is the restriction of the invariant distance-difference curvature to
the saturated surface.  The apparent enhancement near
$|\eta_\star|=a$ reflects the endpoint foci; those points lie on the boundary
of the physical flow domain rather than in its regular interior.

\subsection*{Asymptotic limit and physical interpretation}
\label{app:cutoff-fermi-asymptotic}

For fixed $(\eta,\lambda)$ and $a\gg1$, the potential has the expansion
\begin{equation}
	\Psi_A
	=-\eta+\mathrm e^{-2a}\sinh(2\eta)\tanh^2\!\lambda
	+O(\mathrm e^{-4a}).
	\label{eq:cutoff-fermi-potential-large-a}
\end{equation}
Correspondingly,
\begin{align}
	4Gv_A^\lambda
	&=\operatorname{sech}\lambda
	\left[1-2\mathrm e^{-2a}\cosh(2\eta)
	\tanh^2\!\lambda+O(\mathrm e^{-4a})\right],
	\label{eq:cutoff-fermi-vlambda-large-a}\\
	4Gv_A^\eta
	&=2\mathrm e^{-2a}\sinh(2\eta)
	\tanh\lambda\operatorname{sech}^3\!\lambda
	+O(\mathrm e^{-4a}).
	\label{eq:cutoff-fermi-veta-large-a}
\end{align}
The PEE-selected field thus approaches the geodesic flow
\eqref{eq:cutoff-fermi-normal-comparator} when the two finite foci are pushed
to the ideal endpoints of the complete RT geodesic.  At finite cutoff the
$v_A^\eta$ term shears neighboring flux tubes relative to the normal
foliation.  It changes the boundary-to-RT and horizon-to-RT calibration but does
not create additional flux: both flows continue to saturate the same
bottleneck and carry the same total entropy.

The large-$a$ behavior also reproduces the expected order of the cutoff
correction.  For example, in Poincar\'e AdS,
\begin{equation}
	\mathrm e^{-2a}
	=\frac{z_c^2}{(\sqrt{b^2+z_c^2}+b)^2}
	=\frac{z_c^2}{4b^2}+O\!\left(\frac{z_c^4}{b^4}\right),
	\label{eq:cutoff-fermi-poincare-suppression}
\end{equation}
so the tangential drift and the geodesic curvature begin at order $z_c^2$.

\subsection*{Support and the spectator continuation}
\label{app:cutoff-fermi-spectators}

The Fermi chart also makes precise the status of the additional
complement-to-complement curves that appear in the normal construction.  The
physical bottleneck is only the segment $|\eta|\leq a$ of the complete
geodesic $\lambda=0$.  Nevertheless, the smooth formula
\eqref{eq:cutoff-fermi-normal-comparator} defines a normal congruence for all
$\eta$.  Its curves with $|\eta|>a$ never cross the physical RT segment.
When they intersect the cutoff boundary twice, both endpoints lie in $A^c$.
They are therefore spectator curves: they complete the smooth normal field
but carry no flux out of $A$ and do not contribute to $S_A$.

Keeping the spectators is optional.  Since $\eta=\pm a$ are themselves
streamlines, one may instead use the supported representative
\begin{equation}
	v_{\rm geo}
	=\frac{\Theta(a-|\eta|)}{4G\cosh\lambda}\,\partial_\lambda .
	\label{eq:cutoff-fermi-supported-normal}
\end{equation}
No flux crosses the support boundary, and the distributional derivative of
the step function does not enter the divergence because the field has no
$\eta$ component.  Equation \eqref{eq:cutoff-fermi-supported-normal} is
therefore divergenceless in the weak sense.  The untruncated and truncated
normal fields have identical flux through $A$ and through the RT segment;
they differ only in an entropy-invisible completion outside the active
bundle.

The PEE-selected field has a different organization.  Every regular level
$-a<\Psi_A<a$ of the distance-difference potential intersects the physical RT
segment once, because $\Psi_A(\eta,0)=-\eta$ there.  Its continuation to the
physical boundary---the cutoff boundary, or the horizon for the corresponding
BTZ levels---therefore belongs to the active bundle rather than to a second
PEE sector.  In particular, boundary-returning PEE-selected curves can use the
immediate complementary region left unused by the normal geodesic bundle.
The distinction is a concrete manifestation of max-flow nonuniqueness: RT
saturation fixes the measure at the bottleneck, while the off-bottleneck
landing map and the presence or absence of spectators depend on the chosen
representative.

\subsection*{Specialization to the three cutoff geometries}
\label{app:cutoff-fermi-specializations}

All local formulas above apply after substituting the proper half-length of
the physical RT segment:
\begin{equation}
	a=
	\begin{cases}
		\displaystyle
		s_b^{\rm P}=\operatorname{arcsinh}\!\left(\frac b{z_c}\right),
		&\text{Poincar\'e cutoff},\\[8pt]
		\displaystyle
		s_b^{\rm G}=\operatorname{arcsinh}\!\left(\frac{\sin b}{z_c}\right),
		&\text{global-AdS cutoff, direct branch},\\[8pt]
		\displaystyle
		\sigma_b=\operatorname{arcsinh}\!\left[
		\frac{z_h}{z_c}\sinh\!\left(\frac b{z_h}\right)\right],
		&\text{planar-BTZ cutoff}.
	\end{cases}
	\label{eq:cutoff-fermi-a-specializations}
\end{equation}
The same universal vector field is therefore being expressed in three
different physical truncations of $\mathbb H^2$.  For Poincar\'e and global
AdS the cutoff boundary determines where a given level set
\eqref{eq:cutoff-fermi-level} begins and ends.  In BTZ, the cutoff boundary and horizon
also divide the levels into distinct endpoint sectors.

For the Poincar\'e cutoff, we set
\begin{equation}
	K=b^2+2z_c^2.
	\label{eq:cutoff-fermi-poincare-K}
\end{equation}
A normal geodesic entering at $x\in A$ returns to the cutoff boundary at $y=K/x$.
Hence the active normal bundle does not use the adjacent complementary
strips
\begin{equation}
	b<|y|<\frac{K}{b}.
	\label{eq:cutoff-fermi-poincare-unused-strip}
\end{equation}
The spectator family fills these strips pairwise.  Its limiting member is
tangent to the cutoff boundary at $|x|=\sqrt K$, and its Fermi labels obey
\begin{equation}
	s_b^{\rm P}<|\eta|<\eta_{\rm tan}^{\rm P},
	\qquad
	\eta_{\rm tan}^{\rm P}
	=\operatorname{arcsinh}\!\left(
	\frac{\sqrt{b^2+z_c^2}}{z_c}\right).
	\label{eq:cutoff-fermi-poincare-spectator-band}
\end{equation}
Both the unused strip and this finite label interval collapse to the ideal
endpoint as $z_c/b\to0$.

In the BTZ case, the conserved label $\eta_\star=-\Psi_A$ identifies the RT element crossed
by a PEE-selected streamline.  The endpoint measure fixes
\begin{equation}
	|\eta_\star|<\frac{b}{z_h}: \text{horizon ending},
	\qquad
	\frac{b}{\zh}<|\eta_\star|<\sigma_b: \text{cutoff-boundary returning}.
	\label{eq:cutoff-fermi-BTZ-sectors}
\end{equation}
Accordingly, the two PEE separatrices are obtained simply by inserting
$\eta_\star=\pm b/\zh$ into the exact curve
\eqref{eq:cutoff-fermi-exact-curve}.  Although those curves bend away from
the RT surface, their conserved label cannot change, so the endpoint-sector
assignment is stable throughout the flow.

The independent normal-geodesic representative instead follows
$\eta=\text{constant}$.  Its horizon threshold is
\begin{equation}
	\eta_N
	=\operatorname{arccosh}\!\left[
	\frac{\cosh \left(\frac{b}{\zh}\right)}{\sqrt{1-q^2}}\right],
	\label{eq:cutoff-fermi-normal-threshold}
\end{equation}
with the physical normal horizon segment bounded by
$|\eta|<\min(\sigma_b,\eta_N)$.  At nonzero cutoff one has
$\eta_N>b/\zh$ whenever both cutoff-boundary-returning and horizon-ending normal sectors
are present.  The normal field therefore assigns a larger portion of the
same RT bottleneck to the horizon than the PEE-selected field.  This is not
an entropy mismatch: it is an explicit demonstration that maximality fixes
the total bottleneck flux but not its endpoint-resolved decomposition.

The finite-cutoff BTZ normal congruence has a spectator band as well.  The two boundary intersections $x$ and $x'$ of a constant-$\eta$ normal obey
\begin{equation}
	\tanh\left(\frac{x}{2\zh}\right)\,\tanh \left(\frac{x'}{2\zh}\right) =\frac{\cosh \left(\frac{b}{\zh}\right)-(1-q^2)}{\cosh \left(\frac{b}{\zh}\right)+(1-q^2)}.
	\label{eq:cutoff-fermi-BTZ-normal-partner}
\end{equation}
The normal through the right RT endpoint $x=b$ therefore returns at
\begin{equation}
	x_{\rm gap}=2\zh\operatorname{arctanh}\!\left[
	\frac{\cosh \left(\frac{b}{\zh}\right)-(1-q^2)}{\cosh \left(\frac{b}{\zh}\right)+(1-q^2)}\coth\left(\frac{b}{2\zh}\right)\right]>b.
	\label{eq:cutoff-fermi-BTZ-gap}
\end{equation}
The interval $b<x<x_{\rm gap}$, and its reflection on the left, is not used
by the active normal geodesic bundle. Spectator geodesics have both endpoints in
these strips and meet at the boundary-tangent curve
\begin{align}
	x_{\rm tan}
	&=\zh\operatorname{arccosh}\!\left[\frac{\cosh \left(\frac{b}{\zh}\right)}{1-q^2}\right],
	\nonumber\\
	\sigma_b<|\eta|<\eta_{\rm tan}^{\rm BTZ},
	\qquad
	\eta_{\rm tan}^{\rm BTZ}
	&=\operatorname{arctanh}\!\sqrt{
		\frac{\sinh^2\left(\frac{b}{\zh}\right)+q^2}{\sinh^2\left(\frac{b}{\zh}\right)+2q^2-q^4}}.
	\label{eq:cutoff-fermi-BTZ-spectator-band}
\end{align}
These curves neither end on the horizon nor cross the RT segment.  They are
the BTZ counterpart of the Poincar\'e complement-to-complement completion,
not an additional thermal PEE contribution.  As $q\to0$,
$x_{\rm gap}\to b$ and the spectator strip disappears.

In the limit $z_c/z_h\to0$, both $\sigma_b$ and the endpoint distance from a
fixed interior RT point become large, while
$\eta_N\to b/\zh$.  Equations
\eqref{eq:cutoff-fermi-potential-large-a}--\eqref{eq:cutoff-fermi-veta-large-a}
then show that the PEE and normal fields coincide locally, and their BTZ
separatrices agree.  This provides a single Fermi-coordinate explanation of
both the finite-cutoff distinction and its disappearance at the asymptotic
boundary.

Because the derivation uses only the local hyperbolic metric and the two
endpoint foci, the same finite-focus mechanism extends to mixed
finite--ideal endpoint configurations and to geodesic boundaries.  Those
applications alter the global truncation and the admissible endpoint
support, but not the universal field
\eqref{eq:cutoff-fermi-vector-compact}; the physical interpretation of the
resulting sectors must be fixed by the chosen boundary condition or
purification.

\nocite{BasuWen:BCFTMinPurification}
\bibliographystyle{JHEP}
\bibliography{Threads-arXiv}

\end{document}